\documentclass[prd,twocolumn,nofootinbib,superscriptaddress,amsmath,amssymb,eqsecnum]{revtex4-1}
\usepackage[utf8]{inputenc}
\usepackage[T1]{fontenc}
\usepackage{mathtools}
\usepackage{graphics}
\usepackage{graphicx}
\usepackage{dcolumn}
\usepackage{bm}
\usepackage{dsfont} 
\usepackage{amsmath,amssymb}
\usepackage{hyperref}
\usepackage{tabularx}
\usepackage{epstopdf}
\usepackage{tensor}
\usepackage{mathrsfs}
\usepackage[normalem]{ulem}
\usepackage[usenames]{color}
\hypersetup{
    colorlinks=true,
    linkcolor=blue,
    filecolor=magenta,      
    urlcolor=blue,
    citecolor=blue
}
\usepackage[dvipsnames]{xcolor}
\usepackage{mathrsfs}
\usepackage{microtype}

\newcommand\be{\begin{equation}}
\newcommand\ba{\begin{eqnarray}}
\newcommand\ee{\end{equation}}
\newcommand\ea{\end{eqnarray}}
\newcommand\bw{\begin{widetext}}
\newcommand\ew{\end{widetext}}
\newcommand{\lb}{\left(}
\newcommand{\rb}{\right)}
\newcommand{\ETH}{\text{\DH}}

\newcommand{\UVA}{Department of Physics, University of Virginia, P.O.~Box 400714, 382 McCormick Road, Charlottesville, VA 22904-4714, USA}

\begin{document}
\allowdisplaybreaks

\title{Brans-Dicke theory in Bondi-Sachs form: Asymptotically flat solutions, asymptotic symmetries and gravitational-wave memory effects}

\author{Shammi Tahura}
\affiliation{\UVA}

\author{David A.~Nichols}
\email{david.nichols@virginia.edu}
\affiliation{\UVA}

\author{Alexander Saffer}
\affiliation{\UVA}

\author{Leo C. Stein}
\affiliation{
 Department of Physics and Astronomy, The University of Mississippi, University, MS 38677, USA
}

\author{Kent Yagi}
\affiliation{\UVA}

\date{\today}

\begin{abstract} 

Gravitational-wave memory effects are identified by their distinctive effects on families of freely falling observers: after a burst of waves pass by their locations, memory effects can cause lasting relative displacements of the observers. 
These effects are closely related to the infrared properties of gravity, including its asymptotic symmetries and conserved quantities. In this paper, we investigate the connection between memory effects, symmetries, and conserved quantities in Brans-Dicke theory. We compute the field equations in Bondi coordinates, and we define a set of boundary conditions that represent asymptotically flat solutions in this context. Next, we derive the asymptotic symmetry group of these spacetimes, and we find that it is the same as the Bondi-Metzner-Sachs group in general relativity. Because there is an additional polarization of gravitational waves in Brans-Dicke theory, we compute the memory effects associated with this extra polarization (the  so-called ``breathing'' mode). 
This breathing mode produces a uniform expansion (or contraction) of a ring of freely falling observers.
After these breathing gravitational waves pass by the observers' locations, there are two additional memory effects that depend on their initial displacements and relative velocities. Neither of these additional memory effects seems to be related to asymptotic symmetries or conserved quantities; rather, they are determined by the properties of the nonradiative region before and after the bursts of the scalar field and the gravitational waves. We discuss the properties of these regions necessary to support nontrivial breathing-mode-type memory effects.

\end{abstract}

\maketitle

\section{Introduction} \label{sec:intro}

Since the first detection of gravitational waves (GWs) in 2015 from the merger of a binary black hole~\cite{Abbott:2016blz}, ten additional mergers of compact objects were discovered during the first two observing runs of LIGO and Virgo~\cite{LIGOScientific:2018mvr}.
During the third observing runs of LIGO and Virgo, compact-binary-merger candidates were announced at a rate of roughly one per week~\cite{GraceDB}.
These discoveries, and the rapid announcement of GW candidates, have opened the new field of GW astronomy.
Along with the discoveries, numerous tests of gravity with GWs have been performed to determine the consistency of the observed gravitational waves with the predictions of general relativity (see, e.g.,~\cite{Abbott_IMRcon2,Yunes:2016jcc,TheLIGOScientific:2016pea,Abbott:2018lct,LIGOScientific:2019fpa,Berti:2018cxi,Berti:2018vdi}).
Compact binary mergers opened a new parameter space of general relativity to be tested (the region of strong curvature and high GW luminosities) which was less well probed by tests of general relativity in the Solar System or with binary pulsars. In this parameter space, there are some types of relativistic phenomena that are only likely to be measured for strongly curved and highly radiating systems.
One such class of effects that has yet to be detected, but are under active investigation (see, e.g.,~\cite{Hubner:2019sly,Ebersold:2020zah,Boersma:2020gxx,Aggarwal:2019ypr,Islo:2019qht}), are gravitational-wave memory effects.

The best known GW memory effect (sometimes referred to as \textit{the} GW memory effect) is characterized by lasting change in the GW strain after a burst of GWs pass by a GW detector. 
One of the earliest explicit calculations of the GW memory effect from gravitational scattering was performed in Ref.~\cite{Zeldovich:1974gvh} (see also \cite{1979ApJ,Smarr1977}), though the possibility of a nonvanishing GW strain at late times was discussed previously (e.g.,~\cite{Newman:1966ub}).
It was subsequently noted that massless (or nearly massless) fields could also produce the GW memory  effect~\cite{1978Natur.274..565T,1978ApJ...223.1037E} including the nonlinear effective stress-energy of gravitational waves themselves~\cite{PhysRevLett.67.1486,Blanchet:1992br}.
The GW memory effect has a distinctive observational signature, in that it causes a constant, enduring displacement between nearby freely falling observers after a burst of gravitational waves have passed.
A number of generalizations of the GW memory effect have been found by considering asymptotic changes in burst of other fields (such as electromagnetism~\cite{Bieri:2013hqa} or massless Yang-Mills theory~\cite{Pate:2017vwa}) or in time integrals of the GW strain (e.g.,~\cite{Pasterski:2015tva,Nichols:2018qac}).
Other GW memories have been found from examining other kinds of lasting kinematical effects on freely falling observers (like lasting relative velocities~\cite{Bondi:1957dt,Grishchuk:1989qa}, relative changes in proper time~\cite{Strominger:2014pwa,Flanagan:2014kfa}, relative rotations of parallel transported tetrads~\cite{Flanagan:2014kfa}) or through other types of measurement procedures~\cite{Flanagan:2016oks,Flanagan:2018yzh}.
Also important in the discovery of new GW memory effects was the understanding of how certain GW memories are closely related to symmetries, conserved quantities, and soft theorems (see, e.g.,~\cite{Strominger:2017zoo}).

For understanding the relationship between memory effects and the asymptotic structure of spacetime, two approaches have been taken to study asymptotic flatness: a covariant conformal completion of spacetime~\cite{Penrose:1962ij,Penrose:1965am} and calculations in particular coordinate systems adapted to the spacetime geometry by Bondi, van der Burg, and Metzner~\cite{Bondi:1962px} and Sachs~\cite{Sachs:1962wk} or Newman and Unti~\cite{Newman:1962cia}.
We will focus on the Bondi-Sachs approach to asymptotic flatness.
In this approach, coordinates are chosen that are well suited to the null hypersurfaces and the null geodesics of the spacetime. 
Boundary conditions can then be imposed on the metric to determine a reasonable notion of a spacetime that becomes asymptotically Minkowskian as the light rays travel an infinite distance from an isolated source. 
Although spacetime can be cast in an asymptotically Minkowskian form at large Bondi radius $r$, the asymptotic symmetry group of this spacetime does not reduce to the Poincar\'e group of flat spacetime; rather, it becomes the infinite-dimensional Bondi-Metzner-Sachs (BMS) group~\cite{Bondi:1962px,PhysRev.128.2851}.

The structure of the BMS group is in some ways similar to the Poincar\'e group: it contains the Lorentz transformations, but rather than having an additional four spacetime translations as the remaining group elements, it has an infinite-dimensional commutative group called the supertranslations~\cite{PhysRev.128.2851} (the usual Poincar\'{e} translations are a normal finite subgroup of the supertranslations).
It is possible to associate charges conjugate to these asymptotic symmetries (see, e.g.,~\cite{Geroch:1977jn,Ashtekar:1981bq,Geroch:1981ut,Wald:1999wa}).
These charges are conserved in the sense that the difference in the charges between two times is equal to the flux of the quantity between these two times.
Associated with the Lorentz symmetries are the six components of the relativistic angular momentum [which can be divided into center-of-mass (CM) and spin parts] and corresponding to the supertranslations are conserved quantities called supermomenta.
Note that there also have been proposals to extend the Lorentz part of the symmetry algebra to include all conformal Killing vectors on the 2-sphere called superrotations~\cite{Barnich:2009se,Barnich:2010eb,Barnich:2011mi} (see also~\cite{Banks:2003vp}) or all smooth vector fields on the 2-sphere~\cite{Campiglia:2014yka,Campiglia:2015yka} (sometimes called super-Lorentz symmetries~\cite{Compere:2018ylh}). 
The additional charges of these extended BMS algebras are the super CM and super spin charges~\cite{Flanagan:2015pxa} or the super-angular momentum~\cite{Nichols:2018qac}. 

The connection between asymptotic symmetries, conserved quantities, and GW memory can now be more clearly stated with the nomenclature now set.
Changes in the supermomentum charges, generated by both massive particles and massless fields, induce a nonzero GW memory effect; in addition, when the GW memory effect is present, the final state of the system is supertranslated from a certain canonical asymptotic rest frame for the system (see, e.g.,~\cite{Flanagan:2015pxa}).
Changes in the super-angular momentum charges can induce two additional types of GW memory effects called spin~\cite{Pasterski:2015tva} and CM~\cite{Nichols:2018qac} memory.
These memory effects are not necessarily related to a spacetime that has been superrotated or super-Lorentz transformed from a certain canonical frame, since such solutions often are not asymptotically flat in the usual sense~\cite{Strominger:2016wns,Compere:2018ylh}.

While GW memory effects and their analogues for other matter fields have now been much more carefully studied in a number of contexts, they have not been studied as systematically in modified theories of gravity.
Modified theories can have additional GW polarizations~\cite{Will:2014kxa,Chatziioannou:2012rf,Zhang:2019iim}, which could allow for additional types of GW memory effects (see, e.g.,~\cite{Lang:2013fna,Lang:2014osa,Du:2016hww,Koyama:2020vfc}).
In addition, as far as we are aware, there is not a standard definition of asymptotic flatness in these theories, nor is the set of asymptotic symmetries of these solution clearly understood.
It is not obvious, \textit{a priori}, that modified theories of gravity generically have the same asymptotic properties as in general relativity, or that their memory effects would be related to symmetries and conserved quantities as in general relativity.
A main aim of this paper is to develop a better understanding of these relationships in a relatively simple modification of general relativity known as Brans-Dicke theory~\cite{Brans:1961sx}.

Brans-Dicke theory is one example of a scalar-tensor theory, i.e., a theory in which there is a scalar field that couples to gravity nonminimally (see, e.g., the review~\cite{Berti:2015itd}).
Scalar-tensor theories have appeared in the contexts of string theory, inflation~\cite{Clifton:2011jh,Barrow:1990nv}, and the accelerated expansion of the Universe~\cite{Brax:2004qh,Baccigalupi:2000je,Riazuelo:2001mg}. 
In this paper, we will focus on Brans-Dicke theory, with a massless scalar field. 
It is known from calculations in linearized gravity and post-Newtonian (PN) theory, the scalar field generates an additional polarization of gravitational waves sometimes called a ``breathing mode''~\cite{Will:1989sk,Will:1994fb,Will:2014kxa} (it produces a transverse uniform expansion and contraction of a ring of freely falling test masses). 
It was also noted (from the 2PN calculation in~\cite{Lang:2013fna,Lang:2014osa}) that the GW memory effect differs in scalar-tensor theory from in general relativity.\footnote{Specifically, the energy radiated from the dipole moment of the scalar field gives rise to a formally 1.5PN-order effect in the tensor gravitational waveform that would appear at Newtonian order in the waveform for nonspinning compact binaries, which are inspiraling because of the emission of dipole radiation.
This is analogous to how the energy radiated in gravitational waves gives rise to a 2.5PN-order effect that appears at Newtonian order in the waveform for nonspinning compact-binary sources in GR~\cite{Wiseman:1991ss,Favata:2008yd}.
Because stationary black holes in Brans-Dicke theory do not support scalar fields~\cite{Hawking:1972qk,Sotiriou:2011dz}, the compact binary can have at most one black hole to have this new scalar-dipole-sourced GW memory effect.}
It was also shown in~\cite{Lang:2014osa} that the scalar, breathing polarization of the GWs does not have a nonlinear-type memory effect at 2PN order.
Finally, it was observed that there is a new type of nonhereditary, nonlinear term in the tensor waveform arising from the scalar field that took on an analogous form to the nonhereditary and nonoscillatory term found in~\cite{Arun:2004ff} (and discussed in~\cite{Favata:2008yd}), which was shown to be related to the spin memory effect in~\cite{Nichols:2017rqr}.
Our calculations in Brans-Dicke theory in Bondi-Sachs coordinates allow us to compute the memory effects using the fully nonlinear field equations.
This will provide us with the framework to understand the presence (and absence) of the memory effects computed at 2PN order in~\cite{Lang:2013fna,Lang:2014osa} (though we leave the explicit calculations for future work) and to determine the relevant radiative and nonradiative data needed to compute these effects.

Scalar-tensor theories are frequently studied in two different conformal frames, called the Jordan and Einstein frames, respectively.
In this paper, we find that the Einstein frame is more convenient for determining the asymptotic boundary conditions on the scalar field and metric, because the field equations have the same form as the Einstein-Klein-Gordon equations for a massless scalar field.
The statement of stress-energy conservation is more complicated in the Einstein frame, however, because the stress-energy tensor of all matter fields besides the scalar field is no longer divergence free, but equals a nontrivial right-hand side involving gradients of the scalar field.
Consequently, test particles follow accelerated curves in the Einstein frame (with an acceleration related to the gradients of the scalar field in this frame) rather than following the geodesics of the Einstein-frame metric.
In the Jordan frame, the modified Einstein equations are more complicated than in the Einstein frame, but the stress-energy tensor of all matter fields besides the scalar field is divergence free, and thus test particles follow the geodesics of the Jordan-frame metric.
It is therefore much simpler to compute the response of a gravitational-wave detector to any impinging gravitational waves in the Jordan frame. 
Flanagan~\cite{Flanagan:2004bz} has argued that all classical physical predictions (such as gravitational-wave memory effects) are conformal-frame invariants. 
This allows us to compute the memory effects in the Jordan frame, in which the computation is simpler, but to obtain a result that is independent of the choice of conformal frame (after properly identifying any potentially different conventions between the frames, as discussed further in~\cite{Flanagan:2004bz}).

The rest of the paper is organized as follows: In Sec.~\ref{sec:framework}, we describe the conditions we use to define  asymptotic flatness in Brans-Dicke theory, by examining the theory in both Einstein and Jordan frames~\cite{Fujii:2003pa}. 
This includes deriving the field equations of the theory in Bondi-Sachs coordinates. 
In Sec.~\ref{sec:symmetries}, we compute the asymptotic symmetries that preserve our definition of asymptotic flatness in the previous part.
We describe how the functions in the metric must transform to maintain the Bondi gauge conditions and the asymptotically flat boundary conditions.
In Sec.~\ref{sec:memory}, we describe how the memory effects can be measured through geodesic deviation and how the changes in the charges related to (extended) BMS symmetries constrain the different GW memory effects in Brans-Dicke theory.
We discuss our results and some future directions in Sec.~\ref{sec:conclusions}.

Throughout this paper, we use units in which $c=1$, and we use the conventions for the metric and curvature tensors given in~\cite{Misner:1974qy}.
Greek indices ($\mu,\nu,\alpha, \dots$) represent four-dimensional spacetime indices, and uppercase Latin indices $(A, B, C, \dots)$ represent indices on the 2-sphere.
Indices with a circumflex diacritic (e.g., $\hat \alpha$) represent those of an orthonormal tetrad.

While we were completing this work, there appeared a closely related pre-print~\cite{Hou:2020tnd} investigating asymptotically flat solutions and GW memory effects in scalar-tensor theories. 
Our work and that of~\cite{Hou:2020tnd} agree in the boundary conditions used to define asymptotically flat solutions in Brans-Dicke theory and the leading-order symmetry vectors that preserve these conditions and our gauge choices (though not subleading corrections to extend these symmetries into the spacetime).
Our works differ in the choices of gauges, the classes of spacetimes in which we compute memory effects, and the procedures by which we compute the scalar-type memory effect. 
We will comment in more detail on the similarities and differences between our works at a few points throughout the text.

\section{Bondi-Sachs Framework}\label{sec:framework}

In this section, we impose the Bondi-Sachs coordinate conditions in Brans-Dicke theory, and we solve the field equations in both the Einstein and the Jordan frames. 
We begin with the Einstein frame, where it is easier to identify a set of asymptotic boundary conditions that can be imposed on the scalar field and on the metric that we use to define an asymptotically flat solution in Brans-Dicke theory. 
We next perform conformal transformation to the Jordan frame (in which the stress-energy tensor of all other matter fields besides the scalar field is divergence free), and we find the corresponding boundary conditions on the scalar field and metric. We then solve the field equations of Brans-Dicke theory in this frame. 
Our notation and conventions for the Bondi-Sachs framework will parallel the ones used in Ref.~\cite{Madler:2016xju}, which treats general relativity.

\subsection{Einstein frame}

We begin by investigating the Brans-Dicke theory in the Einstein frame. 
The action in the Einstein frame in the absence of additional matter fields is given by~\cite{Damour:1992we}
\ba
S=\int d^{4} x \sqrt{-\tilde{g}}\left[\frac{\tilde{R}}{16 \pi}-\frac{1}{2} \tilde{g}^{\rho \sigma}\left(\tilde{\nabla}_{\rho} \Phi\right)\left(\tilde{\nabla}_{\sigma} \Phi\right)\right]\,,
\ea
where $\tilde g$ is the metric in the Einstein frame, $\tilde R$ is the Ricci scalar and $\Phi$ is a real scalar field. 
We also use units where the gravitational constant in the Einstein frame $G_E$ satisfies $G_E=1$.
We use $\tilde \nabla_\mu$ to denote the covariant derivative compatible with $\tilde g_{\mu\nu}$. 
Varying the action with respect to the metric and the scalar field leads to the following equations of motion for the theory:
\begin{subequations}
\ba \label{eq:Einstein-field}
\tilde{\mathcal E}_{\mu\nu} \equiv \tilde R_{\mu\nu}-\frac{1}{2}\tilde R \tilde g_{\mu\nu}-8\pi \tilde{T}_{\mu \nu}^{(\Phi)}&=& 0\,,
\\
\label{eq:Scalar-Wave}
\tilde \nabla_{\mu}\tilde \nabla^{\mu} \Phi&=&0\, .
\ea
The quantity $\tilde{T}_{\mu \nu}^{(\Phi)}$ is the stress-energy tensor for the scalar field, which is given by
\ba \label{eq:Einstein-stress-energy}
\tilde{T}_{\mu \nu}^{(\Phi)} = \tilde{\nabla}_{\mu} \Phi \tilde{\nabla}_{\nu} \Phi-\tilde{g}_{\mu \nu}\left[\frac{1}{2} \tilde{g}^{\rho \sigma} \tilde{\nabla}_{\rho} \Phi \tilde{\nabla}_{\sigma} \Phi\right]\,.
\ea
\end{subequations}
The field equations, therefore, have the same form as in Einstein-Klein-Gordon theory for a real scalar field $\Phi$, so
their solutions will also have the same form as in Einstein-Klein-Gordon theory in general relativity.
We will review the solutions of these equations in Bondi coordinates next.

\subsubsection{Bondi gauge and field equations}
\label{subsubsec:BondiEquations}

First, we introduce Bondi-Sachs coordinates $\tilde x^\mu = (\tilde u,\tilde r, \tilde x^A)$.
The quantity $\tilde u$ is the retarded time, $\tilde r$ is an areal coordinate (and $\vec \partial_{\tilde r}$ is a null vector field), and $\tilde x^A$ are coordinates on the 2-sphere (with $A=1,2$)~\cite{Bondi:1962px,Madler:2016xju}. 
The conditions that define Bondi gauge are given by~\cite{Bondi:1962px,Madler:2016xju}
\be 
\tilde g_{\tilde r A}=\tilde g_{\tilde r \tilde r}=0, \quad \det\left[\tilde g_{A B}\right] = \tilde r^{4} q\left(\tilde x^{C}\right)\,.
\ee
The function $q$ is the determinant of a metric on the 2-sphere, $q_{AB}(x^C)$, which is restricted to be independent of $\tilde u$ and $\tilde r$.
The Bondi gauge conditions fix four of the ten functions in the metric, leaving six free functions.
It is conventional to write these six degrees of freedom as follows:
\begin{align}
\tilde g_{\mu \nu} d \tilde x^{\mu} d \tilde x^{\nu}=&-\frac{\tilde V}{\tilde r} e^{2 \tilde \beta} d \tilde u^{2}-2 e^{2 \tilde \beta} d \tilde u d \tilde r \nonumber \\
&+\tilde r^{2} \tilde h_{A B}\left(d \tilde x^{A}-\tilde U^{A} d \tilde u\right)\left(d \tilde x^{B}-\tilde U^{B} d \tilde u\right)\, .
\label{eq:Bondi_metric}
\end{align}
The functions $\tilde V$, $\tilde \beta$, $\tilde U^A$ and $\tilde h_{AB}$ here depend on all four Bondi coordinates $\tilde x^\mu = (\tilde u,\tilde r, \tilde x^A)$. 

The modified Einstein equations \eqref{eq:Einstein-field} and the scalar-field equation~\eqref{eq:Scalar-Wave} satisfy an interesting hierarchy in Bondi coordinates~\cite{Bondi:1962px,Madler:2016xju}, which we will now further elaborate.
The functions $\tilde V$, $\tilde \beta$, and $\tilde U^A$ satisfy the so-called ``hypersurface equations.'' 
The equations were given this name because they do not involve derivatives with respect to $\tilde u$, which in turn allows the functions $\tilde V$, $\tilde \beta$, and $\tilde U^A$ to be determined on surfaces of constant $\tilde u$ in terms of the 2-metric $\tilde h_{AB}$, the scalar field $\Phi$, and ``functions of integration'' (i.e., functions of $\tilde u$ and $\tilde x^A$ that will be constrained by $\tilde u\tilde u$ and $\tilde u\tilde A$ components of the modified Einstein equations).
The concrete form of the hypersurface equations can be obtained from substituting the metric~\eqref{eq:Bondi_metric} into the field equations in Eq.~\eqref{eq:Einstein-field}, using the definition of the stress-energy tensor in Eq.~\eqref{eq:Einstein-stress-energy}, and considering the appropriate components of the modified Einstein equations.
The $\tilde r \tilde r$ component yields the equation 
\begin{subequations}
\begin{align}
    \partial_{\tilde r} \tilde \beta - \frac{\tilde r}{16}\tilde h^{AC} \tilde h^{BD} \partial_{\tilde r} \tilde h_{AB} \partial_{\tilde r} \tilde h_{CD} = {} & 2\pi \tilde r \partial_{\tilde r} \Phi \partial_{\tilde r} \Phi\,.
    \label{eq:Einstein-Field-1}
\end{align}
where $\tilde h^{AB}$ is the inverse of $\tilde h_{AB}$.  
Once $\tilde \beta$ is determined in terms of $\tilde h_{AB}$ (and its inverse), $\Phi$, and their derivatives, then it is also possible to use the $\tilde r A$ components of the field equations to solve for $\tilde U^A$ in terms of the same quantities from the following equation:
\begin{align}
    \partial_{\tilde r}\left[\tilde r^4 e^{-2 \tilde \beta} \tilde h_{AB}\partial_{\tilde r} \tilde U^{B}\right] -2 \tilde r^4\partial_{\tilde r} \lb\frac{1}{\tilde r^2}\tilde D_A \tilde \beta \rb & \nonumber \\
    +\tilde r^2\tilde h^{BC}\tilde D_B \partial_{\tilde r} \tilde h_{AC} - 16\pi \tilde r^2\partial_{\tilde r} \Phi \partial_A \Phi & = 0 \, .
\end{align}
where $\tilde D_{A}$ is the covariant derivative compatible with the 2-metric $\tilde h_{AB}$. Finally, from the trace of the $AB$ components of the field equations, it is then possible to solve for $\tilde V$ in terms of the same data:
\begin{align}
    \label{eq:Einstein-Field-2}2 e^{-2 \tilde \beta}\left(\partial_{\tilde r} \tilde V\right)-\tilde{\mathscr{R}} -\frac{e^{-2 \tilde \beta}}{\tilde r^{2}} \tilde D_{A}\left[\partial_{\tilde r}\left(\tilde r^{4} \tilde U^{A}\right)\right] & \nonumber\\
    + 2\tilde h^{A B}\left[\tilde D_{A}\tilde  D_{B} \tilde \beta-\left(\tilde D_{A}\tilde \beta\right)\left(\tilde D_{B} \tilde \beta\right)\right] & \nonumber \\
    +\frac{1}{2} \tilde r^{4} e^{-4 \tilde \beta} \tilde h_{A B}\left(\partial_{\tilde r} \tilde U^{A}\right)\left(\partial_{\tilde r} \tilde U^{B}\right) - 8\pi \tilde h^{AB}\partial_A \Phi \partial_B \Phi & = 0\,.
\end{align}
Here $\tilde{\mathscr{R}}$ is the Ricci scalar of 2-metric $\tilde h_{AB}$. 
The remaining two independent components of the modified Einstein equations are called the evolution equations, and they arise from the trace-free (with respect to $h_{AB}$) part of $\tilde{\mathcal E}_{AB}$.
It is convenient to write this expression using a complex polarization dyad composed of $\tilde m^A = {\delta^A}_\mu \tilde m^\mu$ (which satisfies $\tilde m^{\mu}\tilde \nabla_{\mu}\tilde u=0$) and $\bar{\tilde m}^A$ (the complex conjugate of $\tilde m^A$). 
The evolution is given by $\tilde m^A \tilde m^B \tilde{\mathcal E}_{AB}=0$, which can be written in terms of the metric functions as
\bw
\ba
\label{eq:GAB_Ein}
\tilde m^{A} &  \tilde m^{B}&\left\{\tilde r \partial_{\tilde r}\left[\tilde r\left(\partial_{\tilde u} \tilde h_{A B}\right)\right]-\frac{1}{2} \partial_{\tilde r}\left[\tilde r \tilde V\left(\partial_{\tilde r} \tilde h_{A B}\right)\right]-2 e^{\tilde \beta} \tilde D_{A} \tilde D_{B} e^{\tilde \beta} 
+\tilde h_{C A} \tilde D_{B}\left[\partial_{\tilde r}\left(\tilde r^{2} \tilde U^{C}\right)\right]\right.\nonumber\\
&-&\frac{1}{2} \tilde r^{4} e^{-2 \tilde \beta} \tilde h_{A C} \tilde h_{B D}\left(\partial_{\tilde r} \tilde U^{C}\right)\left(\partial_{\tilde r} \tilde U^{D}\right)
+\frac{\tilde r^{2}}{2}\left(\partial_{\tilde r} \tilde h_{A B}\right)\left(\tilde D_{C} \tilde U^{C}\right)+\tilde r^{2} \tilde U^{C} \tilde D_{C}\left(\partial_{\tilde r} \tilde h_{A B}\right) \nonumber\\
&-&\left. \tilde r^{2}\left(\partial_{\tilde r} \tilde h_{A C}\right) \tilde h_{B E}\left(\tilde D^{C} \tilde U^{E}-\tilde D^{E} \tilde U^{C}\right)-8 \pi e^{2 \tilde \beta} \partial_A\Phi \partial_B\Phi\right\}=0\,.
\ea
\ew
We will discuss the evolution equations in more detail in Sec.~\ref{sec:Jordan} on the Jordan frame.

In vacuum general relativity, once the metric functions $\tilde\beta$, $\tilde U^A$ and $\tilde V$ are determined on a hypersurface of constant $\tilde u$, they can be used in the evolution equation for the 2-metric $\tilde h_{AB}$, to evolve $h_{AB}$ to the next hypersurface; the hypersurface equations can then be solved again in an iterative process. In Brans-Dicke theory, however, one must jointly evolve the evolution equation for $\tilde h_{AB}$ with the scalar field equation to obtain the data $\tilde h_{AB}$ and $\Phi$ needed to solve the hypersurface equations.
For convenience, we give the scalar wave equation~\eqref{eq:Scalar-Wave} when written in terms of the Bondi metric functions below:
\begin{align} \label{eq:scalar-wave-einstein}
& 2\partial_{\tilde u} \partial_{\tilde r} \Phi + \tilde D_{A}(\tilde U^{A} \partial_{\tilde r}\Phi) + \partial_{\tilde r}(\tilde U^{A} \tilde D_{A} \Phi)  \nonumber \\
& - \frac{1}{\tilde r} \left(-2\tilde  U^{A} \tilde D_{A} \Phi-2\partial_{\tilde u} \Phi
+ \partial_{\tilde r} \tilde V\partial_{\tilde r} \Phi+\tilde V\partial_{\tilde r} \partial_{\tilde r} \Phi\right) \nonumber \\
& -\frac{1}{\tilde r^2}\left[e^{2 \tilde \beta} \tilde h^{AB}\left(2\tilde D_A \tilde \beta \tilde D_B\Phi+\tilde D_B \tilde D_A\Phi\right) + \tilde V \left(\partial_{\tilde r} \Phi\right)\right] = 0 \, .
\end{align}
\end{subequations}
Aside from the additional complication that the scalar-wave equation and evolution equation for $\tilde h_{AB}$ must be solved as a coupled system, the form and the hierarchy of the modified Einstein and scalar field equations in the Einstein frame is similar to that of the Einstein equations in vacuum general relativity.

\subsubsection{Conditions for asymptotic flatness}

We next study the asymptotic behavior of the metric and the scalar field at large Bondi radius $r$.
Because $\tilde V$, $\tilde \beta$, and $\tilde U^A$ are determined by $\tilde h_{AB}$ and $\Phi$, we must posit boundary conditions on $\tilde h_{AB}$ and $\Phi$; we can then deduce the remaining conditions on the metric from the hypersurface equations~\eqref{eq:Einstein-Field-1}--\eqref{eq:Einstein-Field-2} up to functions of integration.
There are well-established definitions for asymptotic flatness for the Einstein equations~\cite{Bondi:1962px,Sachs:1962wk}.
For the scalar field, we will assume that it satisfies the following scaling as $\tilde r\rightarrow\infty$:
\be\label{eq:Phi-fall-off}
\Phi\left(\tilde u, \tilde r, \tilde x^{A}\right) = \Phi_{0} + \frac{\Phi_{1}\left(\tilde u, \tilde x^{A}\right)}{\tilde r} + O(\tilde r^{-2}) \, ,
\ee
where $\Phi_0$ is a constant.\footnote{With the expansion of $\tilde h_{AB}$ in Eq.~\eqref{eq:hAB fall-off} and with a polynomial expansion of $\tilde \beta, \tilde U^{A}$ and $\tilde V$ in $\tilde r^{-1}$ consistent with Eq.~\eqref{eq:boundary-einstein}, one can prove from the $\tilde r^{-1}$ piece of Eq.~\eqref{eq:scalar-wave-einstein} that $\Phi_0$ is independent of $\tilde u$; from the $\tilde r^{-2}$ piece of Eq.~\eqref{eq:scalar-wave-einstein}, one can show that $\Phi_0$ is independent of $\tilde x^A$.}

In GR, the action for a massless scalar field (and hence the stress-energy tensor and equations of motion) is independent of the value of $\Phi_0$ in Eq.~\eqref{eq:Phi-fall-off}. 
Thus, there is no loss of generality by requiring that the constant value of the scalar field is zero.
The boundary condition on the massless scalar field as $r$ goes to infinity can then be given by ``Sommerfeld's radiation condition'': $\lim_{r\rightarrow\infty} r\Phi$ is finite.
In Brans-Dicke theory, the constant value of the scalar field is related to the asymptotic value of Newton's constant $G$. 
Setting $\Phi_0$ to zero, therefore, does have a physical effect in Brans-Dicke theory (note, however, that the precise value of the constant does not affect the stress-energy tensor for the scalar field, nor does it affect the equation of motion for the scalar field in vacuum).
We thus require a nonzero $\Phi_0$ in Eq.~\eqref{eq:Phi-fall-off}, and we do not employ Sommerfeld's radiation condition to write the limit of the scalar field as $r$ goes to infinity.

Similarly, we adopt the same expansion of the 2-metric $\tilde h_{AB}$ as $\tilde r\rightarrow\infty$ as in GR:
\be\label{eq:hAB fall-off}
\tilde h_{AB} = q_{AB}(\tilde x^C) + \frac{\tilde c_{AB}(\tilde u,\tilde x^C)}{\tilde r} + O(\tilde r^{-2}) \, .
\ee
The determinant condition of Bondi gauge requires that $q^{AB} \tilde c_{AB}=0$.
It is then convenient to define a covariant derivative operator compatible with $q_{AB}$, which will be denoted by $\eth_A$.
In addition, it is also helpful to raise (or lower) capital Latin indices on 2-spheres of constant $\tilde u$ and $\tilde r$ with the 2-metric $q^{AB}$ (or $q_{AB}$). 

Next, we assume the functions $\tilde \beta$, $\tilde U^A$, $\tilde V$ and $\tilde h_{AB}$ have the following limits as $\tilde r$ approaches infinity:\footnote{We assume that it is possible to impose these conditions at all retarded times $u$.
Given the structure of the Bondi-Sachs equations as described in Sec.~\ref{subsubsec:BondiEquations}, these conditions can be imposed on an initial hypersurface $u={}$const., but they will not necessarily be preserved under evolution to future hypersurfaces. 
It is possible to construct coordinate transformations that reimpose the conditions Eq.~\eqref{eq:boundary-einstein} after evolution (see, e.g.,~\cite{Handmer:2015dsa,Handmer:2016mls} for more details).}
\be\label{eq:boundary-einstein}
\lim _{\tilde r \rightarrow \infty} \tilde\beta=\lim _{\tilde r \rightarrow \infty} \tilde U^{A}=0, \quad 
\lim _{\tilde r \rightarrow \infty} \frac{\tilde V}{\tilde r}=1, 
\quad \lim _{\tilde r \rightarrow \infty} \tilde h_{A B}=q_{A B}\,. 
\ee
The metric thus reduces to Minkowski spacetime in inertial Bondi coordinates in this limit. 
Hou and Zhu independently proposed similar conditions in~\cite{Hou:2020tnd}. 
Imposing these conditions and radially integrating the hypersurface equations in Eqs.~\eqref{eq:Einstein-Field-1}--\eqref{eq:Einstein-Field-2} further, we then arrive at the solutions\footnote{Note that in the expression for $U^A$ in Eq.~\eqref{eq:UA_Ein}, the remainder contains a term of order $\tilde r^{-3}\log \tilde r$.
The coefficient of the term that scales as $\tilde r^{-3}\log \tilde r$ is proportional to $\eth_B \tilde D^{AB}$, where we have denoted the $\tilde r^{-2}$ the part of $\tilde h_{AB}$ that is trace-free with respect to the metric $q^{AB}$ by $\tilde D_{AB}$. 
The order $1/\tilde r$ part of the Einstein equation~\eqref{eq:GAB_Ein} imposes that $\tilde D_{AB}$ satisfies $\partial_{\tilde u} \tilde D_{AB} = 0$ (i.e., that it is nondynamical, as the analogous quantity in general relativity is). 
This will not be true of the analogous function in the Jordan frame, as we discuss in Sec.~\ref{sec:Jordan}.}
\begin{subequations}
\ba
\label{eq:beta_Ein}
\tilde \beta &=&-\frac{1}{32\tilde r^2}\tilde c^{AB}\tilde c_{AB}-\frac{1}{\tilde r^2}\pi \Phi_1^2 + O(\tilde r^{-3}) \, ,\\ 
\label{eq:UA_Ein}
\tilde U^A&=&-\frac{1}{2 \tilde r^2}\eth_B \tilde c^{AB} + O(\tilde r^{-3} \log \tilde r ) \, ,\\
\label{eq:V_Ein}
\tilde V&=& \tilde r-2\tilde M + O(\tilde r^{-1}) \, .
\ea
\end{subequations}
The function $\tilde M(\tilde u, \tilde x^A)$ is called the Bondi mass aspect and is one of the functions of integration that arises from integrating the hypersurface equations.


\subsection{Jordan frame}\label{sec:Jordan}

Having determined the asymptotic fall-off conditions in the Einstein frame, we now consider the asymptotic properties of the solutions in the Jordan frame, in which it is more straightforward to understand the response of a detector to the gravitational waves emitted from an isolated system (because test particles follow geodesics of the Jordan-frame metric).
A solution in the Jordan frame can be found from one in the Einstein frame by performing a conformal transformation~\cite{Carroll:2004st}
\be \label{eq:Phi-to-lambda}
g_{\mu \nu}=\frac{1}{\lambda}\tilde{g}_{\mu \nu}\,,
\ee 
where
\be
\label{eq:lambda_Phi}
\lambda=\exp{\lb \Phi/\Omega\rb}\,,
\quad \quad \Omega\equiv\sqrt{\frac{2 \omega+3}{16 \pi}} \,.
\ee
The scalar field is called $\lambda$ in this frame, and $\omega$ is the Brans-Dicke parameter. 
In the limits in which $\omega \to \infty$ and $\lambda$ becomes nondynamical, general relativity is recovered.
The Brans-Dicke action in the Jordan frame is given by~\cite{Brans:1961sx}
\be\label{eq:action-jordan}
S=\int d^{4} x \sqrt{-g}\left[\frac{\lambda}{16 \pi} R-\frac{\omega}{16 \pi} g^{\mu \nu} \frac{\left(\partial_{\mu} \lambda\right)\left(\partial_{\nu} \lambda\right)}{\lambda}\right]\,,
\ee
where $R$ is the Ricci scalar of the Jordan-frame metric $g_{\mu\nu}$. 
\begin{subequations}
The field equations are given by
\begin{align}
\label{eq:Field}
G_{\mu\nu}
&=\frac{1}{\lambda}\lb 8\pi T_{\mu\nu}^{(\lambda)}+\nabla_{\mu}\nabla_{\nu}\lambda-g_{\mu\nu}\Box \lambda\rb \,, \\
\label{eq:lambda}
\Box \lambda&=0\,, 
\end{align}
where $G_{\mu\nu}$ is the Einstein tensor, $ \Box= g^{\mu\nu} \nabla_\mu \nabla_\nu$ is the covariant wave operator, and 
\be 
T_{\mu\nu}^{(\lambda)}=\frac{\omega}{8 \pi\lambda} \left(\nabla_{\mu}\lambda\nabla_{\nu}\lambda-\frac{1}{2}g_{\mu\nu}\nabla^{\alpha}\lambda\nabla_{\alpha}\lambda\right)\,.
\ee
\end{subequations}
is the stress-energy tensor of the scalar field.
It is also convenient to define a tensor $\mathcal{E}_{\mu\nu}$ by
\begin{equation}
    \mathcal{E}_{\mu\nu}\equiv G_{\mu\nu} - \frac{1}{\lambda}\lb 8\pi T_{\mu\nu}^{(\lambda)}+\nabla_{\mu}\nabla_{\nu}\lambda-g_{\mu\nu}\Box \lambda\rb \, ,
\end{equation}
which vanishes when the equations of motion are satisfied.

\subsubsection{Bondi gauge and asymptotic boundary conditions}

We would now like to compute a metric in Bondi-Sachs coordinates in the Jordan frame that is consistent with our definition of asymptotic flatness in the Einstein frame.
The transformation in Eq.~\eqref{eq:lambda_Phi} implies that $\lambda$ admits an expansion in $1/\tilde r$, in which the leading-order term is constant: i.e.,
\begin{align}
\lambda(\tilde u,\tilde r,\tilde x^A) = {} & \exp\left(\frac{\Phi_0}{\Omega}\right) \left(1+\frac{\Phi_1}{\Omega} \frac 1{\tilde r} \right) + O(\tilde r^{-2}) \,,
\end{align}
The conformal transformation of the metric in Eq.~\eqref{eq:Phi-to-lambda} preserves the Bondi gauge conditions $g_{rr} = g_{rA} = 0$, but the determinant condition becomes $\det[g_{AB}] = \tilde r^4 \lambda^{-2} q(x^C)$.
Consequently, the final condition of Bondi gauge will not be satisfied in general (i.e., $\tilde r$ is not an areal radius in the Jordan frame).
It is possible to work in a set of coordinates that do not impose the determinant condition in Bondi gauge (as was done in~\cite{Hou:2020tnd}); however, when $\lambda$ is positive (as it is expected to be far from an isolated source, since $\lambda$ is related to the gravitational constant~\cite{Will:2014kxa}), it is also possible to redefine the radial coordinate so as to make it an areal coordinate.
The transformation that effects this change is 
\begin{equation}
    u = \frac{\tilde u}{\sqrt{\lambda_0}} \, , \quad r = \tilde r \lambda^{-1/2} \, , \quad x^A = \tilde x^A \, ,
\end{equation}
where we have introduced the notation $\lambda_0=\exp(\Phi_0/\Omega)$. 
The retarded time $\tilde u$ is rescaled by $\lambda_0$ so that the metric coefficient $-g_{ur}$ becomes one as $r\rightarrow\infty$.
In the coordinates $(u,r,x^A)$, the metric takes the Bondi form, 
\begin{align}
g_{\mu \nu} d x^{\mu} d x^{\nu}=&-\frac{V}{r} e^{2 \beta} d u^{2}-2 e^{2 \beta} d u d r \nonumber \\
&+r^{2} h_{A B}\left(d x^{A}-U^{A} d u\right)\left(d x^{B}-U^{B} d u\right)\,,
\end{align}
where $V$, $\beta$, $U^A$ and $h_{AB}$ are functions of coordinates $x^\mu=(u,r,x^A)$. 
The metric satisfies all the Bondi gauge conditions
\begin{equation}
    g_{r A}=g_{r r}=0\,, \quad \det\left[g_{A B}\right] = r^{4} q\left(x^{C}\right)\,.
\end{equation}

By performing the conformal and coordinate transformation on the solutions of the field equations in the Einstein frame [Eqs.~\eqref{eq:beta_Ein}--\eqref{eq:V_Ein}], we find that the functions $\beta$, $U^A$, and $V$ should have the following forms:
\begin{subequations}
\begin{align}
\beta = & -\frac{1}{2r}\frac{\Phi_1}{\Omega\sqrt{\lambda_0}} + O\lb r^{-2} \rb\,,\\
U^{A} = & -\frac{1}{2\sqrt{\lambda_0}{r}^2}\lb \eth_B c^{AB}-\eth^A \Phi_1\rb + O\lb r^{-3} \log r\rb \, , \\
V = {} &\lb 1+ \frac{\partial_u \Phi_1}{\Omega}\rb r + O\lb {r}^{0}\rb.
\end{align}
\end{subequations}
Interestingly, in the limit as $r$ goes to infinity, $V/r$ goes as $1+\partial_u \Phi_1/\Omega$ (i.e., when $\Phi_1$ is dynamical, the leading-order Minkowski part of the metric is expressed in noninertial coordinates when the Bondi gauge conditions are imposed).
This occurs because the component of the Ricci tensor, $R_{uu}$, scales as $1/r$ when $\partial_u \Phi_1$ in nonvanishing, as we discuss in more detail below and in Sec.~\ref{sec:memory}.
In addition, $\beta$ scales as $O\left(r^{-1}\right)$ instead of $O\left(r^{-2}\right)$, as in the Einstein frame (or in general relativity).
Based on these considerations, we expect that the metric functions will have the following scaling with $r$ in the Jordan frame:
\begin{equation} \label{eq:JframeBCs}
\beta = O\left(r^{-1}\right)\,, \quad V =  O\left(r\right)\,, \quad U^A =  O\left(r^{-2}\right)\,.
\end{equation}
We explicitly verify this by solving the field equations in the next part.

\subsubsection{Asymptotically flat solutions}

The Bondi-Sachs field equations for Brans-Dicke theory in the Jordan frame have a similar hierarchy as in the Einstein frame.
The trace-free part of $h_{AB}$ satisfies an evolution equation, and the scalar field satisfies the curved-space wave equation (also an evolution equation).
The remaining metric functions $\beta$, $U^A$, and $V$ can be solved from hypersurface equations on surfaces of constant $u$ in terms of $h_{AB}$, $\lambda$, and functions of integration known as the Bondi mass aspect and angular momentum aspect.
The mass and angular momentum aspects satisfy the conservation equations.
The full expressions for these equations are rather lengthy, though we give the expressions for the scalar wave equation and the hypersurface equations in Appendix~\ref{sec:app}.
Thus, we will focus on determining the metric functions and the evolution equations satisfied by these functions when these quantities are expanded in a series in $1/r$. 

As in the Einstein frame, it is necessary to assume an expansion of the 2-metric $h_{AB}$ and the scalar field $\lambda$ as series in $1/r$, and the expansions of the remaining quantities will follow from the field equations and boundary conditions in Eq.~\eqref{eq:JframeBCs}.
For the scalar field, we have
\begin{align}
   \label{eq:lambda_exp}
\lambda(u,r,x^A) = {} & \lambda_0 + \frac{\lambda_1\left(u,x^A\right)}{r}
+ \frac{\lambda_2\left(u,x^A\right)}{r^2}+ \frac{\lambda_3\left(u,x^A\right)}{r^3} \nonumber \\
& + O (r^{-4}) \, ,
\end{align}
The constant $\lambda_0$ is related to the gravitational constant\footnote{The relation between the gravitational constant and the scalar field in Brans-Dicke theory is given by $G(\lambda)=\frac{4+2\omega}{3+2\omega}\frac{1}{\lambda}$. If one assumes the experimentally measured value of $G$ at infinity to be $1$, $\lambda_0$ can be written in terms of the Brans-Dicke parameter $\omega$ as $\lambda_0=\frac{3+2\omega}{4+2\omega}$.}, and $\lambda_1$ is the leading-order non-constant part of the scalar field, which is closely connected to the additional polarization of the gravitational waves in Brans-Dicke theory. 
That $\lambda$ in Eq.~\eqref{eq:lambda_exp} has a similar expansion in $1/r$ as $\Phi$ in Eq.~\eqref{eq:Phi-fall-off} follows from the relation between $\lambda$ and $\Phi$ in Eq.~\eqref{eq:lambda_Phi}.

For the two-metric, we take the expansion to be
\begin{subequations}
\begin{align}
\label{eq:h_AB_expand}
h_{AB} = {} & q_{AB}(x^C) + \frac{c_{AB}(u,x^C)}{r} +\frac{d_{AB}(u,x^C)}{r^2} \nonumber \\
& +\frac{e_{AB}(u,x^C)}{r^3}+O\lb r^{-4}\rb \, .
\end{align}
The determinant condition in Bondi gauge fixes the part of $d_{AB}$ and $e_{AB}$ that is proportional to $q_{AB}$.
Thus, we write them as
\begin{align}
d_{AB} = {} & D_{AB}+\frac{1}{4}c_{FG} c^{FG}q_{AB} \, ,\\
e_{AB} = {} & E_{AB}+\frac{1}{2}c_{FG} D^{FG} q_{AB} \, ,
\end{align}
\end{subequations}
where $c_{AB}$, $D_{AB} $, and $E_{AB}$ are trace-free with respect to $q_{AB}$. 
The tensor $c_{AB}(u,x^A)$ is closely related to the shear of outgoing null geodesics at large $r$, and is thus also related to the gravitational waves. 

We now substitute the expansion of $\lambda$ and $h_{AB}$ in Eqs.~\eqref{eq:lambda_exp} and~\eqref{eq:h_AB_expand} into the field equations, solve order by order in $r^{-1}$, and compute the metric functions and their corresponding evolution equations.  
We begin with the curved-space, scalar wave equation in Eq.~\eqref{eq:lambda}. The explicit forms, in Bondi coordinates, of Eq.~\eqref{eq:lambda} and the hypersurface equations in Eq.~\eqref{eq:Field} are given in Appendix~\ref{sec:app}.
We find that the assumption of $\lambda_0=\,$constant is consistent with these field equations; at $O(r^{-2})$, the wave equation reduces to the expression $\partial_r(\partial_u\lambda_1)=0$, which implies that $\partial_u \lambda_1 = N_{(\lambda)}(u,x^A)$ is an arbitrary function.
An analogous equation arises for the evolution of the tensor $c_{AB}$, which leads to $\partial_u c_{AB}$ being unconstrained (and equal to an arbitrary symmetric, trace-free tensor that gets called the Bondi news tensor, which is defined below).
To obtain higher-order terms in the wave equation, we need to first solve for some functions in the Bondi metric.

Next, integrating the \emph{rr, rA,} and the trace of the \emph{AB} components of the modified Einstein equations presented in Appendix~\ref{sec:app}, we find
\begin{subequations}
\begin{align}
\beta =&-\frac{\lambda_1}{2\lambda_0 r}-\frac{1}{r^2}\left(\frac{1}{32}c^{AB}c_{AB} + \frac{\omega-1}{8\lambda_0^2} \lambda_1^2 + \frac{3\lambda_2}{4\lambda_0}\right) \nonumber \\
& +O\lb r^{-3}\rb\,,\\
\label{eq:UA_sol}
U^{A}=&-\frac{1}{2r^2}\lb\eth_F c^{AF}-\frac{\eth^A\lambda_1}{\lambda_0}\rb \nonumber \\
& +\frac{1}{3 r^3}\Bigg[ c^{AD}\eth^F c_{DF} - \frac{1}{\lambda_0}c^{AD}\eth_{D}\lambda_1 + \frac{\lambda_1}{\lambda_0}\eth_B c^{AB} \nonumber \\
& - \frac{\lambda_1}{\lambda_0^2}\eth^A\lambda_1+\mathcal{U}^A\lb1+3 \log{r}\rb+6L^A \Bigg] +O\lb r^{-4}\rb\,, \\
\label{eq:V_sol}
V = {} & \left(1+\frac{\partial_{u} \lambda_{1}}{\lambda_{0}}\right) r - 2 M + O\lb r^{-1}\rb \,,
\end{align}
\end{subequations}
respectively. 
Here $M(u,x^A)$ is a function of integration.
While an analogous quantity is defined to be the Bondi mass aspect in the Einstein frame or in GR, here we find it convenient to define a slightly different quantity to be the mass aspect (which is defined shortly below). 
The second function of integration, the angular-momentum aspect $L^A$, can be obtained from the expression 
\begin{align}
L_{A}(u,x^A)=&-\frac{1}{6} \lim _{r \rightarrow \infty}\left(r^{4} e^{-2\beta}h_{A B} \partial_{r} U^{B}\right. \nonumber \\
& \left. -r \eth^B c_{A B}  +r \frac{\eth_A \lambda_1}{\lambda_0} +3\mathcal{U}_A\log{r}\right) \,.
\end{align}
The integration procedure allows for a term proportional to $\log r/r^3$ in $U^A$.
The term $\mathcal U^A$ is given by
\begin{equation}
\label{eq:divD_AB}
\mathcal{U}_A = - \frac 23 \eth^B \left(D_{AB} + \frac{1}{2\lambda_0}  \lambda_1 c_{AB}\right) \, .
\end{equation}
We will only consider solutions with $\mathcal U^A = 0$ in this paper for reasons which we discuss below Eq.~\eqref{eq:dotD_AB}.

We can now return to the scalar wave equation to solve for the higher-order terms.
The $O(r^{-3})$ and $O(r^{-4})$ parts of the scalar wave equation determine that $\lambda_2$ and $\lambda_3$ evolve via the equations
\begin{subequations}
\label{eq:lambda23ev}
\begin{align}
\partial_u \lambda_2 = & -\frac{1}{2}\ETH^2\lambda_1\, , \\
\partial_u \lambda_3 = & -\frac 1{2\lambda_0} \partial_u (\lambda_1 \lambda_2) + \frac 12 M \lambda_1 - \frac 14 (\ETH^2 +2) \lambda_2 \nonumber \\
& - \frac 1{8\lambda_0} \lambda_1 \ETH^2 \lambda_1 + \frac 14 c^{AB} \eth_A \eth_B \lambda_1 + \frac 18 \lambda_1 \eth_A \eth_B c^{AB} \nonumber \\
& + \frac 12 \eth_B c^{AB} \eth_A \lambda_1 \, .
\label{eq:lambda2ev}
\end{align}
\end{subequations}
To simplify the notation slightly, we have introduced the quantity $\ETH^2=\eth_A \eth^A$ to denote the Laplacian on the 2-sphere.

The evolution equations for $h_{AB}$ come from the trace-free part of the \textit{AB} components of the field equations 
\begin{equation}
    \mathcal{E}_{AB}-\frac{1}{2}g_{AB}g^{CD}\mathcal{E}_{CD} = 0 \, .
    \label{eq:EABstf}
\end{equation}
Because we have already imposed the field equation $h^{CD}\mathcal{E}_{CD}=0$ to determine $V$, the term proportional to $g_{AB}$ in Eq.~\eqref{eq:EABstf} does not contribute.
As a practical computational matter, it can be more convenient to contract Eq.~\eqref{eq:EABstf} into a complex polarization dyad $m^A = {\delta^A}_\mu m^\mu$ (and its complex conjugate) where $m^\mu$ satisfies $m^\mu \nabla_\mu u = 0 $~\cite{winicour1983,Winicour:2008vpn,Madler:2016xju} (a similar procedure was performed in the Einstein frame).
Then the two degrees of freedom in the evolution equation can be recast in terms of a single complex equation
\be\label{eq:evolution}
m^Am^B\mathcal{E}_{AB}=0 \, .
\ee
The $O(r^0)$ part of Eq.~\eqref{eq:evolution} reduces to the equation proportional to $\partial_r (\partial_u c_{AB}) = 0$. 
This implies that
\begin{equation}
    \partial_u c_{AB} = N_{AB} \, ,
    \label{eq:news_def}
\end{equation} 
where $N_{AB}$ is an arbitrary symmetric trace-free tensor, called the news tensor.
In GR, spacetimes with a vanishing news tensor contain no gravitational waves~\cite{Geroch:1977jn}. 
The $O(r^{-1})$ terms of Eq.~\eqref{eq:evolution} lead to the equation
\begin{equation}
\label{eq:dotD_AB}
\partial_u \left(D_{AB} + \frac{1}{2\lambda_0} \lambda_1 c_{AB}\right) = 0 \, .
\end{equation}
By taking $\partial_u$ of Eq.~\eqref{eq:divD_AB} and $\eth_A$ of Eq.~\eqref{eq:dotD_AB}, then one can see that one must have 
$\partial_u \mathcal{U}^A=0$. 
Thus the choice $\mathcal U^A = 0$ made above will not affect the dynamics of $D_{AB}$, but it does impose a constraint on the allowed initial data for the quantity $D_{AB}+\lambda_1 c_{AB}/(2\lambda_0)$.
The $O\lb r^{-2} \rb$ part of Eq.~\eqref{eq:evolution} is a significantly more complicated expression, which we give below:
\bw
\ba\label{eq:Eev}
\partial_u E_{AB}&=&-\frac{1}{2}D_{AB}+\frac{1}{2}c_{AB}\mathcal{M}-\eth_{(B} L_{A)}+\frac{1}{2}q_{AB}\eth_C L^C+\frac{1}{4}c_{AB}c^{CD}N_{CD}+\frac{1}{32}\left(\eth_A\eth_B -\frac{1}{2}q_{AB}\ETH^2 \right) (c^{ED} c_{ED})\nonumber\\
&&+\frac{1}{6}\left[\eth_{(B}\lb c_{A)}^C\eth^Dc_{CD}\rb-\frac{1}{2}q_{AB}\eth^E\lb c_E^C\eth^Dc_{CD}\rb\right]+\frac 18\epsilon_{C(A} {c_{B)}}^C(\epsilon^{DE}\eth_E \eth^F c_{DF})-\frac{1}{4\lambda_0^2}\lb 4\lambda_0 D_{AB}-\lambda_1 c_{AB}\rb \partial_u\lambda_1\nonumber
\\
&&-\frac{1}{12\lambda_0^2}\lb 3\omega+ 7\rb\lb\eth_A \lambda_1 \eth_B \lambda_1-\frac{1}{2}q_{AB}\eth^C\lambda_1\eth_C\lambda_1\rb
+\frac{1}{12\lambda_0^2}\lb 3\omega+2\rb\lambda_1\lb\eth_B\eth_A-\frac{1}{2}q_{AB}\ETH^2\rb \lambda_1 \nonumber\\
&&+\frac{1}{4\lambda_0}\lb\eth_B\eth_A-\frac{1}{2}q_{AB}\ETH^2\rb \lambda_2
+\frac{1}{12\lambda_0}c_{AB} \ETH^2 \lambda_1
+\frac{3\lambda_1^2}{8\lambda_0^2}N_{AB} -\frac{1}{2\lambda_0} \lambda_2 N_{AB} -\frac 1{3\lambda_0} \lambda_1 c_{AB}
\nonumber \\
&&-\frac 1{6\lambda_0}\lb\eth^C\lambda_1\eth_{(B} c_{A)C}-\frac 12 q_{AB} \eth_C c^{CD} \eth_D \lambda_1\rb +\frac{ 1}{12\lambda_0}\eth^C\lambda_1 \eth_C c_{AB} + \frac 1{24\lambda_0} \lambda_1 \ETH^2 c_{AB} \,.
\ea
\ew
We use this expression to understand the properties of the angular momentum aspect $L_A$ in nonradiative regions of spacetime in Sec.~\ref{sec:memory}.

To complete our treatment of the field equations, we must consider the conservation equations in $\mathcal E_{uu}$ and $\mathcal E_{uA}$.
These equations result in conservation equations for the mass and angular-momentum aspects.
The equation for the mass aspect comes from the $O(r^{-2})$ part of $\mathcal{E}_{uu}$, and it is given by
\begin{subequations}
\begin{align}
\label{eq:mass_aspect_evolution}
\partial_u \mathcal{M}=&-\frac{1}{8}N_{AB}N^{AB}+\frac{1}{4}\eth_A\eth_BN^{AB} \nonumber \\
&-\lb 3+2\omega\rb\frac{1}{4\lambda_0^2}\lb\partial_u\lambda_1\rb^2+\frac{1}{4\lambda_0}\partial_u \ETH^2\lambda_1\,,
\end{align}
where we have defined
\be \label{eq:CalMdef}
\mathcal{M}\lb u,x^A\rb=M\lb u,x^A\rb-\frac{1}{4\lambda_0^2}\lambda_1\partial_u\lambda_1\,,
\ee
\end{subequations}
to be the Bondi mass aspect in the Jordan frame. 
With this definition of $\mathcal{M}$ the average of the right-hand side of Eq.~\eqref{eq:mass_aspect_evolution} over the 2-sphere is a non-positive number: i.e., the average value of $\mathcal M$ is a strictly decreasing quantity.
This makes $\mathcal M$ more closely analogous to the Bondi mass aspect in general relativity, in which the average value of mass aspect gives rise to the well known Bondi mass-loss formula~\cite{Bondi:1962px}.
Note that $M$ would not necessarily satisfy this property, because $\lambda_1 \partial_u \lambda_1 = \partial_u (\lambda_1^2/2)$ is not necessarily a decreasing quantity. 
The calculations of symplectic fluxes and charges in Sec.~\ref{sec:memory} would suggest one might also include the $\ETH^2\lambda_1$ term in the definition of the mass aspect, though we do not do that above.

Finally, from the $O(r^{-2})$ part of $\mathcal{E}_{uA}$, the angular momentum aspect satisfies a conservation equation of the form 
\begin{widetext}
\begin{align}
-3 \partial_{u} L_{A}= {} &\eth_A \mathcal{M}-\frac{1}{4} \eth^E\left(\eth_E \eth^F c_{A F}-\eth_{A} \eth^{F} c_{E F}\right)+\frac{1}{16} \eth_{A}\left(c_{E F} N^{E F}\right)-\frac{1}{2}\eth_{C}\left(c^{C F} N_{AF}\right)+\frac{1}{4} c^{E F}\left(\eth_{A} N_{E F}\right)\nonumber \\  &+\frac{1}{8\lambda_0}\eth_A\ETH^2\lambda_1-\frac{1}{4\lambda_0^2}\lb 2+3\omega\rb \eth_A\lambda_1\partial_u\lambda_1+\frac{\lambda_1}{4\lambda_0^2}\lb 4+\omega\rb \eth_A\partial_u\lambda_1 +\frac{1}{4\lambda_0}\partial_u\lb c_{AB}\eth^B\lambda_1-\lambda_1\eth^B c_{AB}\rb\,.
\label{eq:LAdot}
\end{align}
\end{widetext}
To summarize, the structure of the Einstein equations is very much like the Bondi-Sachs formalism for general relativity~\cite{Madler:2016xju} (though with an additional massless field).
There are unconstrained functions $N_{AB} = \partial_u c_{AB}$ and $N_{(\lambda)} = \partial_u \lambda_1$ that determine the evolution of the different functions in the expansion of the metric and scalar field.
Then initial data must be given for $\lambda_1$, $\lambda_2$, $\lambda_3$, $\mathcal M$, $L_A$, $c_{AB}$, and $E_{AB}$.
Initial data also must be given for $D_{AB}$, but because we did not allow log terms in our expansion, this initial data is not independent of that of $\lambda_1$ and $c_{AB}$.
Our field equations have a slightly different form than those given in~\cite{Hou:2020tnd}, because of the different gauge conditions that we use (note also that~\cite{Hou:2020tnd} did not compute the evolution equations for $E_{AB}$ or $\lambda_3$).

\section{Asymptotic Symmetries}\label{sec:symmetries}

We now turn to computing the infinitesimal vector fields $\vec\xi$ that preserve the Bondi gauge conditions and the asymptotic form of the metric and the scalar field in Brans-Dicke theory.
Our treatment parallels that given in~\cite{Barnich:2010eb} for general relativity.
The scalar field is Lie dragged along the generators of these asymptotic symmetries $\vec\xi$, so it transforms as $\lambda \rightarrow \lambda + \mathcal L_{\xi} \lambda$ (where we use $\mathcal L_\xi$ to denote the Lie derivative along $\vec\xi$). 
To preserve the Bondi gauge conditions, the following components of the metric must be left invariant when the metric is Lie dragged along $\vec\xi$:
\be 
\label{eq:preserveBondi}
\mathcal{L}_{\xi} g_{r r}=0, \quad \mathcal{L}_{\xi} g_{r A}=0, \quad g^{A B} \mathcal{L}_{\xi} g_{A B}=0\,.
\ee
The four differential equations in Eq.~\eqref{eq:preserveBondi} constrain the four components of $\vec\xi$.
Because these conditions rely only upon Bondi gauge and not the underlying theory, we can expand the solution in Eq.~(4.7) of~\cite{Barnich:2010eb} using our solutions for $\beta$, $U^A$, and $h_{AB}$ in the Jordan frame of Brans-Dicke theory, which were computed in the Sec.~\ref{sec:framework}.
The results are that
\begin{subequations}
\label{eq:xi_sol}
\begin{align}
\xi^{u} = {} & f\lb u,x^A\rb\,, \\
\xi^{r} = & -\frac{1}{2}r\eth_A Y^A+\frac{1}{2}\eth^A\eth_A f-\frac{1}{4r}\bigg( c^{AB}\eth_B\eth_A f \nonumber \\
& + 2 \eth_A f \eth_B c^{AB} +\frac{\lambda_1}{\lambda_0}\ETH^2 f\bigg) + O\lb r^{-2}\rb\,, \\
\xi^{A}= {} & Y^A\lb u, x^A \rb-\frac{1}{r}\eth^A f + \frac{1}{2 r^2}\lb c^{AB}\eth_B f+\frac{1}{\lambda_0}\lambda_1\eth^A f \rb \nonumber \\
& +\frac{1}{r^3} \bigg[\frac{1}{3}D^{AB}\eth_B f -\frac{1}{16}c^{BC}c_{BC}\eth^A f - \frac{\lambda_1}{3\lambda_0}c^{AB}\eth_B f
\nonumber \\
& +\frac{\lambda_2}{2\lambda_0}\eth^A f +\frac{\lambda_1^2}{12\lambda_0^2}\lb \omega-3\rb\eth^A f\bigg]+O\lb r^{-4}\rb\,.
\end{align} 
\end{subequations}
The functions of integration $f(u,x^A)$ and $Y^A(u,x^A)$ come from radially integrating Eq.~\eqref{eq:preserveBondi}.

To maintain the asymptotic fall-off conditions that we have determined, we require that the remaining metric components transform as follows:
\ba 
\mathcal L_{\xi} g_{ur} = O(r^{-1}) \,, \quad \mathcal L_{\xi} g_{uA} = O(r^0)\,,\nonumber\\
\quad \mathcal L_{\xi} g_{AB} = O(r)\,, \quad\mathcal L_{\xi} g_{uu} = O(r^{0})\,.
\ea
Note that in GR $\mathcal{L}_{\xi}g_{uu}=O(r^{-1})$; however, because in Brans-Dicke theory in the Jordan frame $g_{uu}$ is given by $g_{uu}=-1+\partial_u \lambda_1/\lambda_0+O(r^{-1})$, we allow a change in $g_{uu}$ at $O(r^0)$ (which occurs from the change in $\partial_u \lambda_1$). 
To express the conditions that we use to constrain $\vec\xi$ and the change in the metric coefficients, it is convenient to expand $\mathcal L_{\xi}g_{\mu\nu}$ in a series in $r$ as
\be 
 \mathcal L_{\xi}g_{\mu\nu} = \sum_n r^{n} l_{\mu\nu}^{(n)}\,,
\ee
where $n$ can be an integer, and the coefficients $l_{\mu\nu}^{(n)}$ in the expansion are functions of $u$ and $x^A$.
Then one can show from $l_{uA}^{(2)}=0$ that $Y^A$ is independent of $u$, and from $l_{AB}^{(2)}=0$ that it is a conformal Killing vector on a 2-sphere: i.e.,
\be 
\eth_A Y_B + \eth_B Y_A=\psi q_{AB}\,,
\label{eq:YckvS2}
\ee
where $\psi=\eth_A Y^A$.
The coefficient $l_{ur}^{(0)}=0$ restricts $f$ to be given by
\be 
f=\frac{1}{2}u \psi+\alpha\lb x^A \rb \,.
\ee

The functions $f$ and $Y^A$ have the same form as in general relativity. Thus, the different fall-off conditions of components of the the metric in Brans-Dicke theory do not alter the symmetries of the spacetime.
The interpretation of $Y^A$ and $\alpha$ will, therefore, be the same as in GR: the globally defined $Y^A$ span a six-parameter algebra isomorphic to the proper, isochronous Lorentz algebra (and the locally defined $Y^A$ will be the infinite-dimensional group of super-rotation symmetries~\cite{Barnich:2009se}) and $\alpha$ span the infinite-dimensional commutative algebra of supertranslations~\cite{Bondi:1962px,PhysRev.128.2851}.
How the asymptotic Killing vectors $\vec \xi$ are extended into the interior of the spacetime from future null infinity is different in GR from in Brans-Dicke theory in the Jordan frame.
This will lead to the functions in the metric transforming differently between the two theories. 

Before we compute the transformation of the metric functions, it is necessary to determine how the functions $\lambda_1$ and $\lambda_2$ in the expansion of the scalar field transform as they are Lie dragged along $\vec\xi$.
We denote this transformation as $\delta_{\xi}\lambda_1$ and $\delta_\xi\lambda_2$ and they can be computed from the $O(r^{-1})$ and $O(r^{-2})$ of $\mathcal L_{\xi}\lambda$, respectively.
The result is
\begin{align} \label{eq:lambda1}
\delta_{\xi}\lambda_1 = {} &\frac{1}{2}\lambda_1 \psi+Y^A \eth_A \lambda_1+f \partial_u \lambda_1\,, \\
\delta_{\xi}\lambda_2 = {} &\lambda_2 \psi-\frac{1}{2}\lambda_1 \ETH^2 f -\eth^C f \eth_C \lambda_1+Y^D\eth_D \lambda_2+f \partial_u \lambda_2\,. 
\end{align}
Next, we can compute how the functions $c_{AB}$, $D_{AB}$, $\mathcal M$, and $L_A$ transform when Lie dragged along $\vec \xi$ given in Eq.~\eqref{eq:xi_sol}.
We denote these quantities $\delta_\xi c_{AB}$ and similarly for the other three functions.
The term $\delta_\xi c_{AB}$ can be obtained directly from the appropriate coefficients and components of $l_{\mu\nu}^{(n)}$, but other terms require also removing the transformation of combinations of $\delta_\xi\lambda_1$ and $\delta_\xi c_{AB}$ that appear at the same order in the metric.
The expressions used to compute these quantities are given below:
\begin{subequations}
\label{eq:delta_xis}
\begin{align}
    \delta_{\xi}c_{AB} = {} & l_{AB}^{(1)} \\
    \delta_{\xi} \mathcal{M} = {} & \frac{1}{2}l_{uu}^{(-1)} -\frac{1}{2}\left[\frac{\delta_{\xi}\lambda_1}{\lambda_0} +\frac{3}{2\lambda_0^2}\delta_{\xi}\lb\lambda_1\partial_u\lambda_1\rb\right] \\
    \delta_{\xi}D_{AB} = {} & l_{AB}^{(0)}-\frac{1}{4}q_{AB}\delta_{\xi}\lb c^{CD} c_{CD} \rb \\
    \delta_{\xi} L_A = & -\frac{1}{2} l_{uA}^{(-1)} + \frac{1}{12}\delta_{\xi}\lb c_{AB}\eth_C c^{BC} \rb \nonumber \\
    & -\frac{1}{6\lambda_0}\delta_{\xi}\lb \lambda_1 \eth_B c^B_A\rb -\frac{1}{12\lambda_0}\delta_{\xi} \lb c_{AB}\eth^B\lambda_1 \rb \nonumber \\
    & +\frac{1}{6\lambda_0^2}\delta_{\xi}\lb \lambda_1\eth_A\lambda_1\rb
\end{align}
\end{subequations}
Thus, we can compute $\delta_\xi \mathcal M$ and $\delta_\xi c_{AB}$ from the relevant $l_{\mu\nu}^{(n)}$ and $\delta_\xi \lambda_1$ to be
\begin{subequations}
\label{eq:delta_xi_metric}
\begin{equation}
 \label{eq:deltacAB}
\delta_{\xi}c_{AB} = \mathcal{L}_{Y}c_{AB}+ f N_{AB}-\frac{1}{2}\psi c_{AB}-2\eth_A\eth_B f+q_{AB}\ETH^2 f\,,
\end{equation}
\begin{align}
\label{eq:deltaM}
\delta_{\xi} \mathcal{M} = {} & f\partial_u \mathcal{M} +\frac{3}{2}\mathcal{M}\psi +Y^A \eth_A \mathcal{M}+ \frac{1}{8}c^{AB}\eth_A\eth_B \psi \nonumber \\
& +\frac{1}{2}\eth_A f \eth_B N^{AB}+\frac{1}{4}N^{AB}\eth_A\eth_B f -\frac{\lambda_1 \psi}{4 \lambda_0} \nonumber\\
&+\frac{1}{4\lambda_0}\eth_A \psi \eth^A \lambda_1 + \frac{1}{4 \lambda_0}\ETH^2 f\partial_u\lambda_1  +\frac{1}{2\lambda_0}\eth^A f \eth_A \partial_u \lambda_1\,.
\end{align}
Then with the expression for $\delta_\xi c_{AB}$, it is possible to compute the remaining two terms for $\delta_\xi D_{AB}$ and $\delta_\xi L_A$.
They are given by 
\begin{align}
\delta_{\xi}D_{AB} = {}  &\mathcal{L}_{Y}D_{AB}+\frac{\lambda_1}{\lambda_0}\lb \eth_A \eth_B -\frac{1}{2} q_{AB}\ETH^2 \rb f \nonumber \\ & -\frac{1}{2\lambda_0}f\partial_u\lb \lambda_1 c_{AB}\rb \,,
\end{align} 
for $\delta_\xi D_{AB}$ and
\bw
\ba
\delta_{\xi} L_A &=& f\partial_u L_A+\mathcal{L}_Y L_A+L_A \psi+\frac{1}{96}c^{CD}c_{CD}\eth_A \psi+\frac{1}{6}D_{AB}\eth^B\psi-\mathcal{M}\eth_A f+\frac{1}{12} \lb \eth^C\eth_C f \eth^B c_{AB} -c_{AB}\eth^B \eth^C\eth_C f\rb\nonumber\\
&&-\frac{1}{8}\eth_A\lb c^{BC}\eth_B\eth_C f\rb+\frac{1}{4}\lb\eth^D\eth_C c_{AD}-\eth_A\eth^B c_{BC} \rb\eth^C f-\frac{1}{6}c_{AB}\eth^B f-\frac{1}{6}\eth_B\eth_Af\eth_C c^{BC}-\frac{5}{48}c^{BC}N_{BC}\eth_A f \nonumber\\
&&+\frac{1}{6}c^{BC}N_{AB}\eth_C f- \frac{1}{12\lambda_0^2}\left[c_{AB}\lambda_0 \eth^B f+\lambda_1 \lb \omega+4\rb\eth_A f\right]\partial_u \lambda_1+\frac{1}{24\lambda_0^2}\left[6\lambda_0 \lambda_2+\lb \omega-1\rb \lambda_1^2\right]\eth_A \psi\nonumber\\
&&-\frac{1}{12\lambda_0}\lb 2\lambda_1+3\partial_u \lambda_2 \rb\eth_A f-\frac{5}{24\lambda_0}\eth_A \lb \lambda_1 \eth_C\eth^C f\rb+\frac{\lambda_1}{12\lambda_0}N_{AB}\eth^B f-\frac{1}{12\lambda_0}\lb \eth_B\eth_A f \eth^B \lambda_1+3\eth_B\eth_A\lambda_1\eth^B f\rb\,, \nonumber \\
\ea
\ew
\end{subequations}
for $\delta_\xi L_A$.
In deriving the expression for $\delta_\xi \mathcal M$, we used the properties $\ETH^2\psi = -2\psi$ and $\ETH^2 Y^A=-Y^A$. 
To derive $\delta_\xi L_A$, we also used the identities in~\cite{Barnich:2010eb}
\begin{subequations}
\begin{align}
& 2c_{C(A} \eth_{B)} \eth^C f  - q_{AB} c^{CD}\eth_C\eth_D f -c_{AB} \ETH^2 f =0\,, \\
&2\eth^C c_{C(A} \eth_{B)} f + 2 \eth_{(A} c_{B)C} \eth^C f \nonumber\\
&-2\eth_C c_{AB}\eth^C f-2 q_{AB}\eth_C f\eth_D c^{CD}=0\,.
\end{align}
\end{subequations}
The expressions in Eq.~\eqref{eq:delta_xi_metric} will be useful for understanding the properties of metric in nonradiative regions, which we discuss in Sec.~\ref{sec:memory} soon hereafter.
The GR limit of our expressions agrees with the equivalent results in~\cite{Barnich:2011mi} after taking into account differences in conventions.
Our results are similar to those in~\cite{Hou:2020tnd}, but not identical, because of the different gauge conditions that we use.

Before concluding this part, we note that because the scalar field appears in the metric, we can check whether the transformation of the metric is consistent with requiring that the scalar field is Lie dragged along $\vec\xi$.
We can obtain $\delta_{\xi}\lambda_1$ from $2\lambda_0 l_{ur}^{(-1)}$, and we find that it agrees with Eq.~\eqref{eq:lambda1}. 
We can also obtain $\delta_{\xi}\lb \partial_u \lambda_1 \rb$ from $-\lambda_0 l_{uu}^{(0)}$ and we find that it is equivalent to $\partial_u (\delta_{\xi} \lambda_1)$, as it should be.

We can also explicitly compute the quantities $\delta_\xi(c^{AB}c_{AB})$ and $\delta_\xi(\eth_B c^{AB})$ from Lie dragging the metric.
The relevant expressions for computing this are
\begin{subequations}
\begin{align}
    \delta_{\xi}\lb c^{AB}c_{AB}\rb = {} & 16 l_{ur} ^{(-2)} + \frac{4}{\lambda_0^2}\lb 3 -\omega\rb \delta_{\xi}(\lambda_1)^2 -\frac{24}{\lambda_0} \delta_{\xi}\lambda_2 \\
    \delta_{\xi}\lb \eth^B c_{AB} \rb = {} & 2 l_{uA}^{(0)} + \frac{1}{\lambda_0}\delta_{\xi}\lb \eth_A\lambda_1 \rb 
\end{align}
\end{subequations}
Not surprisingly, we find that 
\begin{subequations}
\begin{align}
    \delta_{\xi}\lb c^{AB}c_{AB}\rb = {} & \delta_{\xi} c^{AB} c_{AB} +  c^{AB} \delta_{\xi}c_{AB}\, , \\
    \delta_{\xi}\lb \eth^B c_{AB} \rb = {} &  \eth^B \lb \delta_{\xi} c_{AB} \rb \, ,
\end{align}
\end{subequations}
as the latter relation was proven in GR in~\cite{Barnich:2010eb}.
For completeness, we give the  expressions for these terms here
\begin{subequations}
\begin{align}
\delta_{\xi}\lb c^{AB}c_{AB}\rb = {} & 2f N_{AB}c^{AB}+c_{AB}c^{AB}\psi \nonumber \\
& +2 c^{BC}Y^A\eth_A c_{BC}-4 c^{AB}\eth_A\eth_B f\,, \\
\delta_{\xi}\lb \eth^B c_{AB} \rb = & -\eth_A(\ETH^2 + 2) f -\frac{1}{2}c_{AB}\eth^{B}\psi \nonumber \\
& +\frac{1}{2}\psi \eth^B c_{AB}+\mathcal{L}_{Y}\eth^C c_{AC}+ \eth^B \lb f N_{AB} \rb\, .
\end{align}
\end{subequations}
It does not seem possible to verify these types of relationships with all of the terms that appear in Eq.~\eqref{eq:delta_xis}.
Thus, for example, for the term $\delta_{\xi}\lb c_{AB}\eth_C c^{BC} \rb$, we assumed it can be written as the sum of $\delta_{\xi} c_{AB} \eth_C c^{BC}$ and $ c_{AB} \delta_{\xi} \eth_C c^{BC}$.

\section{Gravitational-wave memory effects}\label{sec:memory}

Gravitational-wave (GW) memory effects are commonly defined for bursts of gravitational waves of finite duration between two nonradiative regions before and after the burst; they are also defined for sources of gravitational waves that become asymptotically nonradiative in the limits as $u\rightarrow\pm\infty$ at large Bondi radius $r$.
In either case, discussing GW memory effects requires a notion of a nonradiative region of spacetime.
In this section, we first describe the properties of nonradiative regions in Brans-Dicke theory, we then discuss the measurement of GW memory effects through geodesic deviation, and we finally discuss how the conservation equations constrain the GW memory effects (thereby allowing them to be computed approximately).

\subsection{Nonradiative and stationary regions}

\paragraph*{Nonradiative regions}
For general relativity, it is typical to consider regions of vanishing Bondi news $N_{AB}$, and vanishing stress-energy tensor.
In the context of Brans-Dicke theory, we will instead consider regions where $N_{AB}=0$, $\partial_u \lambda_1=0$, and any other stress-energy from matter fields vanishes.
These two equations imply that $\lambda_1$ and $c_{AB}$ must be independent of $u$.
Integrating Eqs.~\eqref{eq:lambda23ev},~\eqref{eq:divD_AB},~\eqref{eq:mass_aspect_evolution} and~\eqref{eq:LAdot}, we can then show that $\lambda_2$, $D_{AB}$, $\mathcal M$, $L_A$, and $\lambda_3$ are given by 
\begin{subequations}
\label{eq:nonradsols}
\begin{align}
    \lambda_1 = {} & \lambda_1^{(0)}(x^A) \, ,\\
    c_{AB} = {} & c_{AB}^{(0)}(x^C) \, ,\\
    \label{eq:lambda2nonrad}
    \lambda_2 = & -\frac u2 \ETH^2 \lambda_1^{(0)} + \lambda_2^{(0)}(x^B) \, ,\\
    D_{AB} = & -\frac 1{2\lambda_0} \lambda_1^{(0)} c_{AB}^{(0)} \, ,\\
    \mathcal M = {} & \mathcal M^{(0)}(x^A) \, ,\\
    \label{eq:LAnonrad}
    L_A = & -\frac u3 \eth_A \mathcal M^{(0)} 
    - \frac u{24\lambda_0} \eth_A \ETH^2 \lambda_1^{(0)} \nonumber \\
    & + \frac u{12} \eth^D(\eth_D \eth^B c^{(0)}_{AB} - \eth_A \eth^B c^{(0)}_{DB})
    + L_A^{(0)}(x^E) \, , \\
    \label{eq:lambda3nonrad}
    \lambda_3 = {} & \frac{u^2}{16\lambda_0} (\ETH^2+2)\ETH^2 \lambda_1^{(0)} + \frac u2 \bigg[- \frac 12 (\ETH^2 +2) \lambda_2^{(0)} \nonumber \\
    & + \mathcal M^{(0)} \lambda_1^{(0)} + \frac 1{4\lambda_0} \lambda_1^{(0)} \ETH^2 \lambda_1^{(0)} + \frac 12 c^{AB}_{(0)} \eth_A \eth_B \lambda_1^{(0)} \nonumber \\
    & + \frac 14 \lambda_1^{(0)} \eth_A \eth_B c^{AB}_{(0)} + \eth_B c^{AB}_{(0)} \eth_A \lambda^{(0)}_1 \bigg]  + \lambda_3^{(0)}(x^C) \, .
\end{align}
\end{subequations}
In a nonradiative region, $E_{AB}$ has the form
\begin{subequations}
\begin{equation}
\label{eq:EABnonrad}
E_{AB} = u^2 E_{AB}^{(2)} + u E_{AB}^{(1)} + E_{AB}^{(0)}(x^C) \, ,
\end{equation}
where the coefficients $E_{AB}^{(2)}$ and $E_{AB}^{(1)}$ are given by
\begin{align}
\label{eq:EAB2nonrad}
E_{AB}^{(2)} = {} & \frac 16 \left(\eth_A \eth_B - \frac 12 q_{AB} \ETH^2\right)
\left(\mathcal M^{(0)} - \frac 1{4\lambda_0} \ETH^2 \lambda_1^{(0)} \right) \nonumber \\
& - \frac 1{24} \eth_{(A} \epsilon_{B)C}\eth^C 
(\epsilon^{DE} \eth_E \eth^F c_{DF}^{(0)}) \, ,\\
\label{eq:EAB1nonrad}
E_{AB}^{(1)} = & -\eth_{(A} L_{B)}^{(0)} + \frac 12 q_{AB} \eth^C L_C^{(0)} + \frac 12 \mathcal M^{(0)} c_{AB}^{(0)} \nonumber \\
& + \frac 18 {\epsilon^C}_{(A}c^{(0)}_{B)C}
(\epsilon^{DE} \eth_E \eth^F c_{DF}^{(0)}) \nonumber \\
& + \frac 1{6} \eth_{(A}\left(c^{(0)}_{B)C} \eth_D c^{DC}_{(0)}\right) 
- \frac 1{12} q_{AB} \eth^D (c^{(0)}_{DC} \eth_E c^{EC}_{(0)}) \nonumber \\
& + \frac 1{32} \left(\eth_A \eth_B - \frac 12 q_{AB} \ETH^2 \right)
(c_{CD}^{(0)} c^{CD}_{(0)}) \nonumber \\
& - \frac 1{12\lambda_0} \lambda_1^{(0)} c_{AB}^{(0)} + \frac 1{24\lambda_0} \lambda_1^{(0)} \ETH^2 c_{AB}^{(0)} \nonumber \\
& + \frac 1{12\lambda_0} \ETH^2 \lambda_1^{(0)}  c_{AB}^{(0)} +
\frac 1{12\lambda_0} \eth^C \lambda_1^{(0)} \eth_C c_{AB}^{(0)} \nonumber \\
& - \frac 1{6\lambda_0}\eth_{(A} c^{(0)}_{B)C} \eth^C \lambda_1^{(0)} + \frac 1{12\lambda_0} q_{AB} \eth^D c_{DC}^{(0)} \eth^C\lambda_1^{(0)} \nonumber \\
& +\frac{1}{4\lambda_0}\left(\eth_A \eth_B - \frac 12 q_{AB} \ETH^2 \right) \lambda_2^{(0)} \nonumber \\
& + \frac{2+3\omega}{12(\lambda_0)^2} \lambda_1^{(0)} \left(\eth_A \eth_B - \frac 12 q_{AB} \ETH^2 \right) \lambda_1^{(0)} \nonumber \\
& - \frac{3\omega+7}{12(\lambda_0)^2} \left(\eth_A \lambda_1^{(0)} \eth_B \lambda_1^{(0)} - \frac 12 q_{AB} \eth_C \lambda_1^{(0)} \eth^C \lambda_1^{(0)}
\right) \, ,\\
E_{AB}^{(0)} = {} & E_{AB}^{(0)}(x^C) \, .
\end{align}
\end{subequations}
This expression will simplify considerably in some more restrictive classes of nonradiative solutions that we discuss next.

\paragraph*{Stationary regions and the canonical frame}
In general relativity, there are frames for stationary regions in which the Bondi metric functions are independent of $u$.
We next discuss how the metric functions and scalar field in nonradiative regions in Brans-Dicke theory [given in Eq.~\eqref{eq:nonradsols}] simplify when there exist such frames in the Jordan frame.
To discuss this, it is useful to recall that a vector field, such as $L_A$, on the 2-sphere can be decomposed into divergence- and curl-free parts as follows
\begin{equation}
    L_A = \eth_A \rho + \epsilon_{AB} \eth^B \sigma \, .
\end{equation}
Similarly, a symmetric trace-free tensor like $c_{AB}$ can be decomposed as~\cite{madler:2016ggp,Newman:1966ub}
\begin{equation}
    c_{AB} = \left(\eth_A \eth_B - \frac 12 q_{AB} \eth_C \eth^C \right) \Theta + \epsilon_{C(A} \eth_{B)} \eth^C \Psi \, .
    \label{eq:cABdecomp}
\end{equation}
The terms without the antisymmetric tensor $\epsilon_{AB}$ in the last two equations are often called the ``electric (parity)'' part and the terms with $\epsilon_{AB}$ are called the ``magnetic (parity)'' part.

If we require that the scalar field is also independent of $u$ in these regions, then the expression for $\lambda_2$ in Eq.~\eqref{eq:lambda2nonrad} requires that $\ETH^2 \lambda_1 = 0$, or namely $\lambda_1$ is constant.
The expression for $L_A$ in Eq.~\eqref{eq:LAnonrad} shows that the magnetic part of $c_{AB}^{(0)}$, $\Psi$, vanishes (see, e.g.,~\cite{Flanagan:2015pxa}).
Together with the fact that $\lambda_1^{(0)}$ is constant, it also follows from  Eq.~\eqref{eq:LAnonrad} that $\mathcal M^{(0)} = M^{(0)}$ is a constant.
Because $c_{AB}$ is an electric-parity tensor field, then it can be set to zero by performing a supertranslation with $\alpha = \Theta/2$ [see Eq.~\eqref{eq:deltacAB}].
With $c_{AB}=0$ as well as $\lambda_1^{(0)}$ and $\mathcal M^{(0)}$ being constant, then by requiring $\lambda_3$ is independent of time Eq.~\eqref{eq:lambda3nonrad} gives the following condition on $\lambda_2^{(0)}$:
\begin{equation}
    (\ETH^2+2) \lambda_2^{(0)} = 2\mathcal M^{(0)} \lambda_1^{(0)} \, .
\end{equation}
This nonhomogeneous linear elliptic equation can be written as the sum of the particular solution $\lambda_2^{(0)} = \mathcal M^{(0)} \lambda_1^{(0)}$ and a linear combination of the solutions to the homogeneous equation
\begin{equation}
    (\ETH^2+2) \lambda_2^{(0)} = 0 \, .
\end{equation}
The solution of the homogeneous equation is a superposition of $\ell=1$ spherical harmonics.
Finally, with these conditions on the metric functions, this greatly simplifies the form of $E_{AB}$ in Eq.~\eqref{eq:EABnonrad}. 
That $\lambda_1^{(0)}$ and $\mathcal M^{(0)}$ are constants and that $c_{AB}^{(0)}$ vanishes cause the coefficient in Eq.~\eqref{eq:EAB2nonrad} to vanish; similarly, the lengthy expression in Eq.~\eqref{eq:EAB1nonrad} reduces to the following much simpler equation:
\begin{align}
    \eth_{(A} L_{B)}^{(0)} - \frac 12 q_{AB} \eth^C L_C^{(0)} = 0 \, .
\end{align}
To have smooth solutions $L_A^{(0)}$, then it must be a superposition of the six electric-parity and magnetic-parity $\ell=1$ vector spherical harmonics.
The electric part of $L_A^{(0)}$ can be set to zero by performing a translation with $\alpha=\kappa/[\mathcal M^{(0)} - \lambda_1^{(0)}/(4\lambda_0)]$
The magnetic part could be chosen to align with a particular axis by performing a rotation if desired.

Like in general relativity, this class of stationary regions in Brans-Dicke theory admit a ``canonical'' frame, in which $\mathcal M$ and $\lambda_1$ are constant, $c_{AB}=0$, and $L_A$ is composed of $\ell=1$ magnetic-parity vector harmonics.
Furthermore, the scalar-field function $\lambda_1$ is also constant, and the function $\lambda_2$ is equal to the constant $\mathcal M \lambda_1$ plus a superposition of $l=1$ spherical harmonics.
For bursts of gravitational and scalar radiation, there can be transitions between such stationary regions where the initial stationary region is in the canonical frame, but the final stationary region is supertranslated from its canonical frame, so that $c_{AB}$ is nonzero.
This nonzero $c_{AB}$ at late times is, in essence, the GW memory effect (see e.g.,~\cite{Strominger:2014pwa}); thus, transitions between these stationary regions provide a sufficiently general arena in which to study certain types of GW memory effects in general relativity (these transitions were called ``BMS vacuum transitions in~\cite{Strominger:2014pwa}).
Note that these types of transitions also do not allow ``ordinary'' memory~\cite{Bieri:2013ada}, so they do not admit memory effects of full generality (see, e.g.,~\cite{Satishchandran:2019pyc}).

However, in Brans-Dicke theory, because $\lambda_1$ must be a constant in both stationary regions in the canonical frames, such a transition would significantly restrict the types of possible memory effects that could occur.
For the memory effects related to the scalar radiation (discussed in greater detail in the next part) such a transition would only allow these scalar-type memory effects to have a uniform sky pattern.
As a result, considering only these types of transitions between these frames will not be sufficiently general to explore the full range of possible memory effects in Brans-Dicke theory. 
A slight generalization would be to consider transitions between stationary regions in which one of the regions is both boosted and supertranslated from the canonical frame.
However, this still seems to be a somewhat restrictive scenario, as it does not seem to admit solutions that are superpositions of boosted massive bodies with a scalar charge.
As a result, we will next focus on a slightly more general set of frames, that is still somewhat simpler than the nonradiative regions without restrictions.

\paragraph*{Nonradiative regions with vanishing magnetic-parity shear}
For a slightly more general set of solutions, though which lack the full generality of the nonradiative regions, we will consider regions of spacetime with vanishing stress-energy (not including the scalar field), $N_{AB}$, $\partial_u \lambda_1$, and $\Psi$ (the magnetic-parity part of the shear).
Like the stationary regions, it will again be possible to set $c_{AB}=0$ by a supertranslation to produce a ``semi-canonical'' frame; however, the mass-aspect and scalar-field functions $\mathcal M^{(0)}$ and $\lambda_1^{(0)}$ will no longer be constants, and will remain arbitrary functions of $x^A$ as in Eq.~\eqref{eq:nonradsols} in this frame.
This will imply that $\lambda_2$ depends linearly on $u$ as in Eq.~\eqref{eq:lambda2nonrad}, $D_{AB}$ will vanish in this semi-canonical frame, and the electric part of $L_A$ will depend linearly on $u$, whereas the part independent of $u$ will contain both electric and magnetic parts.
Thus, transitions between nonradiative regions of this type should be sufficiently general to capture both the usual tensor-type and scalar-type memory effects, which will be discussed in greater detail below. 
This was also the scenario considered by~\cite{Hou:2020tnd}.

\subsection{Geodesic deviation and GW memory effects}

GW memory effects are frequently described by their effects on families of nearby freely falling observers at large distances $r$ from a source of gravitational waves~\cite{Zeldovich:1974gvh,Braginsky:1986ia,Zhang:2017rno}.
The deviation vector $\vec X$ between a geodesic with tangent $\vec u$ and a nearby geodesic satisfies the equation of geodesic deviation
\begin{equation}
    u^\gamma \nabla_\gamma (u^\beta \nabla_\beta X^\alpha) = - {R_{\beta \gamma \delta}}^\alpha u^\beta X^\gamma u^\delta \, ,
\end{equation}
to linear order in the deviation vector $X^\alpha$.
It is then useful to expand the vector $X^\alpha$ in terms of an orthonormal triad $e_{\hat i}^\alpha$ with $e_{\hat i}^\alpha u_\alpha = 0$ that is parallel transported along the geodesic with tangent $u^\alpha$.
If $u^\alpha$ is denoted by $e^\alpha_{\hat 0}$, then $e_{\hat \mu}^\alpha = \{e_{\hat 0}^\alpha, e_{\hat i}^\alpha \}$ forms an orthonormal tetrad.
It is also convenient to introduce a dual triad $e_\alpha^{\hat j}$ with $e_{\hat i}^\alpha e_\alpha^{\hat j} = \delta^{\hat j}_{\hat i}$.
The vector can then be written in the form $X^\alpha = X^{\hat i}(\tau) e_{\hat i}^\alpha$, where $\tau$ is the proper time along the geodesic worldline.
The equation of geodesic deviation then reduces to the expression
\begin{equation}
    \ddot X^{\hat i} = - {R_{\hat 0 \hat j \hat 0}}^{\hat i} X^{\hat j} \, 
\end{equation}
where the dot denotes $d/d\tau$.
Given a set of tetrad coefficients $X^{\hat i}_0 = X^{\hat i}(\tau_0)$ and $\dot X^{\hat i}_0 = \dot X^{\hat i}(\tau_0)$ that represent the initial separation and relative velocity of the nearby geodesics, it is possible to solve for the change in the final values of the tetrad coefficients of the separation vector, which we denote by 
\begin{equation} 
\Delta X^{\hat i} = X^{\hat i}(\tau_f) - X^{\hat i}(\tau_0) \, .
\end{equation}
We then expand this vector in a series to linear order in the Riemann tensor as 
\begin{equation}
    \Delta X^{\hat i} = \Delta X^{\hat i}_{(0)} + \Delta X^{\hat i}_{(1)} \, ,
\end{equation} 
where the value $\Delta X^{\hat i}_{(0)}$ is identical to the expected result in flat spacetime
\begin{equation}
    \Delta X^{\hat i}_{(0)} = (\tau_f - \tau_0) \dot X^{\hat i}_0 \, . 
    \label{eq:DeltaXi_flat}
\end{equation}
The correction $\Delta X^{\hat i}_{(1)}$ to the deviation vector to linear order in the Riemann tensor is given by~\cite{Flanagan:2018yzh}
\begin{align}
    \Delta X^{\hat i}_{(1)} = & - X^{\hat j}_0 \int_{\tau_0}^{\tau_f} \! d\tau \! \int_{\tau_0}^{\tau} \! d\tau'{R_{\hat 0 \hat j \hat 0}}^{\hat i} \nonumber \\ 
    & - \dot X^{\hat j}_0 \int_{\tau_0}^{\tau_f} \! d\tau \! \int_{\tau_0}^{\tau} \! d\tau' \! \int_{\tau'}^{\tau} \! d\tau'' {R_{\hat 0 \hat j \hat 0}}^{\hat i} \, .
\end{align}
Note that in the triple integral, the limits of integration on the innermost integral run from $\tau'$ to $\tau$.

To compute $\Delta X^{\hat i}$ associated with a burst of gravitational waves at large distances $r$ from a source of GWs (and thereby compute the GW memory effects), it will be necessary to compute the leading $1/r$ parts of the Riemann tensor components ${R_{\hat 0 \hat j \hat 0}}^{\hat i}$, the geodesic with tangent $u^\alpha$, the infinitesimal element of proper time $d\tau$, and the orthonormal triad $e_{\hat i}^\alpha$.
In Bondi coordinates, with $V$ given by Eq.~\eqref{eq:V_sol}, a vector $\vec u = \vec e_{\hat 0}$ that is tangent to a timelike geodesic to leading order in $1/r$ is given by
\begin{subequations}
\begin{equation}
    \vec u = \vec \partial_u - \frac 1{2\lambda_0} \dot \lambda_1 \vec\partial_r + O(r^{-1}) \, .
\end{equation}
The retarded time $u$ is the proper time $\tau$ along the geodesic at this order. 
A useful triad is given by
\begin{align}
\vec e_{\hat r} = {} & \vec \partial_u - \left(1 + \frac 1{2\lambda_0} \dot \lambda_1 \right) \vec\partial_r + O(r^{-1}) \, , \\
\vec e_{\hat A} = {} & \frac 1r \vec{\mathsf e}_{\hat A} + O(r^{-2}) \, ,
\end{align}
\end{subequations}
where $\vec{\mathsf e}_{\hat A}$ is an orthonormal dyad associated with the metric $q_{AB}$.
The nonzero tetrad components of the Riemann tensor at $O(r^{-1})$ are given by
\be 
R_{\hat 0\hat A\hat 0 \hat B}=-\frac{1}{2r} \ddot  c_{\hat A\hat B} + \frac{1}{2\lambda_0 r} \delta_{\hat A\hat B} \ddot \lambda_1 + O\lb r^{-2}\rb \, .
\ee
Note that if the Riemann tensor is decomposed into its Weyl and Ricci parts, the relevant nonzero components are given by
\begin{subequations}
\begin{align}
    C_{\hat 0\hat A\hat 0 \hat B} = &-\frac{1}{2r} \ddot c_{\hat A\hat B} + O\lb r^{-2}\rb \, ,\\
    R_{\hat 0\hat 0} = & \frac 1{\lambda_0 r} \ddot \lambda_1 + O\lb r^{-2}\rb \, .
\end{align}
\end{subequations}
It then follows that the Ricci scalar, $R$, satisfies 
$R= O\lb r^{-2}\rb$.

Putting these results together, we find that the $O(r^0)$ part of $\Delta X^{\hat i}$ is the same as the flat-space result in Eq.~\eqref{eq:DeltaXi_flat}, and the $O(r^{-1})$ part is given by
\begin{align}
    \Delta X_{\hat A}^{(1)} = {} & \frac 1{2r} \left(\Delta c_{\hat A\hat B} - \frac 1{\lambda_0} \Delta \lambda_1 \delta_{\hat A\hat B}\right) X^{\hat B}_0 \nonumber \\
    & - \frac 1{r} \left(\Delta \mathcal C_{\hat A\hat B} - \frac 1{\lambda_0} \Delta \Lambda_1 \delta_{\hat A\hat B}\right) \dot X^{\hat B}_0 \nonumber \\
    & + \frac {1}{2r} \Delta \left[ u c_{\hat A\hat B}(u) - \frac 1{\lambda_0} u \lambda_1 (u) \delta_{\hat A\hat B}\right] \dot X^{\hat B}_0  \nonumber \\
    & + \frac {\Delta u}{2r} \left[ c_{\hat A\hat B}(u_0) - \frac 1{\lambda_0} \lambda_1 (u_0) \delta_{\hat A\hat B}\right] \dot X^{\hat B}_0 \nonumber \\
   &  -\frac{u_0}{2r} \left(\Delta c_{\hat A\hat B} - \frac 1{\lambda_0} \Delta \lambda_1 \delta_{\hat A\hat B}\right) \dot X^{\hat B}_0 \, .
    \label{eq:DeltaXsol}
\end{align}
We have defined $\Delta u = u_f - u_0$,
\begin{equation}
    \Delta \mathcal C_{\hat A\hat B} = \int_{u_0}^{u_f} du \, c_{\hat A\hat B} \, ,
    \quad \mbox{and} \quad 
    \Delta \Lambda_1 = \int_{u_0}^{u_f} du \, \lambda_1 \, ; 
\end{equation}
in the third line of Eq.~\eqref{eq:DeltaXsol}, the $\Delta$ of the quantity in square brackets means to take the difference of the quantity within the brackets at $u=u_f$ and $u=u_0$.
Equation~\eqref{eq:DeltaXsol} contains (in addition to initial and final data) six memory effects, which we will now discuss in greater detail (or in the language of~\cite{Flanagan:2018yzh,Flanagan:2019ezo} six persistent observables, three of which are memory effects).

The first two collections of effects, $\Delta c_{\hat A\hat B}$ and $\Delta \mathcal C_{\hat A\hat B}$, have the same type of effect on nearby freely falling observers as GW memory effects in GR: namely, they produce a shearing (transverse to the propagation direction of the gravitational waves) of an initially circular congruence of geodesics after a burst of GWs pass.
The tensor $\Delta c_{\hat A\hat B}$ was the first type of GW memory effect identified in calculations, and it produces a lasting change in the deviation vector between initially comoving observers.
When $\Delta c_{\hat A\hat B}$ is nonvanishing, then the tensor $\Delta \mathcal C_{\hat A\hat B}$ will be the sum of a term that grows with $u$ after the burst passes and a term $\Delta \mathcal C_{\hat A\hat B}^{(0)}$ that is independent of $u$.
For observers with an initial relative velocity, this will cause $\Delta X_{\hat A}^{(1)}$ to have a shearing part that grows linearly with $u$ after the GWs pass (this effect is also sometimes called the ``subleading displacement memory'').
The electric- and magnetic-parity parts of the tensor $\Delta \mathcal C_{\hat A\hat B}^{(0)}$ are closely related to the spin and center-of-mass (CM) GW memory effects, that were more recently identified.
The tensor $\Delta c_{\hat A\hat B}$ was frequently described as being of electric parity, but it was shown that there are sources of stress-energy that can produce a magnetic-parity $\Delta c_{\hat A\hat B}$~\cite{madler:2016ggp,Satishchandran:2019pyc,Bieri:2020pee}.

The second two terms, $\Delta \lambda_1$ and $\Delta \Lambda_1$, are memory effects related to the passage of the scalar field.
These effects cause an initially circular congruence of geodesics to undergo a relative uniform expansion (or contraction) in the direction transverse to the propagation direction of the scalar radiation.\footnote{It is possible to define a suitably adapted Newman-Penrose tetrad~\cite{Newman:1961qr} with $l_\mu = \nabla_\mu u$ and with a complex dyad chosen to have only its 2-sphere indices nonvanishing and to be normalized to one: $m_A\bar m^A=1$.
The spin coefficient $\rho = -m^\mu \bar m^\nu \nabla_\nu l_\mu$ then can be expanded at large Bondi radius $r$ in this tetrad as
\begin{equation}
\rho=-\frac{1}{r^2}m^A\bar m^B\nabla_B \nabla_A u=\frac{1}{r}+\frac{\lambda_1}{2\lambda_0 r^2}+ O (r^{-3}) \, .
\end{equation}
As $\rho$ is one of the ``optical scalars'' and its real part corresponds to the expansion of a congruence to which $l^\mu$ is tangent, this provides a second geometrical viewpoint on how $\lambda_1$ causes a type of expansion at large $r$.}  
Thus, for initially comoving observers, a nonzero $\Delta \lambda_1$ would cause a uniform, transverse change in $\Delta X_{\hat A}^{(1)}$.
When $\Delta \lambda_1$ is nonvanishing, then $\Delta \Lambda_1$ would also be a sum of term that grows with $u$ after the burst of scalar field and a term $\Delta \Lambda_1^{(0)}$ that is independent of $u$; thus, the deviation vector $\Delta X_{\hat A}^{(1)}$ would have an expanding (or contracting) part that grows linearly with $u$ for observers with an initial relative velocity.
The scalar-field memory effect $\Delta \lambda_1$ had been discussed in the context of post-Newtonian theory in~\cite{Lang:2013fna} or in gravitational collapse in~\cite{Du:2016hww,Koyama:2020vfc}, for example. The quantity $\Delta \Lambda_1^{(0)}$ is the scalar-field analog of the CM memory, and it seems to have not been discussed previously.

We turn in the next part of this section to how the different memory scalars and tensors---$\Delta c_{\hat A\hat B}$, $\Delta \mathcal C_{\hat A\hat B}$, $\Delta \lambda_1$ and $\Delta \Lambda_1$---are constrained (or not constrained) by the asymptotic field equations of Brans-Dicke theory and the properties of the nonradiative regions before and after the passages of the gravitational waves and the radiative scalar field.

\subsection{Constraints on GW memory effects from fluxes of conserved quantities}

Memory effects were defined in~\cite{Flanagan:2018yzh} to be the subset of the persistent observables that are associated with symmetries and conserved quantities at spacetime boundaries, like null infinity.
A commonly used procedure for computing these conserved quantities related to symmetries is due to Wald and Zoupas~\cite{Wald:1999wa}, who computed the ``conserved'' quantities associated with BMS symmetries at null infinity in vacuum general relativity.
The word ``conserved'' is used in quotes, because these quantities (also called ``charges'') are not constant along cross-sections (or ``cuts'') of null infinity, but change so that the difference of the charges between the cuts is equal to the flux of the charge integrated over the region of null infinity between the cuts.
The flux had been computed previously by Ashtekar and Streubel~\cite{Ashtekar:1981bq}, and it is consistent with the result in~\cite{Wald:1999wa}.
In Bondi coordinates and in general relativity, the change in the charges, $\Delta Q_\xi$, can be concisely expressed by the expression
\begin{equation}
    \Delta Q_\xi = -\frac 1{32\pi G} \int du \, d^2\Omega \, N^{AB} \delta_\xi c_{AB} \, ,
\end{equation}
where $\delta_\xi c_{AB}$ is given in Eq.~\eqref{eq:deltacAB} and $d^2\Omega = \sqrt{q} dx^1 dx^2$ is the two-dimensional volume element associated with the metric $q_{AB}$ (see, e.g.~\cite{Flanagan:2015pxa}).
The charge is given by the Komar formula~\cite{PhysRev.113.934}, with an additional prescription needed to make the charge integrable in radiative regions.

The formalism for computing conserved quantities outlined in~\cite{Wald:1999wa} can be applied to a large class of gravitational theories that can be derived from a Lagrangian, such as Brans-Dicke theory.
In the Einstein frame, the action has the form of the Einstein-Klein-Gordon theory.
Wald and Zoupas noted in~\cite{Wald:1999wa} that having a minimally coupled scalar field causes stress-energy terms to be added to the flux, but will otherwise not greatly change the charges.
However, they posited that $r \Phi$ has a finite limit to null infinity, which would require that $\Phi_0 = 0$.
We checked whether having a constant $\Phi_0$ that is nonzero would alter the flux, and because this nonzero $\Phi_0$ is constant, and we found that it did not.
Wald and Zoupas also mentioned in~\cite{Wald:1999wa} that a conformally coupled scalar field, such as in Brans-Dicke theory in the Jordan frame, would also only add terms to the flux.
However, they did not specify whether the kinetic term for the scalar field must have the canonical form (as in the Einstein frame), which it does not in the Jordan frame.
Consequently, we computed the flux of the charges associated with a BMS symmetry in the Jordan frame in Bondi coordinates.
We found that the integral of the flux over a region of future null infinity is given by
\begin{align}
    \Delta Q_{\vec\xi} = -\frac{\lambda_0}{32\pi} \int du \, d^2\Omega \, \bigg[
    & N^{AB} \delta_\xi c_{AB} \nonumber \\
    & + \frac{6+4\omega}{(\lambda_0)^2} \partial_u \lambda_1 \delta_\xi \lambda_1 \bigg] \, .
    \label{eq:fluxBD}
\end{align}
Note that by combining Eqs.~\eqref{eq:Einstein-stress-energy} and~\eqref{eq:lambda_Phi}, expanding $\lambda$ as in Eq.~\eqref{eq:lambda_exp}, and using Eq.~\eqref{eq:lambda1}, then we find that the term $(3+2\omega)\partial_u \lambda \delta_\xi\lambda/(16\pi)$ in Eq.~\eqref{eq:fluxBD} is the $O(r^{-2})$ part of $\lambda_0 T^{(\Phi)}_{u \nu}\xi^\nu$.
Thus, the result is consistent with the expectations of Wald and Zoupas for a conformally coupled scalar field, despite the noncanonical form of the kinetic term for $\lambda$ (and the flux is then conformally invariant as required in~\cite{Wald:1999wa}).

\subsubsection{Displacement memory and electric-parity part of $\Delta c_{AB}$}

For computing the GW memory effect connected with the electric-parity part of $\Delta c_{\hat A\hat B}$, we should specialize the flux expression for a supertranslation vector field $\vec\xi = \alpha(x^A) \vec \partial_u$.
Restricting Eqs.~\eqref{eq:deltacAB} and~\eqref{eq:lambda1} to a supertranslation, and integrating by parts for terms involving $\eth_A$ (there are no boundary terms on the 2-sphere), we can show that the expression in Eq.~\eqref{eq:fluxBD} can be written as
\begin{align}
    \Delta Q_{(\alpha)} = -\frac{\lambda_0}{32\pi} \int du \, d^2\Omega \, \alpha \bigg[ &
    N_{AB} N^{AB} - 2 \eth_A \eth_B N^{AB} \nonumber \\
    & + \frac{6+4\omega}{(\lambda_0)^2} (\partial_u \lambda_1)^2 \bigg] \, .
    \label{eq:fluxBDalpha}
\end{align}
It will next be useful to make a few definitions and to relate some of the quantities in Eq.~\eqref{eq:fluxBDalpha} to quantities that we have computed earlier in the paper.

The term proportional to $\eth_A \eth_B N^{AB}$ depends only on the electric part of $N_{AB}$ (and thus the electric part of $\Delta c_{AB}$, when the integral with respect to $u$ is performed).
With the definition of $c_{AB}$ in Eq.~\eqref{eq:cABdecomp} and of the news tensor in Eq.~\eqref{eq:news_def}, we can write the term $\eth_A \eth_B N^{AB}$ as
\begin{equation}
    2\eth_A \eth_B N^{AB} = \ETH^2 (\ETH^2+2) \partial_u \Theta \, .
\end{equation}
With the equation for the Bondi mass aspect~\eqref{eq:mass_aspect_evolution}, it is possible to show that the supertranslation charge (i.e., the supermomentum) needed to satisfy Eq.~\eqref{eq:fluxBDalpha} is given by 
\begin{equation}
    Q_{(\alpha)} = \frac{\lambda_0}{4\pi} \int d^2\Omega \, \alpha \left(  \mathcal M - \frac 1{4\lambda_0} \ETH^2 \lambda_1 \right) \, .
\end{equation}
Note that when $\alpha=1$, this corresponds to a time translation, and the associated conserved charge is the energy. 
Solutions of physical interest have non-negative energy.
Because $\ETH^2\lambda_1$ vanishes when integrated over the 2-sphere, this implies that the integral of $\mathcal M$ over $S^2$ must be non-negative, or the 2-sphere integral of $M$ must be greater than or equal to the same integral of $\frac{1}{4\lambda_0^2}\lambda_1\partial_u\lambda_1$.
Finally, it will be helpful to define $\alpha$ times the change in the energy radiated by the gravitational waves and the scalar field $\lambda$ as
\begin{equation}
    \Delta \mathscr E_{(\alpha)} = \frac{\lambda_0}{32\pi} \int du \, d^2\Omega \, \alpha \left[ N_{AB} N^{AB} + \frac{6+4\omega}{(\lambda_0)^2} (\partial_u \lambda_1)^2 \right] \, .
\end{equation}
Then Eq.~\eqref{eq:fluxBDalpha} can be written as
\begin{equation}
    \int d^2\Omega \, \alpha \,  \ETH^2 (\ETH^2+2) \Delta \Theta = \frac{32\pi}{\lambda_0} \left( \Delta \mathscr E_{(\alpha)} + \Delta Q_{(\alpha)} \right) \, .
    \label{eq:memory_alpha}
\end{equation}

The supertranslations $\alpha$ are allowed to be any smooth function on the 2-sphere.
By choosing for $\alpha$ an appropriate basis of functions that span this space of smooth functions on $S^2$ (e.g., spherical harmonics), it is then possible to use Eq.~\eqref{eq:memory_alpha} to determine the coefficients of $\Delta\Theta$ expanded in these basis functions in terms of the expansion coefficients of the energy flux and the change in the supermomentum charges.
In other words, supposing that the energy flux $\Delta\mathscr E_{(\alpha)}$ is known for some given radiative data $N_{AB}$ and $\partial_u \lambda_1$, and that the early- and late-time nonradiative data through $\Delta \mathcal M$ and $\Delta \lambda_1$ are also known, then it is possible to determine the corresponding electric-parity memory effect in $\Delta c_{AB}$.
The computation of this memory effect is not substantially different from in general relativity; the main difference is that it is necessary to provide both radiative ($\partial_u\lambda_1$) and nonradiative ($\Delta\lambda_1$) data for the scalar field, in addition to the radiative ($N_{AB}$) and nonradiative ($\Delta\mathcal M$) gravitational data.
We will use this procedure to calculate the memory effect from compact binaries in Brans-Dicke theory in future work.

The two types of sources of GW memory in Eq.~\eqref{eq:memory_alpha}---i.e., $\Delta \mathscr E_{(\alpha)}$ and $\Delta Q_{(\alpha)}$---are often called ``null'' and ``ordinary'' memory, respectively in general relativity~\cite{Bieri:2013ada}.
The word ``null'' refers to the fact that it is sourced by massless fields (including gravitational waves), and the word ``ordinary'' refers to the fact that it is sourced by ``ordinary'' massive particles (and fields). 
The specific components of the spacetime curvature and matter stress-energy tensor responsible for producing the ordinary and null memory are distinct and distinguishable in GR.
How to classify the contributions of a scalar field to the ordinary and null memory is not as immediately obvious in Brans-Dicke theory as it is in GR, because (i) massive objects can have ``scalar charges'' (nontrivial stationary scalar field configurations of the massless scalar) in Brans-Dicke theory, and (ii) the radiative and the static parts of the scalar field both appear at leading order in $1/r$.
While in GR all terms involving the scalar field would be treated as null memory, in Brans-Dicke theory, we will consider one part of the scalar field to contribute to the null memory and another part to contribute to the ordinary memory.
Specifically, because the term quadratic in $\partial_u \lambda_1$ in the energy flux $\Delta \mathscr E_{(\alpha)}$ has the form of a flux of energy per solid angle, we will consider it to be null memory.
The term proportional to $\ETH^2 \Delta \lambda_1$ enters in the charge $\Delta Q_{(\alpha)}$ and is linear in $\lambda_1$, so we treat it as a source of ordinary memory for the shearing GW memory $\Delta c_{AB}$.
\footnote{There is a second possibility that one might have considered the change in $\ETH^2 \Delta \lambda_1$ to be a scalar GW memory that is constrained at the same time as the tensor-type memory through the flux $\Delta\mathcal E_{(\alpha)}$ and the change in the integral of $\alpha$ times the mass aspect $\mathcal M$.
However, this is not a viable option, because to specify the properties of the initial and final nonradiative states, one has to specify the nonradiative value of the scalar field $\lambda_1$ (analogously to how one has to specify the value of the mass aspect $\mathcal M$.
Thus, there is no freedom to constrain the value of $\lambda_1$ through the memory equation~\eqref{eq:memory_alpha}.
This does have the noteworthy consequence that to determine the tensor-type memory $\Delta\Theta$, one needs to know the scalar memory $\Delta \lambda_1$ to be able to compute the term  $\Delta\ETH^2\lambda_1$ that enters into the ordinary memory $\Delta Q_{(\alpha)}$.}

Because the right-hand side of~\eqref{eq:memory_alpha} is determined by the changes in $\mathcal M$ and $\Delta \lambda_1$ (for the term $\Delta Q_{(\alpha)}$) and the change in the flux of tensor and scalar waves (for the term $\Delta \mathcal E_{(\alpha)}$), then we can solve for $\Delta \Theta$ in Eq.~\eqref{eq:memory_alpha} in terms of a sum of these two contributions.
We will then write this solution for the total potential as a sum of two terms
\begin{equation}
    \Delta \Theta = \Delta \Theta_{(\mathrm n)} + \Delta \Theta_{(\mathrm o)} \, ,
\end{equation}
which correspond to the solutions for the null and ordinary parts, separately.
This splitting will be useful in discussing the CM memory effect in the next part.

Lastly, note that no constraints on the magnetic-parity part of $\Delta c_{AB}$ are found from supermomentum conservation.
Thus, it would be classified as a persistent observable rather than a memory effect in the language of~\cite{Flanagan:2018yzh}.

\subsubsection{Subleading displacement memory and $\Delta \mathcal C_{AB}$}
 
In the BMS group, there are also symmetries parameterized by the vector field on the 2-sphere, $Y^A$.
This vector field is required to be a conformal Killing vector on the 2-sphere from Eq.~\eqref{eq:YckvS2}; the space of such vector fields that are globally defined form a six-dimensional algebra, which is isomorphic to the Lorentz algebra of
3+1 dimensional Minkowski spacetime.
There have also been proposals to consider extensions of the BMS algebra that enlarge the symmetry algebra by including either all the conformal Killing vectors on the 2-sphere that have complex-analytic singularities~\cite{Barnich:2009se,Barnich:2010eb} or all smooth vector fields on the 2-sphere~\cite{Campiglia:2014yka,Campiglia:2015yka}.
When the Wald-Zoupas prescription was applied to these extended BMS algebras, it was found that there needed to be an additional term to the flux (or the change in the charges) to maintain that the difference in the charges was equal to the flux~\cite{Flanagan:2015pxa}.
For the smooth vector fields, it was shown that this related term could be absorbed into the definition of the charges~\cite{Compere:2018ylh}.
This new term was closely related to a new type of GW memory effect called GW spin memory~\cite{Pasterski:2015tva}.
There was also a second type of new GW memory related to these extended symmetries called GW CM memory~\cite{Nichols:2018qac}.
These two new memory effects are related to the electric- and magnetic-parity parts of the subleading displacement memory in $\Delta \mathcal C_{AB}$.
We now discuss the computation of these effects in Brans-Dicke theory.

First, we write the change in the charges associated with an extended BMS algebra element $\vec \xi = Y^A \vec \partial_A $ for a smooth vector field $Y^A$. 
Starting from Eq.~\eqref{eq:fluxBD} and integrating by parts to simplify the expression, we find
\bw
\begin{align}
    \Delta Q_{(Y)} = -\frac{\lambda_0}{32\pi} \int du \, d^2\Omega \, Y^A \Bigg\{ & \frac u2 \eth_A \bigg[ 2 \eth_B \eth_C N^{BC} - N_{BC} N^{BC} -  \frac{6+4\omega}{(\lambda_0)^2} (\partial_u \lambda_1)^2 \bigg] + \frac 12 \eth_A (c_{BC} N^{BC}) \nonumber \\
    & + N^{BC} \eth_A c_{BC} - 2 \eth_B (c_{AC} N^{BC}) + \frac{2\omega+3}{(\lambda_0)^2} (\partial_u \lambda_1 \eth_A \lambda_1 - \lambda_1 \eth_A \partial_u \lambda_1) \Bigg\} + \Delta \mathcal F_{(Y)} \, ,
    \label{eq:fluxBDY}
\end{align}
\ew
where $\Delta \mathcal F_{(Y)}$ is the additional term needed to relate the change in the charges to the flux integral.
It is given by
\begin{equation}
    \Delta \mathcal F_{(Y)} = \frac{\lambda_0}{64\pi} \int d^2\Omega \, Y^A \epsilon_{AB} \eth^B \ETH^2(\ETH^2 + 2) \Delta \Sigma \, ,
\end{equation}
where we introduced the notation of~\cite{Nichols:2017rqr} for the $u$ integral of $\Psi$
\begin{equation}
    \Delta\Sigma = \int du \, \Psi \, ,
\end{equation}
and where $\Psi$ determines the magnetic-parity part of $c_{AB}$ in Eq.~\eqref{eq:cABdecomp}.\footnote{The modification to the charge defined in~\cite{Compere:2018ylh} is similar to the quantity $\Delta \mathcal F_{(Y)}$, but instead of $\Delta\Sigma$, a term proportional to $u\Psi$ was used instead.}
The GW spin memory effect is related to the quantity $\Delta\Sigma$, which determines the magnetic-parity part of $\Delta\mathcal C_{AB}$.
In the absence of magnetic-parity displacement memory $\Delta c_{AB}$, the spin memory will be independent of $u$, and given by just the magnetic-parity part of $\Delta\mathcal C_{AB}^{(0)}$.

Let us now make a few additional definitions.
Note that in Eq.~\eqref{eq:fluxBDY}, there is a term that is linear in the news tensor $N_{AB}$, like the term that gives rise to the displacement memory; however, the term in~\eqref{eq:fluxBDY} is multiplied by $u$.
When this term is integrated over $u$, the resulting quantity has dimensions or strain multiplied by time, like the GW spin memory.
It was argued in~\cite{Nichols:2018qac} that a quantity related to this term is responsible for a new type of GW memory called GW center-of-mass (CM) memory.
Specifically, consider the quantity defined by $u$ times the $u$ integral of $\partial_u \Theta$, with the part of $\partial_u \Theta$ responsible for the ordinary memory $\Delta \Theta_{(\mathrm o)}$; i.e.,
\begin{equation}
    \Delta \mathcal K = \int du \, u \partial_u (\Theta - \Theta_{(\mathrm n)}) \, .
\end{equation}
Then the integral of the term in square brackets in Eq.~\eqref{eq:fluxBD} can be written in the form
\begin{equation}
    \Delta \mathcal C_{(Y)} = - \frac{\lambda_0}{64\pi} \int d^2\Omega \, Y^A \eth_A \ETH^2( \ETH^2 + 2) \Delta \mathcal K \, .
\end{equation}
Finally, define the remaining terms in Eq.~\eqref{eq:fluxBDY} to be \begin{align}
    \Delta \mathcal J_{(Y)} = \frac{\lambda_0}{64\pi} & \int du \, d^2\Omega \, Y^A \bigg[ \eth_A (c_{BC} N^{BC}) \nonumber \\
    & + 2 N^{BC} \eth_A c_{BC} - 4 \eth_B (c_{AC} N^{BC}) \nonumber \\
    & + \frac{4\omega+6}{(\lambda_0)^2} (\partial_u \lambda_1 \eth_A \lambda_1 - \lambda_1 \eth_A \partial_u \lambda_1) \bigg] \, ,
    \label{eq:JofY}
\end{align}
which are the moments of the change in the super angular momentum with respect to the vector field $Y^A$.
With these definitions, Eq.~\eqref{eq:fluxBDY} reduces to the expression
\begin{equation}
    \Delta Q_{(Y)} = -\Delta \mathcal J_{(Y)} + \Delta \mathcal C_{(Y)} +\Delta \mathcal F_{(Y)}
\end{equation}
Using the evolution equation for the Bondi mass aspect~\eqref{eq:LAdot}, we can show that the definition of the charge needed to satisfy Eq.~\eqref{eq:fluxBDY} is given by
\begin{align}
    Q_{(Y)} = & \frac{\lambda_0}{8\pi} \int d^2\Omega \, Y^A \Bigg[ -u \, \eth_A \bigg(\mathcal M - \frac 1{4\lambda_0} \ETH^2 \lambda_1 \bigg) - 3L_A \nonumber \\
    & + \frac{1}{32}\eth_A(c_{BC} c^{BC}) + \frac{1}{4\lambda_0} \eth_A \bigg( 3\lambda_2 + \frac{\omega-1}{2\lambda_0}(\lambda_1)^2 \bigg) \nonumber \\
    & - \frac 1{4\lambda_0}(c_{AB} \eth^B \lambda_1 - \lambda_1 \eth^B c_{AB})\Bigg] \, .
    \label{eq:QofY}
\end{align}
Next, it is useful to consider decomposing the vector field $Y^A$ into gradient and curl parts via the expression
\begin{equation}
    Y^A = \eth^A \beta(x^C) + \epsilon^{AB}\eth_B \gamma(x^C) \, ,
\end{equation}
(for smooth functions $\beta$ and $\gamma$) and to treat the case of divergence- and curl-free vector fields $Y^A$ separately.
This will allow us to isolate the GW spin and CM memory effects.

\paragraph*{CM memory and electric-parity $Y^A$}
Let us first specialize to $Y^A = \eth^A \beta$.
The term $\Delta \mathcal F_{(Y)}$ vanishes for vector fields $Y^A$ of this type.
After integrating by parts, this means that we can determine the CM memory through the equation 
\begin{equation}
    \int d^2\Omega \, \beta \,  \ETH^4 (\ETH^2 +2) \Delta \mathcal K = \frac{64\pi}{\lambda_0} \left( \Delta \mathcal J_{(\beta)} + \Delta Q_{(\beta)} \right) \, .
\end{equation}
In the above equation, we have defined $\ETH^4 = (\ETH^2)^2$, and we have let $\Delta Q_{(\beta)}$ and $\Delta \mathcal J_{(\beta)}$ given in Eqs.~\eqref{eq:QofY} and~\eqref{eq:JofY} be the change in the charges and in a part of the flux associated with the vector field $Y^A = \eth^A \beta$.
The procedure for computing the CM memory works similarly to that for computing the standard GW memory described by the potential $\Delta\Theta$: (i) first pick a basis of functions for the smooth function $\beta$ on $S^2$ to determine the coefficients of $\Delta\mathcal K$ expanded in this basis (perhaps most usefully, spherical harmonics); (ii) then provide radiative (and some nonradiative) data in the functions $\lambda_1$, $c_{AB}$, and their $u$ derivatives to evaluate the basis-function coefficients of the flux term $\Delta \mathcal J_{(\beta)}$; (iii) next specify the nonradiative data in $\mathcal M$, $L_A$, $c_{AB}$, $\lambda_1$, and $\lambda_2$ to evaluate the coefficients of the change in the charges $\Delta Q_{(\beta)}$; (iv) finally, solve for the relevant coefficients of $\Delta\mathcal K$ by acting on it with the elliptic operator $\ETH^4(\ETH^2+2)$ and performing the integral.
Some interesting differences from the standard GW memory are that the flux term involves $c_{AB}$ and $\lambda_1$ in addition to $\partial_u \lambda_1$ and $N_{AB}$, and the charge involves $c_{AB}$, $L_A$, and $\lambda_2$ in addition to $\lambda_1$ and $\mathcal M$.

\paragraph*{Spin memory and magnetic-parity $Y^A$}
Next we shall discuss vector fields given by $Y^A = \epsilon^{AB}\eth_B \gamma$.
In this case, it is the term $\Delta \mathcal C_{(Y)}$ that vanishes, and one can solve for the spin memory through the equation
\begin{equation}
    \int d^2\Omega \, \gamma \,  \ETH^4 (\ETH^2 +2) \Delta \Sigma = - \frac{64\pi}{\lambda_0} \left( \Delta \mathcal J_{(\gamma)} + \Delta Q_{(\gamma)} \right) \, .
\end{equation}
The prescription used to determine the coefficients of the potential $\Delta\Sigma$ when expanded in a basis of functions on $S^2$ works nearly identically to that for the expansion of $\Delta\mathcal K$ for the spin memory.
The main difference is that less nonradiative data is needed to determine the spin memory.
Specifically, because the quantities $\eth_A\lambda_2$ and $\eth_A\mathcal M$ enter into the charge as gradients, then these terms will vanish for a magnetic-parity vector field of the form $Y^A = \epsilon^{AB}\eth_B \gamma$.
Thus, computing the spin memory does not require knowledge of the functions $\lambda_2$ and $\mathcal M$.

\subsection{Summary and discussion}

To summarize, in asymptotically flat general relativity in Bondi coordinates, there are four types of memory effects that are encoded in the electric- and magnetic-parity parts of the tensors $\Delta c_{AB}$ and $\Delta \mathcal C_{AB}$.
All four memory effects can be measured through geodesic deviation, and they produce a type of shearing of a family of deviation vectors pointing from some fiducial timelike worldline far from a source of gravitational waves.
The memory effects encoded in $\Delta c_{AB}$ are related to the dependence of the final deviation vector on the initial deviation vector; the memory effects encapsulated in $\Delta \mathcal C_{AB}$ are connected to the dependence of the final deviation on the initial relative velocity of the deviation vector. 
Three of the four memory effects were constrained by conservation laws for charges associated with the (extended) BMS algebra.
Specifically, the electric-parity part of $\Delta c_{AB}$ is constrained through the statement of supermomentum conservation associated with the supertranslation symmetries of the BMS group.
The electric- and magnetic-parity parts of $\Delta \mathcal C_{AB}$ were determined through the conservation of super-angular momentum conjugate to the super-Lorentz symmetries of the extended BMS algebra.
The magnetic-parity part of $\Delta c_{AB}$ does not seem to have any conservation equation that constrains its value (and thus might be classified as just a persistent observable).

In our treatment of asymptotically flat solutions of Brans-Dicke theory in Bondi coordinates, we observed that there were a total of six types of memory effects: the four that exist in general relativity, and two more that are related to the leading-order dynamical part of the scalar field, $\lambda_1$.
The two new memory effects also could be measured through geodesic deviation, though they would produce an expansion (or contraction) of the family of deviation vectors pointing orthogonally away from a given worldline (a so-called ``breathing'' mode).
The memory effect $\Delta \lambda_1$ was related to the amplitude of this effect which depends on the initial deviation vector, and the effect in $\Delta \Lambda_1$ corresponded to the scale of the effect depending on the initial relative velocity of the nearby worldlines.
The quantities $\Delta \lambda_1$ and $\Delta \Lambda_1$ were not constrained by any conservation laws associated with conserved quantities in asymptotically flat spacetimes in Brans-Dicke theory (so they would also just be persistent observables).
Rather, because the symmetries of asymptotically flat solutions of Brans-Dicke theory are the same as those of general relativity, the same three types of memory effects are constrained by the fluxes of conserved quantities as in general relativity.
Because the definition of the flux and charges includes additional radiative and non-radiative data (namely, $\partial_u \lambda_1$, $\lambda_1$, and $\lambda_2$), the precise expressions used for computing the memory effects and the data necessary to compute these effects differs in Brans-Dicke theory from the expressions used in general relativity.

\section{Conclusions}\label{sec:conclusions}

In this paper, we investigated asymptotically flat solutions of Brans-Dicke theory in Bondi-Sachs coordinates. 
We solved the field equations of this theory, and we found that they have a similar structure to the Bondi-Sachs form of the Einstein equations in general relativity.
The expansions of the metric and the Ricci tensor in series in $1/r$ ($r$ being the areal radius) have somewhat different forms from the equivalent quantities in general relativity. 
Specifically, the Ricci tensor in Bondi coordinates scales like $1/r$, which allows for a scalar (or breathing-mode) gravitational-wave polarization not present in general relativity; other coefficients in the metric also fall off more slowly with $1/r$ in Brans-Dicke theory than in general relativity to accommodate this additional GW polarization. 
Interestingly, this different ``peeling'' property of the Ricci tensor does not affect the asymptotic symmetry group in Brans-Dicke theory, which remains the Bondi-Metzner-Sachs group (though the way in which these symmetries are extended into the interior of the spacetime in Brans-Dicke theory differs from the related extension in general relativity).
We also computed the properties of nonradiative and stationary regions of spacetime in Brans-Dicke theory, which is important for computing and understanding GW memory effects.

We found six types of memory effects generated after a burst of the scalar field and tensorial gravitational waves pass by an observer's location.
Four of these effects are also present in GR: namely, they are the electric- and magnetic-parity parts of displacement and subleading displacement memories.
These effects produce the familiar, lasting shearing of a ring of freely falling test masses, with the displacement part depending on the initial separation of the test masses, and the subleading displacement part depending on the initial relative velocity of the masses (the electric- and magnetic-parity parts refer to the parity properties of the sky pattern of the memory effect over the anti-celestial sphere).
The amplitude of the memory effects in Brans-Dicke and in GR will differ, because in Brans-Dicke theory, there are additional contributions from the fluxes of energy and angular momentum per solid angle from the scalar field.
The two additional GW memory effects in Brans-Dicke theory are related to the breathing-mode polarization of the gravitational waves, and they could also be classified into leading and subleading displacement terms.
These memory effects cause a ring of freely falling test masses to have an enduring, uniform expansion (or contraction) of a circular congruence of geodesics transverse to GW propagation. The leading part that depends upon the initial displacement of the masses had been previously considered, but the subleading part, which depends on the initial relative velocity of the masses appears not to have been. 
The latter can be thought of as the scalar analog of the center-of-mass memory effect.

Half of these memory effects are constrained by fluxes of conserved quantities associated with the extended BMS group (these are the electric-parity displacement memory, the spin memory, and the center-of mass memory).
The other half (the magnetic-parity displacement memory, and both breathing-mode memory effects) are not, and would be described as being persistent observables in the nomenclature of~\cite{Flanagan:2018yzh,Flanagan:2019ezo}.
For all the memory effects, but particularly for the persistent-observable types, understanding the properties of the nonradiative regions before and after the burst of the scalar field and gravitational waves is important for understanding the set of possible memory effects.
For example, in general relativity, stationary-to-stationary transitions in which the two stationary regions differ by only a supertranslation allow for a wide range of possible electric-parity displacement memory effects; however, in Brans-Dicke theory, such transitions would only allow for scalar-type memory effects with constant sky pattern.
More general types of nonradiative regions at early and late times are necessary to have less trivial memory effects.

Let us conclude with a few comments on future applications and directions for our work. 
It would be interesting to explore the post-Newtonian limit of our results for compact binary systems, so as to make contact with some existing results computed by Lang~\cite{Lang:2013fna,Lang:2014osa}. 
Another potential direction is to explore a broader set of modified gravity theories. We note that our formalism can easily be extended to more general massless scalar-tensor theories, such as those proposed by Damour and Esposito-Far\`ese~\cite{Damour:1992we,Damour:1993hw}. 
It would be interesting to understand whether there are similar relationships between symmetries and memory effects in theories where additional polarizations are present, such as the scalar-vector-tensor theories~\cite{Heisenberg:2018acv}. (A generic theory of gravity can have up to six polarizations, and there would typically be additional GW memory effects associated with all such polarizations.)
Other viable theories of gravity such as the higher curvature theories~\cite{Berti:2015itd} (e.g., dynamical Chern-Simon gravity~\cite{Jackiw:2003pm} and Einstein-dilaton-Gauss-Bonnet gravity~\cite{Pani:2009wy,Moura:2006pz}) and massive scalar-tensor theories would also be useful to explore. 

\acknowledgments
S.T.\ and K.Y.\ acknowledge support from the Owens Innovation Fund.
K.Y.\ would like to acknowledge support from NSF Award PHY-1806776, NASA Grant 80NSSC20K0523, a Sloan Foundation Research Fellowship, the COST Action GWverse CA16104, and JSPS KAKENHI Grants No. JP17H06358.
D.A.N.\ thanks Alexander Grant for helpful correspondence about the solutions to the equation of geodesic deviation.
D.A.N.\ acknowledges support from the NSF grant PHY-2011784. 

\appendix
\section{Field Equations in Jordan Frame}\label{sec:app}
In this appendix, we present scalar wave equation and the hypersurface equations in the Jordan frame.
The \emph{rr, rA,} and the trace of the \emph{AB} components of the modified Einstein equations in Eq.~\eqref{eq:Field} give
\bw
\begin{subequations}
\ba 
\left(\frac{4}{r}+\frac{2\partial_r\lambda}{\lambda} \right)\partial_r\beta-\frac{1}{\lambda}\partial_r\partial_r\lambda-\frac{\omega}{\lambda^2}(\partial_r\lambda)^2
-\frac{1}{4}h^{AB}h^{CD}\partial_r h_{AC}\partial_r h_{BD}&=&0\,,\\
\nonumber\\
\frac{1}{2r^2}\partial_r(r^4e^{-2\beta}h_{AB}\partial_rU^B)-r^{2} \partial_r\left(\frac{1}{r^{2}} D_{A} \beta\right)
+\frac{1}{2} h^{B C} D_{B}\left(\partial_r h_{A C}\right)+\frac{D_A\lambda}{\lambda r}-\frac{\omega}{\lambda^2}\partial_r\lambda D_A \lambda&&\nonumber\\
+\frac{1}{\lambda}D_A\beta\partial_r\lambda
+\frac{1}{2\lambda} h^{BC} D_B\lambda\partial_r h_{AC}
+\frac{1}{2\lambda}e^{-2\beta}h_{AB}r^2\partial_r\lambda\partial_rU^B-\frac{1}{\lambda}\partial_rD_A\lambda
&=&0\,,\\
\nonumber\\
2h^{AB}\lb D_A D_B\beta+D_A\beta D_B\beta\rb-\mathscr R-\frac{1}{r^2}e^{-2\beta}D_A\partial_r(r^4U^A)+\frac{1}{2}r^4e^{-4\beta}h_{AB}\partial_rU^A\partial_rU^B&&\nonumber\\
+2e^{-2\beta}\partial_rV+\frac{r^2}{\lambda}\Box\lambda+\frac{r^2}{\lambda}g^{AB}\nabla_A \nabla_B \lambda+\frac{\omega r^2}{\lambda^2}g^{AB}\nabla_A\lambda\nabla_B\lambda&=&0\,,
\ea
\end{subequations}
respectively. On the other hand, the scalar wave equation, $\Box \lambda=0$, is given by
\begin{align}
2\partial_u \partial_{r} \lambda + D_{A}(U^{A} \partial_{r}\lambda) + \partial_{r}(U^{A}D_{A} \lambda) &- \frac{1}{r} \left(-2 U^{A} D_{A} \lambda-2\partial_{u} \lambda
+ \partial_{r} V\partial_{r} \lambda+V\partial_{r} \partial_{r} \lambda\right)\nonumber\\
& -\frac{1}{r^2}\left[e^{2\beta}h^{AB}\left(2D_A\beta D_B\lambda+D_B D_A\lambda\right) + V \left(\partial_{r} \lambda\right)\right] = 0 \,.
\end{align}

\ew
\bibliography{memory}

\begin{thebibliography}{100}%
\makeatletter
\providecommand \@ifxundefined [1]{%
 \@ifx{#1\undefined}
}%
\providecommand \@ifnum [1]{%
 \ifnum #1\expandafter \@firstoftwo
 \else \expandafter \@secondoftwo
 \fi
}%
\providecommand \@ifx [1]{%
 \ifx #1\expandafter \@firstoftwo
 \else \expandafter \@secondoftwo
 \fi
}%
\providecommand \natexlab [1]{#1}%
\providecommand \enquote  [1]{``#1''}%
\providecommand \bibnamefont  [1]{#1}%
\providecommand \bibfnamefont [1]{#1}%
\providecommand \citenamefont [1]{#1}%
\providecommand \href@noop [0]{\@secondoftwo}%
\providecommand \href [0]{\begingroup \@sanitize@url \@href}%
\providecommand \@href[1]{\@@startlink{#1}\@@href}%
\providecommand \@@href[1]{\endgroup#1\@@endlink}%
\providecommand \@sanitize@url [0]{\catcode `\\12\catcode `\$12\catcode
  `\&12\catcode `\#12\catcode `\^12\catcode `\_12\catcode `\%12\relax}%
\providecommand \@@startlink[1]{}%
\providecommand \@@endlink[0]{}%
\providecommand \url  [0]{\begingroup\@sanitize@url \@url }%
\providecommand \@url [1]{\endgroup\@href {#1}{\urlprefix }}%
\providecommand \urlprefix  [0]{URL }%
\providecommand \Eprint [0]{\href }%
\providecommand \doibase [0]{http://dx.doi.org/}%
\providecommand \selectlanguage [0]{\@gobble}%
\providecommand \bibinfo  [0]{\@secondoftwo}%
\providecommand \bibfield  [0]{\@secondoftwo}%
\providecommand \translation [1]{[#1]}%
\providecommand \BibitemOpen [0]{}%
\providecommand \bibitemStop [0]{}%
\providecommand \bibitemNoStop [0]{.\EOS\space}%
\providecommand \EOS [0]{\spacefactor3000\relax}%
\providecommand \BibitemShut  [1]{\csname bibitem#1\endcsname}%
\let\auto@bib@innerbib\@empty
\bibitem [{\citenamefont {Abbott}\ \emph
  {et~al.}(2016{\natexlab{a}})\citenamefont {Abbott} \emph
  {et~al.}}]{Abbott:2016blz}%
  \BibitemOpen
  \bibfield  {author} {\bibinfo {author} {\bibfnamefont {B.}~\bibnamefont
  {Abbott}} \emph {et~al.} (\bibinfo {collaboration} {LIGO Scientific,
  Virgo}),\ }\href {\doibase 10.1103/PhysRevLett.116.061102} {\bibfield
  {journal} {\bibinfo  {journal} {Phys. Rev. Lett.}\ }\textbf {\bibinfo
  {volume} {116}},\ \bibinfo {pages} {061102} (\bibinfo {year}
  {2016}{\natexlab{a}})},\ \Eprint {http://arxiv.org/abs/1602.03837}
  {arXiv:1602.03837 [gr-qc]} \BibitemShut {NoStop}%
\bibitem [{\citenamefont {Abbott}\ \emph
  {et~al.}(2019{\natexlab{a}})\citenamefont {Abbott} \emph
  {et~al.}}]{LIGOScientific:2018mvr}%
  \BibitemOpen
  \bibfield  {author} {\bibinfo {author} {\bibfnamefont {B.}~\bibnamefont
  {Abbott}} \emph {et~al.} (\bibinfo {collaboration} {LIGO Scientific,
  Virgo}),\ }\href {\doibase 10.1103/PhysRevX.9.031040} {\bibfield  {journal}
  {\bibinfo  {journal} {Phys. Rev. X}\ }\textbf {\bibinfo {volume} {9}},\
  \bibinfo {pages} {031040} (\bibinfo {year} {2019}{\natexlab{a}})},\ \Eprint
  {http://arxiv.org/abs/1811.12907} {arXiv:1811.12907 [astro-ph.HE]}
  \BibitemShut {NoStop}%
\bibitem [{\citenamefont {{GraceDB: Gravitational-Wave Candidate Event
  Database}}(2020)}]{GraceDB}%
  \BibitemOpen
  \bibfield  {author} {\bibinfo {author} {\bibnamefont {{GraceDB:
  Gravitational-Wave Candidate Event Database}}},\ }\href@noop {} {}\bibinfo
  {howpublished} {{https://gracedb.ligo.org/}} (\bibinfo {year}
  {2020})\BibitemShut {NoStop}%
\bibitem [{\citenamefont {Abbott}\ \emph
  {et~al.}(2016{\natexlab{b}})\citenamefont {Abbott} \emph
  {et~al.}}]{Abbott_IMRcon2}%
  \BibitemOpen
  \bibfield  {author} {\bibinfo {author} {\bibfnamefont {B.~P.}\ \bibnamefont
  {Abbott}} \emph {et~al.} (\bibinfo {collaboration} {LIGO Scientific,
  Virgo}),\ }\href {\doibase 10.1103/PhysRevLett.116.221101,
  10.1103/PhysRevLett.121.129902} {\bibfield  {journal} {\bibinfo  {journal}
  {Phys. Rev. Lett.}\ }\textbf {\bibinfo {volume} {116}},\ \bibinfo {pages}
  {221101} (\bibinfo {year} {2016}{\natexlab{b}})},\ \bibinfo {note} {[Erratum:
  Phys. Rev. Lett.121,no.12,129902(2018)]},\ \Eprint
  {http://arxiv.org/abs/1602.03841} {arXiv:1602.03841 [gr-qc]} \BibitemShut
  {NoStop}%
\bibitem [{\citenamefont {Yunes}\ \emph {et~al.}(2016)\citenamefont {Yunes},
  \citenamefont {Yagi},\ and\ \citenamefont {Pretorius}}]{Yunes:2016jcc}%
  \BibitemOpen
  \bibfield  {author} {\bibinfo {author} {\bibfnamefont {N.}~\bibnamefont
  {Yunes}}, \bibinfo {author} {\bibfnamefont {K.}~\bibnamefont {Yagi}}, \ and\
  \bibinfo {author} {\bibfnamefont {F.}~\bibnamefont {Pretorius}},\ }\href
  {\doibase 10.1103/PhysRevD.94.084002} {\bibfield  {journal} {\bibinfo
  {journal} {Phys. Rev. D}\ }\textbf {\bibinfo {volume} {94}},\ \bibinfo
  {pages} {084002} (\bibinfo {year} {2016})},\ \Eprint
  {http://arxiv.org/abs/1603.08955} {arXiv:1603.08955 [gr-qc]} \BibitemShut
  {NoStop}%
\bibitem [{\citenamefont {Abbott}\ \emph
  {et~al.}(2016{\natexlab{c}})\citenamefont {Abbott} \emph
  {et~al.}}]{TheLIGOScientific:2016pea}%
  \BibitemOpen
  \bibfield  {author} {\bibinfo {author} {\bibfnamefont {B.~P.}\ \bibnamefont
  {Abbott}} \emph {et~al.} (\bibinfo {collaboration} {LIGO Scientific,
  Virgo}),\ }\href {\doibase 10.1103/PhysRevX.6.041015,
  10.1103/PhysRevX.8.039903} {\bibfield  {journal} {\bibinfo  {journal} {Phys.
  Rev.}\ }\textbf {\bibinfo {volume} {X6}},\ \bibinfo {pages} {041015}
  (\bibinfo {year} {2016}{\natexlab{c}})},\ \bibinfo {note} {[erratum: Phys.
  Rev.X8,no.3,039903(2018)]},\ \Eprint {http://arxiv.org/abs/1606.04856}
  {arXiv:1606.04856 [gr-qc]} \BibitemShut {NoStop}%
\bibitem [{\citenamefont {Abbott}\ \emph
  {et~al.}(2019{\natexlab{b}})\citenamefont {Abbott} \emph
  {et~al.}}]{Abbott:2018lct}%
  \BibitemOpen
  \bibfield  {author} {\bibinfo {author} {\bibfnamefont {B.~P.}\ \bibnamefont
  {Abbott}} \emph {et~al.} (\bibinfo {collaboration} {LIGO Scientific,
  Virgo}),\ }\href {\doibase 10.1103/PhysRevLett.123.011102} {\bibfield
  {journal} {\bibinfo  {journal} {Phys. Rev. Lett.}\ }\textbf {\bibinfo
  {volume} {123}},\ \bibinfo {pages} {011102} (\bibinfo {year}
  {2019}{\natexlab{b}})},\ \Eprint {http://arxiv.org/abs/1811.00364}
  {arXiv:1811.00364 [gr-qc]} \BibitemShut {NoStop}%
\bibitem [{\citenamefont {Abbott}\ \emph
  {et~al.}(2019{\natexlab{c}})\citenamefont {Abbott} \emph
  {et~al.}}]{LIGOScientific:2019fpa}%
  \BibitemOpen
  \bibfield  {author} {\bibinfo {author} {\bibfnamefont {B.~P.}\ \bibnamefont
  {Abbott}} \emph {et~al.} (\bibinfo {collaboration} {LIGO Scientific,
  Virgo}),\ }\href {\doibase 10.1103/PhysRevD.100.104036} {\bibfield  {journal}
  {\bibinfo  {journal} {Phys. Rev.}\ }\textbf {\bibinfo {volume} {D100}},\
  \bibinfo {pages} {104036} (\bibinfo {year} {2019}{\natexlab{c}})},\ \Eprint
  {http://arxiv.org/abs/1903.04467} {arXiv:1903.04467 [gr-qc]} \BibitemShut
  {NoStop}%
\bibitem [{\citenamefont {Berti}\ \emph
  {et~al.}(2018{\natexlab{a}})\citenamefont {Berti}, \citenamefont {Yagi},\
  and\ \citenamefont {Yunes}}]{Berti:2018cxi}%
  \BibitemOpen
  \bibfield  {author} {\bibinfo {author} {\bibfnamefont {E.}~\bibnamefont
  {Berti}}, \bibinfo {author} {\bibfnamefont {K.}~\bibnamefont {Yagi}}, \ and\
  \bibinfo {author} {\bibfnamefont {N.}~\bibnamefont {Yunes}},\ }\href
  {\doibase 10.1007/s10714-018-2362-8} {\bibfield  {journal} {\bibinfo
  {journal} {Gen. Rel. Grav.}\ }\textbf {\bibinfo {volume} {50}},\ \bibinfo
  {pages} {46} (\bibinfo {year} {2018}{\natexlab{a}})},\ \Eprint
  {http://arxiv.org/abs/1801.03208} {arXiv:1801.03208 [gr-qc]} \BibitemShut
  {NoStop}%
\bibitem [{\citenamefont {Berti}\ \emph
  {et~al.}(2018{\natexlab{b}})\citenamefont {Berti}, \citenamefont {Yagi},
  \citenamefont {Yang},\ and\ \citenamefont {Yunes}}]{Berti:2018vdi}%
  \BibitemOpen
  \bibfield  {author} {\bibinfo {author} {\bibfnamefont {E.}~\bibnamefont
  {Berti}}, \bibinfo {author} {\bibfnamefont {K.}~\bibnamefont {Yagi}},
  \bibinfo {author} {\bibfnamefont {H.}~\bibnamefont {Yang}}, \ and\ \bibinfo
  {author} {\bibfnamefont {N.}~\bibnamefont {Yunes}},\ }\href {\doibase
  10.1007/s10714-018-2372-6} {\bibfield  {journal} {\bibinfo  {journal} {Gen.
  Rel. Grav.}\ }\textbf {\bibinfo {volume} {50}},\ \bibinfo {pages} {49}
  (\bibinfo {year} {2018}{\natexlab{b}})},\ \Eprint
  {http://arxiv.org/abs/1801.03587} {arXiv:1801.03587 [gr-qc]} \BibitemShut
  {NoStop}%
\bibitem [{\citenamefont {Hübner}\ \emph {et~al.}(2020)\citenamefont
  {Hübner}, \citenamefont {Talbot}, \citenamefont {Lasky},\ and\ \citenamefont
  {Thrane}}]{Hubner:2019sly}%
  \BibitemOpen
  \bibfield  {author} {\bibinfo {author} {\bibfnamefont {M.}~\bibnamefont
  {Hübner}}, \bibinfo {author} {\bibfnamefont {C.}~\bibnamefont {Talbot}},
  \bibinfo {author} {\bibfnamefont {P.~D.}\ \bibnamefont {Lasky}}, \ and\
  \bibinfo {author} {\bibfnamefont {E.}~\bibnamefont {Thrane}},\ }\href
  {\doibase 10.1103/PhysRevD.101.023011} {\bibfield  {journal} {\bibinfo
  {journal} {Phys. Rev. D}\ }\textbf {\bibinfo {volume} {101}},\ \bibinfo
  {pages} {023011} (\bibinfo {year} {2020})},\ \Eprint
  {http://arxiv.org/abs/1911.12496} {arXiv:1911.12496 [astro-ph.HE]}
  \BibitemShut {NoStop}%
\bibitem [{\citenamefont {Ebersold}\ and\ \citenamefont
  {Tiwari}(2020)}]{Ebersold:2020zah}%
  \BibitemOpen
  \bibfield  {author} {\bibinfo {author} {\bibfnamefont {M.}~\bibnamefont
  {Ebersold}}\ and\ \bibinfo {author} {\bibfnamefont {S.}~\bibnamefont
  {Tiwari}},\ }\href {\doibase 10.1103/PhysRevD.101.104041} {\bibfield
  {journal} {\bibinfo  {journal} {Phys. Rev. D}\ }\textbf {\bibinfo {volume}
  {101}},\ \bibinfo {pages} {104041} (\bibinfo {year} {2020})},\ \Eprint
  {http://arxiv.org/abs/2005.03306} {arXiv:2005.03306 [gr-qc]} \BibitemShut
  {NoStop}%
\bibitem [{\citenamefont {Boersma}\ \emph {et~al.}(2020)\citenamefont
  {Boersma}, \citenamefont {Nichols},\ and\ \citenamefont
  {Schmidt}}]{Boersma:2020gxx}%
  \BibitemOpen
  \bibfield  {author} {\bibinfo {author} {\bibfnamefont {O.~M.}\ \bibnamefont
  {Boersma}}, \bibinfo {author} {\bibfnamefont {D.~A.}\ \bibnamefont
  {Nichols}}, \ and\ \bibinfo {author} {\bibfnamefont {P.}~\bibnamefont
  {Schmidt}},\ }\href {\doibase 10.1103/PhysRevD.101.083026} {\bibfield
  {journal} {\bibinfo  {journal} {Phys. Rev. D}\ }\textbf {\bibinfo {volume}
  {101}},\ \bibinfo {pages} {083026} (\bibinfo {year} {2020})},\ \Eprint
  {http://arxiv.org/abs/2002.01821} {arXiv:2002.01821 [astro-ph.HE]}
  \BibitemShut {NoStop}%
\bibitem [{\citenamefont {Aggarwal}\ \emph {et~al.}(2020)\citenamefont
  {Aggarwal} \emph {et~al.}}]{Aggarwal:2019ypr}%
  \BibitemOpen
  \bibfield  {author} {\bibinfo {author} {\bibfnamefont {K.}~\bibnamefont
  {Aggarwal}} \emph {et~al.} (\bibinfo {collaboration} {NANOGrav}),\ }\href
  {\doibase 10.3847/1538-4357/ab6083} {\bibfield  {journal} {\bibinfo
  {journal} {\apj}\ }\textbf {\bibinfo {volume} {889}},\ \bibinfo {eid} {38}
  (\bibinfo {year} {2020})},\ \Eprint {http://arxiv.org/abs/1911.08488}
  {arXiv:1911.08488 [astro-ph.HE]} \BibitemShut {NoStop}%
\bibitem [{\citenamefont {Islo}\ \emph {et~al.}(2019)\citenamefont {Islo},
  \citenamefont {Simon}, \citenamefont {Burke-Spolaor},\ and\ \citenamefont
  {Siemens}}]{Islo:2019qht}%
  \BibitemOpen
  \bibfield  {author} {\bibinfo {author} {\bibfnamefont {K.}~\bibnamefont
  {Islo}}, \bibinfo {author} {\bibfnamefont {J.}~\bibnamefont {Simon}},
  \bibinfo {author} {\bibfnamefont {S.}~\bibnamefont {Burke-Spolaor}}, \ and\
  \bibinfo {author} {\bibfnamefont {X.}~\bibnamefont {Siemens}},\ }\href@noop
  {} {\  (\bibinfo {year} {2019})},\ \Eprint {http://arxiv.org/abs/1906.11936}
  {arXiv:1906.11936 [astro-ph.HE]} \BibitemShut {NoStop}%
\bibitem [{\citenamefont {Zel'dovich}\ and\ \citenamefont
  {Polnarev}(1974)}]{Zeldovich:1974gvh}%
  \BibitemOpen
  \bibfield  {author} {\bibinfo {author} {\bibfnamefont {Y.~B.}\ \bibnamefont
  {Zel'dovich}}\ and\ \bibinfo {author} {\bibfnamefont {A.~G.}\ \bibnamefont
  {Polnarev}},\ }\href@noop {} {\bibfield  {journal} {\bibinfo  {journal} {Sov.
  Astron.}\ }\textbf {\bibinfo {volume} {18}},\ \bibinfo {pages} {17} (\bibinfo
  {year} {1974})}\BibitemShut {NoStop}%
\bibitem [{\citenamefont {{Bontz}}\ and\ \citenamefont
  {{Price}}(1979)}]{1979ApJ}%
  \BibitemOpen
  \bibfield  {author} {\bibinfo {author} {\bibfnamefont {R.~J.}\ \bibnamefont
  {{Bontz}}}\ and\ \bibinfo {author} {\bibfnamefont {R.~H.}\ \bibnamefont
  {{Price}}},\ }\href {\doibase 10.1086/156880} {\bibfield  {journal} {\bibinfo
   {journal} {\apj}\ }\textbf {\bibinfo {volume} {228}},\ \bibinfo {pages}
  {560} (\bibinfo {year} {1979})}\BibitemShut {NoStop}%
\bibitem [{\citenamefont {Smarr}(1977)}]{Smarr1977}%
  \BibitemOpen
  \bibfield  {author} {\bibinfo {author} {\bibfnamefont {L.}~\bibnamefont
  {Smarr}},\ }\href {\doibase 10.1103/PhysRevD.15.2069} {\bibfield  {journal}
  {\bibinfo  {journal} {Phys. Rev. D}\ }\textbf {\bibinfo {volume} {15}},\
  \bibinfo {pages} {2069} (\bibinfo {year} {1977})}\BibitemShut {NoStop}%
\bibitem [{\citenamefont {Newman}\ and\ \citenamefont
  {Penrose}(1966)}]{Newman:1966ub}%
  \BibitemOpen
  \bibfield  {author} {\bibinfo {author} {\bibfnamefont {E.}~\bibnamefont
  {Newman}}\ and\ \bibinfo {author} {\bibfnamefont {R.}~\bibnamefont
  {Penrose}},\ }\href {\doibase 10.1063/1.1931221} {\bibfield  {journal}
  {\bibinfo  {journal} {J. Math. Phys.}\ }\textbf {\bibinfo {volume} {7}},\
  \bibinfo {pages} {863} (\bibinfo {year} {1966})}\BibitemShut {NoStop}%
\bibitem [{\citenamefont {{Turner}}(1978)}]{1978Natur.274..565T}%
  \BibitemOpen
  \bibfield  {author} {\bibinfo {author} {\bibfnamefont {M.~S.}\ \bibnamefont
  {{Turner}}},\ }\href {\doibase 10.1038/274565a0} {\bibfield  {journal}
  {\bibinfo  {journal} {\nat}\ }\textbf {\bibinfo {volume} {274}},\ \bibinfo
  {pages} {565} (\bibinfo {year} {1978})}\BibitemShut {NoStop}%
\bibitem [{\citenamefont {{Epstein}}(1978)}]{1978ApJ...223.1037E}%
  \BibitemOpen
  \bibfield  {author} {\bibinfo {author} {\bibfnamefont {R.}~\bibnamefont
  {{Epstein}}},\ }\href {\doibase 10.1086/156337} {\bibfield  {journal}
  {\bibinfo  {journal} {\apj}\ }\textbf {\bibinfo {volume} {223}},\ \bibinfo
  {pages} {1037} (\bibinfo {year} {1978})}\BibitemShut {NoStop}%
\bibitem [{\citenamefont {Christodoulou}(1991)}]{PhysRevLett.67.1486}%
  \BibitemOpen
  \bibfield  {author} {\bibinfo {author} {\bibfnamefont {D.}~\bibnamefont
  {Christodoulou}},\ }\href {\doibase 10.1103/PhysRevLett.67.1486} {\bibfield
  {journal} {\bibinfo  {journal} {Phys. Rev. Lett.}\ }\textbf {\bibinfo
  {volume} {67}},\ \bibinfo {pages} {1486} (\bibinfo {year}
  {1991})}\BibitemShut {NoStop}%
\bibitem [{\citenamefont {Blanchet}\ and\ \citenamefont
  {Damour}(1992)}]{Blanchet:1992br}%
  \BibitemOpen
  \bibfield  {author} {\bibinfo {author} {\bibfnamefont {L.}~\bibnamefont
  {Blanchet}}\ and\ \bibinfo {author} {\bibfnamefont {T.}~\bibnamefont
  {Damour}},\ }\href {\doibase 10.1103/PhysRevD.46.4304} {\bibfield  {journal}
  {\bibinfo  {journal} {Phys. Rev. D}\ }\textbf {\bibinfo {volume} {46}},\
  \bibinfo {pages} {4304} (\bibinfo {year} {1992})}\BibitemShut {NoStop}%
\bibitem [{\citenamefont {Bieri}\ and\ \citenamefont
  {Garfinkle}(2013)}]{Bieri:2013hqa}%
  \BibitemOpen
  \bibfield  {author} {\bibinfo {author} {\bibfnamefont {L.}~\bibnamefont
  {Bieri}}\ and\ \bibinfo {author} {\bibfnamefont {D.}~\bibnamefont
  {Garfinkle}},\ }\href {\doibase 10.1088/0264-9381/30/19/195009} {\bibfield
  {journal} {\bibinfo  {journal} {Class. Quant. Grav.}\ }\textbf {\bibinfo
  {volume} {30}},\ \bibinfo {pages} {195009} (\bibinfo {year} {2013})},\
  \Eprint {http://arxiv.org/abs/1307.5098} {arXiv:1307.5098 [gr-qc]}
  \BibitemShut {NoStop}%
\bibitem [{\citenamefont {Pate}\ \emph {et~al.}(2017)\citenamefont {Pate},
  \citenamefont {Raclariu},\ and\ \citenamefont {Strominger}}]{Pate:2017vwa}%
  \BibitemOpen
  \bibfield  {author} {\bibinfo {author} {\bibfnamefont {M.}~\bibnamefont
  {Pate}}, \bibinfo {author} {\bibfnamefont {A.-M.}\ \bibnamefont {Raclariu}},
  \ and\ \bibinfo {author} {\bibfnamefont {A.}~\bibnamefont {Strominger}},\
  }\href {\doibase 10.1103/PhysRevLett.119.261602} {\bibfield  {journal}
  {\bibinfo  {journal} {Phys. Rev. Lett.}\ }\textbf {\bibinfo {volume} {119}},\
  \bibinfo {pages} {261602} (\bibinfo {year} {2017})},\ \Eprint
  {http://arxiv.org/abs/1707.08016} {arXiv:1707.08016 [hep-th]} \BibitemShut
  {NoStop}%
\bibitem [{\citenamefont {Pasterski}\ \emph {et~al.}(2016)\citenamefont
  {Pasterski}, \citenamefont {Strominger},\ and\ \citenamefont
  {Zhiboedov}}]{Pasterski:2015tva}%
  \BibitemOpen
  \bibfield  {author} {\bibinfo {author} {\bibfnamefont {S.}~\bibnamefont
  {Pasterski}}, \bibinfo {author} {\bibfnamefont {A.}~\bibnamefont
  {Strominger}}, \ and\ \bibinfo {author} {\bibfnamefont {A.}~\bibnamefont
  {Zhiboedov}},\ }\href {\doibase 10.1007/JHEP12(2016)053} {\bibfield
  {journal} {\bibinfo  {journal} {JHEP}\ }\textbf {\bibinfo {volume} {12}},\
  \bibinfo {pages} {053} (\bibinfo {year} {2016})},\ \Eprint
  {http://arxiv.org/abs/1502.06120} {arXiv:1502.06120 [hep-th]} \BibitemShut
  {NoStop}%
\bibitem [{\citenamefont {Nichols}(2018)}]{Nichols:2018qac}%
  \BibitemOpen
  \bibfield  {author} {\bibinfo {author} {\bibfnamefont {D.~A.}\ \bibnamefont
  {Nichols}},\ }\href {\doibase 10.1103/PhysRevD.98.064032} {\bibfield
  {journal} {\bibinfo  {journal} {Phys. Rev. D}\ }\textbf {\bibinfo {volume}
  {98}},\ \bibinfo {pages} {064032} (\bibinfo {year} {2018})},\ \Eprint
  {http://arxiv.org/abs/1807.08767} {arXiv:1807.08767 [gr-qc]} \BibitemShut
  {NoStop}%
\bibitem [{\citenamefont {Bondi}(1957)}]{Bondi:1957dt}%
  \BibitemOpen
  \bibfield  {author} {\bibinfo {author} {\bibfnamefont {H.}~\bibnamefont
  {Bondi}},\ }\href {\doibase 10.1038/1791072a0} {\bibfield  {journal}
  {\bibinfo  {journal} {Nature}\ }\textbf {\bibinfo {volume} {179}},\ \bibinfo
  {pages} {1072} (\bibinfo {year} {1957})}\BibitemShut {NoStop}%
\bibitem [{\citenamefont {Grishchuk}\ and\ \citenamefont
  {Polnarev}(1989)}]{Grishchuk:1989qa}%
  \BibitemOpen
  \bibfield  {author} {\bibinfo {author} {\bibfnamefont {L.}~\bibnamefont
  {Grishchuk}}\ and\ \bibinfo {author} {\bibfnamefont {A.}~\bibnamefont
  {Polnarev}},\ }\href@noop {} {\bibfield  {journal} {\bibinfo  {journal} {Sov.
  Phys. JETP}\ }\textbf {\bibinfo {volume} {69}},\ \bibinfo {pages} {653}
  (\bibinfo {year} {1989})}\BibitemShut {NoStop}%
\bibitem [{\citenamefont {Strominger}\ and\ \citenamefont
  {Zhiboedov}(2016)}]{Strominger:2014pwa}%
  \BibitemOpen
  \bibfield  {author} {\bibinfo {author} {\bibfnamefont {A.}~\bibnamefont
  {Strominger}}\ and\ \bibinfo {author} {\bibfnamefont {A.}~\bibnamefont
  {Zhiboedov}},\ }\href {\doibase 10.1007/JHEP01(2016)086} {\bibfield
  {journal} {\bibinfo  {journal} {JHEP}\ }\textbf {\bibinfo {volume} {01}},\
  \bibinfo {pages} {086} (\bibinfo {year} {2016})},\ \Eprint
  {http://arxiv.org/abs/1411.5745} {arXiv:1411.5745 [hep-th]} \BibitemShut
  {NoStop}%
\bibitem [{\citenamefont {Flanagan}\ and\ \citenamefont
  {Nichols}(2015)}]{Flanagan:2014kfa}%
  \BibitemOpen
  \bibfield  {author} {\bibinfo {author} {\bibfnamefont {{\'E}.~{\'E}.}\
  \bibnamefont {Flanagan}}\ and\ \bibinfo {author} {\bibfnamefont {D.~A.}\
  \bibnamefont {Nichols}},\ }\href {\doibase 10.1103/PhysRevD.92.084057}
  {\bibfield  {journal} {\bibinfo  {journal} {Phys. Rev. D}\ }\textbf {\bibinfo
  {volume} {92}},\ \bibinfo {pages} {084057} (\bibinfo {year} {2015})},\
  \bibinfo {note} {[Erratum: Phys. Rev. D 93, 049905 (2016)]},\ \Eprint
  {http://arxiv.org/abs/1411.4599} {arXiv:1411.4599 [gr-qc]} \BibitemShut
  {NoStop}%
\bibitem [{\citenamefont {Flanagan}\ \emph {et~al.}(2016)\citenamefont
  {Flanagan}, \citenamefont {Nichols}, \citenamefont {Stein},\ and\
  \citenamefont {Vines}}]{Flanagan:2016oks}%
  \BibitemOpen
  \bibfield  {author} {\bibinfo {author} {\bibfnamefont {{\'E}.~{\'E}.}\
  \bibnamefont {Flanagan}}, \bibinfo {author} {\bibfnamefont {D.~A.}\
  \bibnamefont {Nichols}}, \bibinfo {author} {\bibfnamefont {L.~C.}\
  \bibnamefont {Stein}}, \ and\ \bibinfo {author} {\bibfnamefont
  {J.}~\bibnamefont {Vines}},\ }\href {\doibase 10.1103/PhysRevD.93.104007}
  {\bibfield  {journal} {\bibinfo  {journal} {Phys. Rev. D}\ }\textbf {\bibinfo
  {volume} {93}},\ \bibinfo {pages} {104007} (\bibinfo {year} {2016})},\
  \Eprint {http://arxiv.org/abs/1602.01847} {arXiv:1602.01847 [gr-qc]}
  \BibitemShut {NoStop}%
\bibitem [{\citenamefont {Flanagan}\ \emph {et~al.}(2019)\citenamefont
  {Flanagan}, \citenamefont {Grant}, \citenamefont {Harte},\ and\ \citenamefont
  {Nichols}}]{Flanagan:2018yzh}%
  \BibitemOpen
  \bibfield  {author} {\bibinfo {author} {\bibfnamefont {{\'E}.~{\'E}.}\
  \bibnamefont {Flanagan}}, \bibinfo {author} {\bibfnamefont {A.~M.}\
  \bibnamefont {Grant}}, \bibinfo {author} {\bibfnamefont {A.~I.}\ \bibnamefont
  {Harte}}, \ and\ \bibinfo {author} {\bibfnamefont {D.~A.}\ \bibnamefont
  {Nichols}},\ }\href {\doibase 10.1103/PhysRevD.99.084044} {\bibfield
  {journal} {\bibinfo  {journal} {Phys. Rev. D}\ }\textbf {\bibinfo {volume}
  {99}},\ \bibinfo {pages} {084044} (\bibinfo {year} {2019})},\ \Eprint
  {http://arxiv.org/abs/1901.00021} {arXiv:1901.00021 [gr-qc]} \BibitemShut
  {NoStop}%
\bibitem [{\citenamefont {Strominger}(2017)}]{Strominger:2017zoo}%
  \BibitemOpen
  \bibfield  {author} {\bibinfo {author} {\bibfnamefont {A.}~\bibnamefont
  {Strominger}},\ }\href@noop {} {\  (\bibinfo {year} {2017})},\ \Eprint
  {http://arxiv.org/abs/1703.05448} {arXiv:1703.05448 [hep-th]} \BibitemShut
  {NoStop}%
\bibitem [{\citenamefont {Penrose}(1963)}]{Penrose:1962ij}%
  \BibitemOpen
  \bibfield  {author} {\bibinfo {author} {\bibfnamefont {R.}~\bibnamefont
  {Penrose}},\ }\href {\doibase 10.1103/PhysRevLett.10.66} {\bibfield
  {journal} {\bibinfo  {journal} {Phys. Rev. Lett.}\ }\textbf {\bibinfo
  {volume} {10}},\ \bibinfo {pages} {66} (\bibinfo {year} {1963})}\BibitemShut
  {NoStop}%
\bibitem [{\citenamefont {Penrose}(1965)}]{Penrose:1965am}%
  \BibitemOpen
  \bibfield  {author} {\bibinfo {author} {\bibfnamefont {R.}~\bibnamefont
  {Penrose}},\ }\href {\doibase 10.1098/rspa.1965.0058} {\bibfield  {journal}
  {\bibinfo  {journal} {Proc. Roy. Soc. Lond. A}\ }\textbf {\bibinfo {volume}
  {284}},\ \bibinfo {pages} {159} (\bibinfo {year} {1965})}\BibitemShut
  {NoStop}%
\bibitem [{\citenamefont {Bondi}\ \emph {et~al.}(1962)\citenamefont {Bondi},
  \citenamefont {van~der Burg},\ and\ \citenamefont {Metzner}}]{Bondi:1962px}%
  \BibitemOpen
  \bibfield  {author} {\bibinfo {author} {\bibfnamefont {H.}~\bibnamefont
  {Bondi}}, \bibinfo {author} {\bibfnamefont {M.}~\bibnamefont {van~der Burg}},
  \ and\ \bibinfo {author} {\bibfnamefont {A.}~\bibnamefont {Metzner}},\ }\href
  {\doibase 10.1098/rspa.1962.0161} {\bibfield  {journal} {\bibinfo  {journal}
  {Proc. Roy. Soc. Lond. A}\ }\textbf {\bibinfo {volume} {A269}},\ \bibinfo
  {pages} {21} (\bibinfo {year} {1962})}\BibitemShut {NoStop}%
\bibitem [{\citenamefont {Sachs}(1962{\natexlab{a}})}]{Sachs:1962wk}%
  \BibitemOpen
  \bibfield  {author} {\bibinfo {author} {\bibfnamefont {R.}~\bibnamefont
  {Sachs}},\ }\href {\doibase 10.1098/rspa.1962.0206} {\bibfield  {journal}
  {\bibinfo  {journal} {Proc. Roy. Soc. Lond. A}\ }\textbf {\bibinfo {volume}
  {A270}},\ \bibinfo {pages} {103} (\bibinfo {year}
  {1962}{\natexlab{a}})}\BibitemShut {NoStop}%
\bibitem [{\citenamefont {Newman}\ and\ \citenamefont
  {Unti}(1962)}]{Newman:1962cia}%
  \BibitemOpen
  \bibfield  {author} {\bibinfo {author} {\bibfnamefont {E.~T.}\ \bibnamefont
  {Newman}}\ and\ \bibinfo {author} {\bibfnamefont {T.~W.~J.}\ \bibnamefont
  {Unti}},\ }\href {\doibase 10.1063/1.1724303} {\bibfield  {journal} {\bibinfo
   {journal} {J. Math. Phys.}\ }\textbf {\bibinfo {volume} {3}},\ \bibinfo
  {pages} {891} (\bibinfo {year} {1962})}\BibitemShut {NoStop}%
\bibitem [{\citenamefont {Sachs}(1962{\natexlab{b}})}]{PhysRev.128.2851}%
  \BibitemOpen
  \bibfield  {author} {\bibinfo {author} {\bibfnamefont {R.}~\bibnamefont
  {Sachs}},\ }\href {\doibase 10.1103/PhysRev.128.2851} {\bibfield  {journal}
  {\bibinfo  {journal} {Phys. Rev.}\ }\textbf {\bibinfo {volume} {128}},\
  \bibinfo {pages} {2851} (\bibinfo {year} {1962}{\natexlab{b}})}\BibitemShut
  {NoStop}%
\bibitem [{\citenamefont {Geroch}(1977)}]{Geroch:1977jn}%
  \BibitemOpen
  \bibfield  {author} {\bibinfo {author} {\bibfnamefont {R.~P.}\ \bibnamefont
  {Geroch}},\ }in\ \href@noop {} {\emph {\bibinfo {booktitle} {{Asymptotic
  Structure of Space-Time}}}},\ \bibinfo {editor} {edited by\ \bibinfo {editor}
  {\bibnamefont {{Esposito, F. P. and Witten, L.}}}}\ (\bibinfo  {publisher}
  {{Plenum Press}},\ \bibinfo {address} {{New York}},\ \bibinfo {year} {1977})\
  pp.\ \bibinfo {pages} {1--105}\BibitemShut {NoStop}%
\bibitem [{\citenamefont {Ashtekar}\ and\ \citenamefont
  {Streubel}(1981)}]{Ashtekar:1981bq}%
  \BibitemOpen
  \bibfield  {author} {\bibinfo {author} {\bibfnamefont {A.}~\bibnamefont
  {Ashtekar}}\ and\ \bibinfo {author} {\bibfnamefont {M.}~\bibnamefont
  {Streubel}},\ }\href {\doibase 10.1098/rspa.1981.0109} {\bibfield  {journal}
  {\bibinfo  {journal} {Proc. Roy. Soc. Lond. A}\ }\textbf {\bibinfo {volume}
  {A376}},\ \bibinfo {pages} {585} (\bibinfo {year} {1981})}\BibitemShut
  {NoStop}%
\bibitem [{\citenamefont {Geroch}\ and\ \citenamefont
  {Winicour}(1981)}]{Geroch:1981ut}%
  \BibitemOpen
  \bibfield  {author} {\bibinfo {author} {\bibfnamefont {R.~P.}\ \bibnamefont
  {Geroch}}\ and\ \bibinfo {author} {\bibfnamefont {J.}~\bibnamefont
  {Winicour}},\ }\href {\doibase 10.1063/1.524987} {\bibfield  {journal}
  {\bibinfo  {journal} {J. Math. Phys.}\ }\textbf {\bibinfo {volume} {22}},\
  \bibinfo {pages} {803} (\bibinfo {year} {1981})}\BibitemShut {NoStop}%
\bibitem [{\citenamefont {Wald}\ and\ \citenamefont
  {Zoupas}(2000)}]{Wald:1999wa}%
  \BibitemOpen
  \bibfield  {author} {\bibinfo {author} {\bibfnamefont {R.~M.}\ \bibnamefont
  {Wald}}\ and\ \bibinfo {author} {\bibfnamefont {A.}~\bibnamefont {Zoupas}},\
  }\href {\doibase 10.1103/PhysRevD.61.084027} {\bibfield  {journal} {\bibinfo
  {journal} {Phys. Rev. D}\ }\textbf {\bibinfo {volume} {61}},\ \bibinfo
  {pages} {084027} (\bibinfo {year} {2000})},\ \Eprint
  {http://arxiv.org/abs/gr-qc/9911095} {arXiv:gr-qc/9911095} \BibitemShut
  {NoStop}%
\bibitem [{\citenamefont {Barnich}\ and\ \citenamefont
  {Troessaert}(2010{\natexlab{a}})}]{Barnich:2009se}%
  \BibitemOpen
  \bibfield  {author} {\bibinfo {author} {\bibfnamefont {G.}~\bibnamefont
  {Barnich}}\ and\ \bibinfo {author} {\bibfnamefont {C.}~\bibnamefont
  {Troessaert}},\ }\href {\doibase 10.1103/PhysRevLett.105.111103} {\bibfield
  {journal} {\bibinfo  {journal} {Phys. Rev. Lett.}\ }\textbf {\bibinfo
  {volume} {105}},\ \bibinfo {pages} {111103} (\bibinfo {year}
  {2010}{\natexlab{a}})},\ \Eprint {http://arxiv.org/abs/0909.2617}
  {arXiv:0909.2617 [gr-qc]} \BibitemShut {NoStop}%
\bibitem [{\citenamefont {Barnich}\ and\ \citenamefont
  {Troessaert}(2010{\natexlab{b}})}]{Barnich:2010eb}%
  \BibitemOpen
  \bibfield  {author} {\bibinfo {author} {\bibfnamefont {G.}~\bibnamefont
  {Barnich}}\ and\ \bibinfo {author} {\bibfnamefont {C.}~\bibnamefont
  {Troessaert}},\ }\href {\doibase 10.1007/JHEP05(2010)062} {\bibfield
  {journal} {\bibinfo  {journal} {JHEP}\ }\textbf {\bibinfo {volume} {05}},\
  \bibinfo {pages} {062} (\bibinfo {year} {2010}{\natexlab{b}})},\ \Eprint
  {http://arxiv.org/abs/1001.1541} {arXiv:1001.1541 [hep-th]} \BibitemShut
  {NoStop}%
\bibitem [{\citenamefont {Barnich}\ and\ \citenamefont
  {Troessaert}(2011)}]{Barnich:2011mi}%
  \BibitemOpen
  \bibfield  {author} {\bibinfo {author} {\bibfnamefont {G.}~\bibnamefont
  {Barnich}}\ and\ \bibinfo {author} {\bibfnamefont {C.}~\bibnamefont
  {Troessaert}},\ }\href {\doibase 10.1007/JHEP12(2011)105} {\bibfield
  {journal} {\bibinfo  {journal} {JHEP}\ }\textbf {\bibinfo {volume} {12}},\
  \bibinfo {pages} {105} (\bibinfo {year} {2011})},\ \Eprint
  {http://arxiv.org/abs/1106.0213} {arXiv:1106.0213 [hep-th]} \BibitemShut
  {NoStop}%
\bibitem [{\citenamefont {Banks}(2003)}]{Banks:2003vp}%
  \BibitemOpen
  \bibfield  {author} {\bibinfo {author} {\bibfnamefont {T.}~\bibnamefont
  {Banks}},\ }\href@noop {} {\  (\bibinfo {year} {2003})},\ \Eprint
  {http://arxiv.org/abs/hep-th/0306074} {arXiv:hep-th/0306074} \BibitemShut
  {NoStop}%
\bibitem [{\citenamefont {Campiglia}\ and\ \citenamefont
  {Laddha}(2014)}]{Campiglia:2014yka}%
  \BibitemOpen
  \bibfield  {author} {\bibinfo {author} {\bibfnamefont {M.}~\bibnamefont
  {Campiglia}}\ and\ \bibinfo {author} {\bibfnamefont {A.}~\bibnamefont
  {Laddha}},\ }\href {\doibase 10.1103/PhysRevD.90.124028} {\bibfield
  {journal} {\bibinfo  {journal} {Phys. Rev. D}\ }\textbf {\bibinfo {volume}
  {90}},\ \bibinfo {pages} {124028} (\bibinfo {year} {2014})},\ \Eprint
  {http://arxiv.org/abs/1408.2228} {arXiv:1408.2228 [hep-th]} \BibitemShut
  {NoStop}%
\bibitem [{\citenamefont {Campiglia}\ and\ \citenamefont
  {Laddha}(2015)}]{Campiglia:2015yka}%
  \BibitemOpen
  \bibfield  {author} {\bibinfo {author} {\bibfnamefont {M.}~\bibnamefont
  {Campiglia}}\ and\ \bibinfo {author} {\bibfnamefont {A.}~\bibnamefont
  {Laddha}},\ }\href {\doibase 10.1007/JHEP04(2015)076} {\bibfield  {journal}
  {\bibinfo  {journal} {JHEP}\ }\textbf {\bibinfo {volume} {04}},\ \bibinfo
  {pages} {076} (\bibinfo {year} {2015})},\ \Eprint
  {http://arxiv.org/abs/1502.02318} {arXiv:1502.02318 [hep-th]} \BibitemShut
  {NoStop}%
\bibitem [{\citenamefont {Compère}\ \emph {et~al.}(2018)\citenamefont
  {Compère}, \citenamefont {Fiorucci},\ and\ \citenamefont
  {Ruzziconi}}]{Compere:2018ylh}%
  \BibitemOpen
  \bibfield  {author} {\bibinfo {author} {\bibfnamefont {G.}~\bibnamefont
  {Compère}}, \bibinfo {author} {\bibfnamefont {A.}~\bibnamefont {Fiorucci}},
  \ and\ \bibinfo {author} {\bibfnamefont {R.}~\bibnamefont {Ruzziconi}},\
  }\href {\doibase 10.1007/JHEP11(2018)200} {\bibfield  {journal} {\bibinfo
  {journal} {J. High Energy Phys.}\ }\textbf {\bibinfo {volume} {11}},\
  \bibinfo {pages} {200} (\bibinfo {year} {2018})},\ \bibinfo {note} {[Erratum:
  J. High Energy Phys. 04, 172 (2020)]},\ \Eprint
  {http://arxiv.org/abs/1810.00377} {arXiv:1810.00377 [hep-th]} \BibitemShut
  {NoStop}%
\bibitem [{\citenamefont {Flanagan}\ and\ \citenamefont
  {Nichols}(2017)}]{Flanagan:2015pxa}%
  \BibitemOpen
  \bibfield  {author} {\bibinfo {author} {\bibfnamefont {E.~E.}\ \bibnamefont
  {Flanagan}}\ and\ \bibinfo {author} {\bibfnamefont {D.~A.}\ \bibnamefont
  {Nichols}},\ }\href {\doibase 10.1103/PhysRevD.95.044002} {\bibfield
  {journal} {\bibinfo  {journal} {Phys. Rev. D}\ }\textbf {\bibinfo {volume}
  {95}},\ \bibinfo {pages} {044002} (\bibinfo {year} {2017})},\ \Eprint
  {http://arxiv.org/abs/1510.03386} {arXiv:1510.03386 [hep-th]} \BibitemShut
  {NoStop}%
\bibitem [{\citenamefont {Strominger}\ and\ \citenamefont
  {Zhiboedov}(2017)}]{Strominger:2016wns}%
  \BibitemOpen
  \bibfield  {author} {\bibinfo {author} {\bibfnamefont {A.}~\bibnamefont
  {Strominger}}\ and\ \bibinfo {author} {\bibfnamefont {A.}~\bibnamefont
  {Zhiboedov}},\ }\href {\doibase 10.1088/1361-6382/aa5b5f} {\bibfield
  {journal} {\bibinfo  {journal} {Class. Quant. Grav.}\ }\textbf {\bibinfo
  {volume} {34}},\ \bibinfo {pages} {064002} (\bibinfo {year} {2017})},\
  \Eprint {http://arxiv.org/abs/1610.00639} {arXiv:1610.00639 [hep-th]}
  \BibitemShut {NoStop}%
\bibitem [{\citenamefont {Will}(2014)}]{Will:2014kxa}%
  \BibitemOpen
  \bibfield  {author} {\bibinfo {author} {\bibfnamefont {C.~M.}\ \bibnamefont
  {Will}},\ }\href {\doibase 10.12942/lrr-2014-4} {\bibfield  {journal}
  {\bibinfo  {journal} {Living Rev. Rel.}\ }\textbf {\bibinfo {volume} {17}},\
  \bibinfo {pages} {4} (\bibinfo {year} {2014})},\ \Eprint
  {http://arxiv.org/abs/1403.7377} {arXiv:1403.7377 [gr-qc]} \BibitemShut
  {NoStop}%
\bibitem [{\citenamefont {Chatziioannou}\ \emph {et~al.}(2012)\citenamefont
  {Chatziioannou}, \citenamefont {Yunes},\ and\ \citenamefont
  {Cornish}}]{Chatziioannou:2012rf}%
  \BibitemOpen
  \bibfield  {author} {\bibinfo {author} {\bibfnamefont {K.}~\bibnamefont
  {Chatziioannou}}, \bibinfo {author} {\bibfnamefont {N.}~\bibnamefont
  {Yunes}}, \ and\ \bibinfo {author} {\bibfnamefont {N.}~\bibnamefont
  {Cornish}},\ }\href {\doibase 10.1103/PhysRevD.86.022004,
  10.1103/PhysRevD.95.129901} {\bibfield  {journal} {\bibinfo  {journal} {Phys.
  Rev.}\ }\textbf {\bibinfo {volume} {D86}},\ \bibinfo {pages} {022004}
  (\bibinfo {year} {2012})},\ \bibinfo {note} {[Erratum: Phys.
  Rev.D95,no.12,129901(2017)]},\ \Eprint {http://arxiv.org/abs/1204.2585}
  {arXiv:1204.2585 [gr-qc]} \BibitemShut {NoStop}%
\bibitem [{\citenamefont {Zhang}\ \emph {et~al.}(2020)\citenamefont {Zhang},
  \citenamefont {Zhao}, \citenamefont {Wang}, \citenamefont {Wang},
  \citenamefont {Yagi}, \citenamefont {Yunes}, \citenamefont {Zhao},\ and\
  \citenamefont {Zhu}}]{Zhang:2019iim}%
  \BibitemOpen
  \bibfield  {author} {\bibinfo {author} {\bibfnamefont {C.}~\bibnamefont
  {Zhang}}, \bibinfo {author} {\bibfnamefont {X.}~\bibnamefont {Zhao}},
  \bibinfo {author} {\bibfnamefont {A.}~\bibnamefont {Wang}}, \bibinfo {author}
  {\bibfnamefont {B.}~\bibnamefont {Wang}}, \bibinfo {author} {\bibfnamefont
  {K.}~\bibnamefont {Yagi}}, \bibinfo {author} {\bibfnamefont {N.}~\bibnamefont
  {Yunes}}, \bibinfo {author} {\bibfnamefont {W.}~\bibnamefont {Zhao}}, \ and\
  \bibinfo {author} {\bibfnamefont {T.}~\bibnamefont {Zhu}},\ }\href {\doibase
  10.1103/PhysRevD.101.044002} {\bibfield  {journal} {\bibinfo  {journal}
  {Phys. Rev.}\ }\textbf {\bibinfo {volume} {D101}},\ \bibinfo {pages} {044002}
  (\bibinfo {year} {2020})},\ \Eprint {http://arxiv.org/abs/1911.10278}
  {arXiv:1911.10278 [gr-qc]} \BibitemShut {NoStop}%
\bibitem [{\citenamefont {Lang}(2014)}]{Lang:2013fna}%
  \BibitemOpen
  \bibfield  {author} {\bibinfo {author} {\bibfnamefont {R.~N.}\ \bibnamefont
  {Lang}},\ }\href {\doibase 10.1103/PhysRevD.89.084014} {\bibfield  {journal}
  {\bibinfo  {journal} {Phys. Rev. D}\ }\textbf {\bibinfo {volume} {89}},\
  \bibinfo {pages} {084014} (\bibinfo {year} {2014})},\ \Eprint
  {http://arxiv.org/abs/1310.3320} {arXiv:1310.3320 [gr-qc]} \BibitemShut
  {NoStop}%
\bibitem [{\citenamefont {Lang}(2015)}]{Lang:2014osa}%
  \BibitemOpen
  \bibfield  {author} {\bibinfo {author} {\bibfnamefont {R.~N.}\ \bibnamefont
  {Lang}},\ }\href {\doibase 10.1103/PhysRevD.91.084027} {\bibfield  {journal}
  {\bibinfo  {journal} {Phys. Rev. D}\ }\textbf {\bibinfo {volume} {91}},\
  \bibinfo {pages} {084027} (\bibinfo {year} {2015})},\ \Eprint
  {http://arxiv.org/abs/1411.3073} {arXiv:1411.3073 [gr-qc]} \BibitemShut
  {NoStop}%
\bibitem [{\citenamefont {Du}\ and\ \citenamefont
  {Nishizawa}(2016)}]{Du:2016hww}%
  \BibitemOpen
  \bibfield  {author} {\bibinfo {author} {\bibfnamefont {S.~M.}\ \bibnamefont
  {Du}}\ and\ \bibinfo {author} {\bibfnamefont {A.}~\bibnamefont {Nishizawa}},\
  }\href {\doibase 10.1103/PhysRevD.94.104063} {\bibfield  {journal} {\bibinfo
  {journal} {Phys. Rev. D}\ }\textbf {\bibinfo {volume} {94}},\ \bibinfo
  {pages} {104063} (\bibinfo {year} {2016})},\ \Eprint
  {http://arxiv.org/abs/1609.09825} {arXiv:1609.09825 [gr-qc]} \BibitemShut
  {NoStop}%
\bibitem [{\citenamefont {Koyama}(2020)}]{Koyama:2020vfc}%
  \BibitemOpen
  \bibfield  {author} {\bibinfo {author} {\bibfnamefont {K.}~\bibnamefont
  {Koyama}},\ }\href {\doibase 10.1103/PhysRevD.102.021502} {\bibfield
  {journal} {\bibinfo  {journal} {Phys. Rev.}\ }\textbf {\bibinfo {volume}
  {D102}},\ \bibinfo {pages} {021502} (\bibinfo {year} {2020})},\ \Eprint
  {http://arxiv.org/abs/2006.15914} {arXiv:2006.15914 [gr-qc]} \BibitemShut
  {NoStop}%
\bibitem [{\citenamefont {Brans}\ and\ \citenamefont
  {Dicke}(1961)}]{Brans:1961sx}%
  \BibitemOpen
  \bibfield  {author} {\bibinfo {author} {\bibfnamefont {C.}~\bibnamefont
  {Brans}}\ and\ \bibinfo {author} {\bibfnamefont {R.}~\bibnamefont {Dicke}},\
  }\href {\doibase 10.1103/PhysRev.124.925} {\bibfield  {journal} {\bibinfo
  {journal} {Phys. Rev.}\ }\textbf {\bibinfo {volume} {124}},\ \bibinfo {pages}
  {925} (\bibinfo {year} {1961})}\BibitemShut {NoStop}%
\bibitem [{\citenamefont {Berti}\ \emph {et~al.}(2015)\citenamefont {Berti}
  \emph {et~al.}}]{Berti:2015itd}%
  \BibitemOpen
  \bibfield  {author} {\bibinfo {author} {\bibfnamefont {E.}~\bibnamefont
  {Berti}} \emph {et~al.},\ }\href {\doibase 10.1088/0264-9381/32/24/243001}
  {\bibfield  {journal} {\bibinfo  {journal} {Class. Quant. Grav.}\ }\textbf
  {\bibinfo {volume} {32}},\ \bibinfo {pages} {243001} (\bibinfo {year}
  {2015})},\ \Eprint {http://arxiv.org/abs/1501.07274} {arXiv:1501.07274
  [gr-qc]} \BibitemShut {NoStop}%
\bibitem [{\citenamefont {Clifton}\ \emph {et~al.}(2012)\citenamefont
  {Clifton}, \citenamefont {Ferreira}, \citenamefont {Padilla},\ and\
  \citenamefont {Skordis}}]{Clifton:2011jh}%
  \BibitemOpen
  \bibfield  {author} {\bibinfo {author} {\bibfnamefont {T.}~\bibnamefont
  {Clifton}}, \bibinfo {author} {\bibfnamefont {P.~G.}\ \bibnamefont
  {Ferreira}}, \bibinfo {author} {\bibfnamefont {A.}~\bibnamefont {Padilla}}, \
  and\ \bibinfo {author} {\bibfnamefont {C.}~\bibnamefont {Skordis}},\ }\href
  {\doibase 10.1016/j.physrep.2012.01.001} {\bibfield  {journal} {\bibinfo
  {journal} {Phys. Rept.}\ }\textbf {\bibinfo {volume} {513}},\ \bibinfo
  {pages} {1} (\bibinfo {year} {2012})},\ \Eprint
  {http://arxiv.org/abs/1106.2476} {arXiv:1106.2476 [astro-ph.CO]} \BibitemShut
  {NoStop}%
\bibitem [{\citenamefont {Barrow}\ and\ \citenamefont
  {Maeda}(1990)}]{Barrow:1990nv}%
  \BibitemOpen
  \bibfield  {author} {\bibinfo {author} {\bibfnamefont {J.~D.}\ \bibnamefont
  {Barrow}}\ and\ \bibinfo {author} {\bibfnamefont {K.-i.}\ \bibnamefont
  {Maeda}},\ }\href {\doibase 10.1016/0550-3213(90)90272-F} {\bibfield
  {journal} {\bibinfo  {journal} {Nucl. Phys. B}\ }\textbf {\bibinfo {volume}
  {341}},\ \bibinfo {pages} {294} (\bibinfo {year} {1990})}\BibitemShut
  {NoStop}%
\bibitem [{\citenamefont {Brax}\ \emph {et~al.}(2004)\citenamefont {Brax},
  \citenamefont {van~de Bruck}, \citenamefont {Davis}, \citenamefont {Khoury},\
  and\ \citenamefont {Weltman}}]{Brax:2004qh}%
  \BibitemOpen
  \bibfield  {author} {\bibinfo {author} {\bibfnamefont {P.}~\bibnamefont
  {Brax}}, \bibinfo {author} {\bibfnamefont {C.}~\bibnamefont {van~de Bruck}},
  \bibinfo {author} {\bibfnamefont {A.-C.}\ \bibnamefont {Davis}}, \bibinfo
  {author} {\bibfnamefont {J.}~\bibnamefont {Khoury}}, \ and\ \bibinfo {author}
  {\bibfnamefont {A.}~\bibnamefont {Weltman}},\ }\href {\doibase
  10.1103/PhysRevD.70.123518} {\bibfield  {journal} {\bibinfo  {journal} {Phys.
  Rev. D}\ }\textbf {\bibinfo {volume} {70}},\ \bibinfo {pages} {123518}
  (\bibinfo {year} {2004})},\ \Eprint {http://arxiv.org/abs/astro-ph/0408415}
  {arXiv:astro-ph/0408415} \BibitemShut {NoStop}%
\bibitem [{\citenamefont {Baccigalupi}\ \emph {et~al.}(2000)\citenamefont
  {Baccigalupi}, \citenamefont {Matarrese},\ and\ \citenamefont
  {Perrotta}}]{Baccigalupi:2000je}%
  \BibitemOpen
  \bibfield  {author} {\bibinfo {author} {\bibfnamefont {C.}~\bibnamefont
  {Baccigalupi}}, \bibinfo {author} {\bibfnamefont {S.}~\bibnamefont
  {Matarrese}}, \ and\ \bibinfo {author} {\bibfnamefont {F.}~\bibnamefont
  {Perrotta}},\ }\href {\doibase 10.1103/PhysRevD.62.123510} {\bibfield
  {journal} {\bibinfo  {journal} {Phys. Rev. D}\ }\textbf {\bibinfo {volume}
  {62}},\ \bibinfo {pages} {123510} (\bibinfo {year} {2000})},\ \Eprint
  {http://arxiv.org/abs/astro-ph/0005543} {arXiv:astro-ph/0005543} \BibitemShut
  {NoStop}%
\bibitem [{\citenamefont {Riazuelo}\ and\ \citenamefont
  {Uzan}(2002)}]{Riazuelo:2001mg}%
  \BibitemOpen
  \bibfield  {author} {\bibinfo {author} {\bibfnamefont {A.}~\bibnamefont
  {Riazuelo}}\ and\ \bibinfo {author} {\bibfnamefont {J.-P.}\ \bibnamefont
  {Uzan}},\ }\href {\doibase 10.1103/PhysRevD.66.023525} {\bibfield  {journal}
  {\bibinfo  {journal} {Phys. Rev. D}\ }\textbf {\bibinfo {volume} {66}},\
  \bibinfo {pages} {023525} (\bibinfo {year} {2002})},\ \Eprint
  {http://arxiv.org/abs/astro-ph/0107386} {arXiv:astro-ph/0107386} \BibitemShut
  {NoStop}%
\bibitem [{\citenamefont {Will}\ and\ \citenamefont
  {Zaglauer}(1989)}]{Will:1989sk}%
  \BibitemOpen
  \bibfield  {author} {\bibinfo {author} {\bibfnamefont {C.~M.}\ \bibnamefont
  {Will}}\ and\ \bibinfo {author} {\bibfnamefont {H.~W.}\ \bibnamefont
  {Zaglauer}},\ }\href {\doibase 10.1086/168016} {\bibfield  {journal}
  {\bibinfo  {journal} {Astrophys. J.}\ }\textbf {\bibinfo {volume} {346}},\
  \bibinfo {pages} {366} (\bibinfo {year} {1989})}\BibitemShut {NoStop}%
\bibitem [{\citenamefont {Will}(1994)}]{Will:1994fb}%
  \BibitemOpen
  \bibfield  {author} {\bibinfo {author} {\bibfnamefont {C.~M.}\ \bibnamefont
  {Will}},\ }\href {\doibase 10.1103/PhysRevD.50.6058} {\bibfield  {journal}
  {\bibinfo  {journal} {Phys. Rev. D}\ }\textbf {\bibinfo {volume} {50}},\
  \bibinfo {pages} {6058} (\bibinfo {year} {1994})},\ \Eprint
  {http://arxiv.org/abs/gr-qc/9406022} {arXiv:gr-qc/9406022} \BibitemShut
  {NoStop}%
\bibitem [{\citenamefont {Wiseman}\ and\ \citenamefont
  {Will}(1991)}]{Wiseman:1991ss}%
  \BibitemOpen
  \bibfield  {author} {\bibinfo {author} {\bibfnamefont {A.~G.}\ \bibnamefont
  {Wiseman}}\ and\ \bibinfo {author} {\bibfnamefont {C.~M.}\ \bibnamefont
  {Will}},\ }\href {\doibase 10.1103/PhysRevD.44.R2945} {\bibfield  {journal}
  {\bibinfo  {journal} {Phys. Rev. D}\ }\textbf {\bibinfo {volume} {44}},\
  \bibinfo {pages} {2945} (\bibinfo {year} {1991})}\BibitemShut {NoStop}%
\bibitem [{\citenamefont {Favata}(2009)}]{Favata:2008yd}%
  \BibitemOpen
  \bibfield  {author} {\bibinfo {author} {\bibfnamefont {M.}~\bibnamefont
  {Favata}},\ }\href {\doibase 10.1103/PhysRevD.80.024002} {\bibfield
  {journal} {\bibinfo  {journal} {Phys. Rev. D}\ }\textbf {\bibinfo {volume}
  {80}},\ \bibinfo {pages} {024002} (\bibinfo {year} {2009})},\ \Eprint
  {http://arxiv.org/abs/0812.0069} {arXiv:0812.0069 [gr-qc]} \BibitemShut
  {NoStop}%
\bibitem [{\citenamefont {Hawking}(1972)}]{Hawking:1972qk}%
  \BibitemOpen
  \bibfield  {author} {\bibinfo {author} {\bibfnamefont {S.~W.}\ \bibnamefont
  {Hawking}},\ }\href {\doibase 10.1007/BF01877518} {\bibfield  {journal}
  {\bibinfo  {journal} {Commun. Math. Phys.}\ }\textbf {\bibinfo {volume}
  {25}},\ \bibinfo {pages} {167} (\bibinfo {year} {1972})}\BibitemShut
  {NoStop}%
\bibitem [{\citenamefont {Sotiriou}\ and\ \citenamefont
  {Faraoni}(2012)}]{Sotiriou:2011dz}%
  \BibitemOpen
  \bibfield  {author} {\bibinfo {author} {\bibfnamefont {T.~P.}\ \bibnamefont
  {Sotiriou}}\ and\ \bibinfo {author} {\bibfnamefont {V.}~\bibnamefont
  {Faraoni}},\ }\href {\doibase 10.1103/PhysRevLett.108.081103} {\bibfield
  {journal} {\bibinfo  {journal} {Phys. Rev. Lett.}\ }\textbf {\bibinfo
  {volume} {108}},\ \bibinfo {pages} {081103} (\bibinfo {year} {2012})},\
  \Eprint {http://arxiv.org/abs/1109.6324} {arXiv:1109.6324 [gr-qc]}
  \BibitemShut {NoStop}%
\bibitem [{\citenamefont {Arun}\ \emph {et~al.}(2004)\citenamefont {Arun},
  \citenamefont {Blanchet}, \citenamefont {Iyer},\ and\ \citenamefont
  {Qusailah}}]{Arun:2004ff}%
  \BibitemOpen
  \bibfield  {author} {\bibinfo {author} {\bibfnamefont {K.}~\bibnamefont
  {Arun}}, \bibinfo {author} {\bibfnamefont {L.}~\bibnamefont {Blanchet}},
  \bibinfo {author} {\bibfnamefont {B.~R.}\ \bibnamefont {Iyer}}, \ and\
  \bibinfo {author} {\bibfnamefont {M.~S.}\ \bibnamefont {Qusailah}},\ }\href
  {\doibase 10.1088/0264-9381/21/15/010} {\bibfield  {journal} {\bibinfo
  {journal} {Classical Quantum Gravity}\ }\textbf {\bibinfo {volume} {21}},\
  \bibinfo {pages} {3771} (\bibinfo {year} {2004})},\ \bibinfo {note}
  {[Erratum: Classical Quantum Gravity 22, 3115 (2005)]},\ \Eprint
  {http://arxiv.org/abs/gr-qc/0404085} {arXiv:gr-qc/0404085} \BibitemShut
  {NoStop}%
\bibitem [{\citenamefont {Nichols}(2017)}]{Nichols:2017rqr}%
  \BibitemOpen
  \bibfield  {author} {\bibinfo {author} {\bibfnamefont {D.~A.}\ \bibnamefont
  {Nichols}},\ }\href {\doibase 10.1103/PhysRevD.95.084048} {\bibfield
  {journal} {\bibinfo  {journal} {Phys. Rev.}\ }\textbf {\bibinfo {volume}
  {D95}},\ \bibinfo {pages} {084048} (\bibinfo {year} {2017})},\ \Eprint
  {http://arxiv.org/abs/1702.03300} {arXiv:1702.03300 [gr-qc]} \BibitemShut
  {NoStop}%
\bibitem [{\citenamefont {Flanagan}(2004)}]{Flanagan:2004bz}%
  \BibitemOpen
  \bibfield  {author} {\bibinfo {author} {\bibfnamefont {E.~E.}\ \bibnamefont
  {Flanagan}},\ }\href {\doibase 10.1088/0264-9381/21/15/N02} {\bibfield
  {journal} {\bibinfo  {journal} {Class. Quant. Grav.}\ }\textbf {\bibinfo
  {volume} {21}},\ \bibinfo {pages} {3817} (\bibinfo {year} {2004})},\ \Eprint
  {http://arxiv.org/abs/gr-qc/0403063} {arXiv:gr-qc/0403063} \BibitemShut
  {NoStop}%
\bibitem [{\citenamefont {Fujii}\ and\ \citenamefont
  {Maeda}(2007)}]{Fujii:2003pa}%
  \BibitemOpen
  \bibfield  {author} {\bibinfo {author} {\bibfnamefont {Y.}~\bibnamefont
  {Fujii}}\ and\ \bibinfo {author} {\bibfnamefont {K.}~\bibnamefont {Maeda}},\
  }\href {\doibase 10.1017/CBO9780511535093} {\emph {\bibinfo {title} {{The
  scalar-tensor theory of gravitation}}}},\ Cambridge Monographs on
  Mathematical Physics\ (\bibinfo  {publisher} {Cambridge University Press},\
  \bibinfo {year} {2007})\BibitemShut {NoStop}%
\bibitem [{\citenamefont {Misner}\ \emph {et~al.}(1973)\citenamefont {Misner},
  \citenamefont {Thorne},\ and\ \citenamefont {Wheeler}}]{Misner:1974qy}%
  \BibitemOpen
  \bibfield  {author} {\bibinfo {author} {\bibfnamefont {C.~W.}\ \bibnamefont
  {Misner}}, \bibinfo {author} {\bibfnamefont {K.}~\bibnamefont {Thorne}}, \
  and\ \bibinfo {author} {\bibfnamefont {J.}~\bibnamefont {Wheeler}},\
  }\href@noop {} {\emph {\bibinfo {title} {{Gravitation}}}}\ (\bibinfo
  {publisher} {W. H. Freeman},\ \bibinfo {address} {San Francisco},\ \bibinfo
  {year} {1973})\BibitemShut {NoStop}%
\bibitem [{\citenamefont {Hou}\ and\ \citenamefont {Zhu}(2020)}]{Hou:2020tnd}%
  \BibitemOpen
  \bibfield  {author} {\bibinfo {author} {\bibfnamefont {S.}~\bibnamefont
  {Hou}}\ and\ \bibinfo {author} {\bibfnamefont {Z.-H.}\ \bibnamefont {Zhu}},\
  }\href@noop {} {\  (\bibinfo {year} {2020})},\ \Eprint
  {http://arxiv.org/abs/2005.01310} {arXiv:2005.01310 [gr-qc]} \BibitemShut
  {NoStop}%
\bibitem [{\citenamefont {Mädler}\ and\ \citenamefont
  {Winicour}(2016)}]{Madler:2016xju}%
  \BibitemOpen
  \bibfield  {author} {\bibinfo {author} {\bibfnamefont {T.}~\bibnamefont
  {Mädler}}\ and\ \bibinfo {author} {\bibfnamefont {J.}~\bibnamefont
  {Winicour}},\ }\href {\doibase 10.4249/scholarpedia.33528} {\bibfield
  {journal} {\bibinfo  {journal} {Scholarpedia}\ }\textbf {\bibinfo {volume}
  {11}},\ \bibinfo {pages} {33528} (\bibinfo {year} {2016})},\ \Eprint
  {http://arxiv.org/abs/1609.01731} {arXiv:1609.01731 [gr-qc]} \BibitemShut
  {NoStop}%
\bibitem [{\citenamefont {Damour}\ and\ \citenamefont
  {Esposito-Farese}(1992)}]{Damour:1992we}%
  \BibitemOpen
  \bibfield  {author} {\bibinfo {author} {\bibfnamefont {T.}~\bibnamefont
  {Damour}}\ and\ \bibinfo {author} {\bibfnamefont {G.}~\bibnamefont
  {Esposito-Farese}},\ }\href {\doibase 10.1088/0264-9381/9/9/015} {\bibfield
  {journal} {\bibinfo  {journal} {Class. Quant. Grav.}\ }\textbf {\bibinfo
  {volume} {9}},\ \bibinfo {pages} {2093} (\bibinfo {year} {1992})}\BibitemShut
  {NoStop}%
\bibitem [{\citenamefont {Handmer}\ \emph {et~al.}(2015)\citenamefont
  {Handmer}, \citenamefont {Szil\'agyi},\ and\ \citenamefont
  {Winicour}}]{Handmer:2015dsa}%
  \BibitemOpen
  \bibfield  {author} {\bibinfo {author} {\bibfnamefont {C.~J.}\ \bibnamefont
  {Handmer}}, \bibinfo {author} {\bibfnamefont {B.}~\bibnamefont {Szil\'agyi}},
  \ and\ \bibinfo {author} {\bibfnamefont {J.}~\bibnamefont {Winicour}},\
  }\href {\doibase 10.1088/0264-9381/32/23/235018} {\bibfield  {journal}
  {\bibinfo  {journal} {Class. Quant. Grav.}\ }\textbf {\bibinfo {volume}
  {32}},\ \bibinfo {pages} {235018} (\bibinfo {year} {2015})},\ \Eprint
  {http://arxiv.org/abs/1502.06987} {arXiv:1502.06987 [gr-qc]} \BibitemShut
  {NoStop}%
\bibitem [{\citenamefont {Handmer}\ \emph {et~al.}(2016)\citenamefont
  {Handmer}, \citenamefont {Szil\'agyi},\ and\ \citenamefont
  {Winicour}}]{Handmer:2016mls}%
  \BibitemOpen
  \bibfield  {author} {\bibinfo {author} {\bibfnamefont {C.~J.}\ \bibnamefont
  {Handmer}}, \bibinfo {author} {\bibfnamefont {B.}~\bibnamefont {Szil\'agyi}},
  \ and\ \bibinfo {author} {\bibfnamefont {J.}~\bibnamefont {Winicour}},\
  }\href {\doibase 10.1088/0264-9381/33/22/225007} {\bibfield  {journal}
  {\bibinfo  {journal} {Class. Quant. Grav.}\ }\textbf {\bibinfo {volume}
  {33}},\ \bibinfo {pages} {225007} (\bibinfo {year} {2016})},\ \Eprint
  {http://arxiv.org/abs/1605.04332} {arXiv:1605.04332 [gr-qc]} \BibitemShut
  {NoStop}%
\bibitem [{\citenamefont {Carroll}(2019)}]{Carroll:2004st}%
  \BibitemOpen
  \bibfield  {author} {\bibinfo {author} {\bibfnamefont {S.~M.}\ \bibnamefont
  {Carroll}},\ }\href@noop {} {\emph {\bibinfo {title} {{Spacetime and
  Geometry}}}}\ (\bibinfo  {publisher} {Cambridge University Press},\ \bibinfo
  {year} {2019})\BibitemShut {NoStop}%
\bibitem [{\citenamefont {Winicour}(1983)}]{winicour1983}%
  \BibitemOpen
  \bibfield  {author} {\bibinfo {author} {\bibfnamefont {J.}~\bibnamefont
  {Winicour}},\ }\href {\doibase 10.1063/1.525796} {\bibfield  {journal}
  {\bibinfo  {journal} {Journal of Mathematical Physics}\ }\textbf {\bibinfo
  {volume} {24}},\ \bibinfo {pages} {1193} (\bibinfo {year}
  {1983})}\BibitemShut {NoStop}%
\bibitem [{\citenamefont {Winicour}(2009)}]{Winicour:2008vpn}%
  \BibitemOpen
  \bibfield  {author} {\bibinfo {author} {\bibfnamefont {J.}~\bibnamefont
  {Winicour}},\ }\href {\doibase 10.12942/lrr-2009-3} {\bibfield  {journal}
  {\bibinfo  {journal} {Living Rev. Rel.}\ }\textbf {\bibinfo {volume} {12}},\
  \bibinfo {pages} {3} (\bibinfo {year} {2009})},\ \Eprint
  {http://arxiv.org/abs/0810.1903} {arXiv:0810.1903 [gr-qc]} \BibitemShut
  {NoStop}%
\bibitem [{\citenamefont {M\"adler}\ and\ \citenamefont
  {Winicour}(2016)}]{madler:2016ggp}%
  \BibitemOpen
  \bibfield  {author} {\bibinfo {author} {\bibfnamefont {T.}~\bibnamefont
  {M\"adler}}\ and\ \bibinfo {author} {\bibfnamefont {J.}~\bibnamefont
  {Winicour}},\ }\href {\doibase 10.1088/0264-9381/33/17/175006} {\bibfield
  {journal} {\bibinfo  {journal} {Class. Quant. Grav.}\ }\textbf {\bibinfo
  {volume} {33}},\ \bibinfo {pages} {175006} (\bibinfo {year} {2016})},\
  \Eprint {http://arxiv.org/abs/1605.01273} {arXiv:1605.01273 [gr-qc]}
  \BibitemShut {NoStop}%
\bibitem [{\citenamefont {Bieri}\ and\ \citenamefont
  {Garfinkle}(2014)}]{Bieri:2013ada}%
  \BibitemOpen
  \bibfield  {author} {\bibinfo {author} {\bibfnamefont {L.}~\bibnamefont
  {Bieri}}\ and\ \bibinfo {author} {\bibfnamefont {D.}~\bibnamefont
  {Garfinkle}},\ }\href {\doibase 10.1103/PhysRevD.89.084039} {\bibfield
  {journal} {\bibinfo  {journal} {Phys. Rev. D}\ }\textbf {\bibinfo {volume}
  {89}},\ \bibinfo {pages} {084039} (\bibinfo {year} {2014})},\ \Eprint
  {http://arxiv.org/abs/1312.6871} {arXiv:1312.6871 [gr-qc]} \BibitemShut
  {NoStop}%
\bibitem [{\citenamefont {Satishchandran}\ and\ \citenamefont
  {Wald}(2019)}]{Satishchandran:2019pyc}%
  \BibitemOpen
  \bibfield  {author} {\bibinfo {author} {\bibfnamefont {G.}~\bibnamefont
  {Satishchandran}}\ and\ \bibinfo {author} {\bibfnamefont {R.~M.}\
  \bibnamefont {Wald}},\ }\href {\doibase 10.1103/PhysRevD.99.084007}
  {\bibfield  {journal} {\bibinfo  {journal} {Phys. Rev. D}\ }\textbf {\bibinfo
  {volume} {99}},\ \bibinfo {pages} {084007} (\bibinfo {year} {2019})},\
  \Eprint {http://arxiv.org/abs/1901.05942} {arXiv:1901.05942 [gr-qc]}
  \BibitemShut {NoStop}%
\bibitem [{\citenamefont {Braginsky}\ and\ \citenamefont
  {Grishchuk}(1985)}]{Braginsky:1986ia}%
  \BibitemOpen
  \bibfield  {author} {\bibinfo {author} {\bibfnamefont {V.}~\bibnamefont
  {Braginsky}}\ and\ \bibinfo {author} {\bibfnamefont {L.}~\bibnamefont
  {Grishchuk}},\ }\href@noop {} {\bibfield  {journal} {\bibinfo  {journal}
  {Sov. Phys. JETP}\ }\textbf {\bibinfo {volume} {62}},\ \bibinfo {pages} {427}
  (\bibinfo {year} {1985})}\BibitemShut {NoStop}%
\bibitem [{\citenamefont {Zhang}\ \emph {et~al.}(2017)\citenamefont {Zhang},
  \citenamefont {Duval}, \citenamefont {Gibbons},\ and\ \citenamefont
  {Horvathy}}]{Zhang:2017rno}%
  \BibitemOpen
  \bibfield  {author} {\bibinfo {author} {\bibfnamefont {P.-M.}\ \bibnamefont
  {Zhang}}, \bibinfo {author} {\bibfnamefont {C.}~\bibnamefont {Duval}},
  \bibinfo {author} {\bibfnamefont {G.}~\bibnamefont {Gibbons}}, \ and\
  \bibinfo {author} {\bibfnamefont {P.}~\bibnamefont {Horvathy}},\ }\href
  {\doibase 10.1016/j.physletb.2017.07.050} {\bibfield  {journal} {\bibinfo
  {journal} {Phys. Lett. B}\ }\textbf {\bibinfo {volume} {772}},\ \bibinfo
  {pages} {743} (\bibinfo {year} {2017})},\ \Eprint
  {http://arxiv.org/abs/1704.05997} {arXiv:1704.05997 [gr-qc]} \BibitemShut
  {NoStop}%
\bibitem [{\citenamefont {Flanagan}\ \emph {et~al.}(2020)\citenamefont
  {Flanagan}, \citenamefont {Grant}, \citenamefont {Harte},\ and\ \citenamefont
  {Nichols}}]{Flanagan:2019ezo}%
  \BibitemOpen
  \bibfield  {author} {\bibinfo {author} {\bibfnamefont {{\'E}.~{\'E}.}\
  \bibnamefont {Flanagan}}, \bibinfo {author} {\bibfnamefont {A.~M.}\
  \bibnamefont {Grant}}, \bibinfo {author} {\bibfnamefont {A.~I.}\ \bibnamefont
  {Harte}}, \ and\ \bibinfo {author} {\bibfnamefont {D.~A.}\ \bibnamefont
  {Nichols}},\ }\href {\doibase 10.1103/PhysRevD.101.104033} {\bibfield
  {journal} {\bibinfo  {journal} {Phys. Rev. D}\ }\textbf {\bibinfo {volume}
  {101}},\ \bibinfo {pages} {104033} (\bibinfo {year} {2020})},\ \Eprint
  {http://arxiv.org/abs/1912.13449} {arXiv:1912.13449 [gr-qc]} \BibitemShut
  {NoStop}%
\bibitem [{\citenamefont {Bieri}(2020)}]{Bieri:2020pee}%
  \BibitemOpen
  \bibfield  {author} {\bibinfo {author} {\bibfnamefont {L.}~\bibnamefont
  {Bieri}},\ }\href@noop {} {\  (\bibinfo {year} {2020})},\ \Eprint
  {http://arxiv.org/abs/2010.07418} {arXiv:2010.07418 [gr-qc]} \BibitemShut
  {NoStop}%
\bibitem [{\citenamefont {Newman}\ and\ \citenamefont
  {Penrose}(1962)}]{Newman:1961qr}%
  \BibitemOpen
  \bibfield  {author} {\bibinfo {author} {\bibfnamefont {E.}~\bibnamefont
  {Newman}}\ and\ \bibinfo {author} {\bibfnamefont {R.}~\bibnamefont
  {Penrose}},\ }\href {\doibase 10.1063/1.1724257} {\bibfield  {journal}
  {\bibinfo  {journal} {J. Math. Phys.}\ }\textbf {\bibinfo {volume} {3}},\
  \bibinfo {pages} {566} (\bibinfo {year} {1962})}\BibitemShut {NoStop}%
\bibitem [{\citenamefont {Komar}(1959)}]{PhysRev.113.934}%
  \BibitemOpen
  \bibfield  {author} {\bibinfo {author} {\bibfnamefont {A.}~\bibnamefont
  {Komar}},\ }\href {\doibase 10.1103/PhysRev.113.934} {\bibfield  {journal}
  {\bibinfo  {journal} {Phys. Rev.}\ }\textbf {\bibinfo {volume} {113}},\
  \bibinfo {pages} {934} (\bibinfo {year} {1959})}\BibitemShut {NoStop}%
\bibitem [{\citenamefont {Damour}\ and\ \citenamefont
  {Esposito-Farese}(1993)}]{Damour:1993hw}%
  \BibitemOpen
  \bibfield  {author} {\bibinfo {author} {\bibfnamefont {T.}~\bibnamefont
  {Damour}}\ and\ \bibinfo {author} {\bibfnamefont {G.}~\bibnamefont
  {Esposito-Farese}},\ }\href {\doibase 10.1103/PhysRevLett.70.2220} {\bibfield
   {journal} {\bibinfo  {journal} {Phys.Rev.Lett.}\ }\textbf {\bibinfo {volume}
  {70}},\ \bibinfo {pages} {2220} (\bibinfo {year} {1993})}\BibitemShut
  {NoStop}%
\bibitem [{\citenamefont {Heisenberg}(2018)}]{Heisenberg:2018acv}%
  \BibitemOpen
  \bibfield  {author} {\bibinfo {author} {\bibfnamefont {L.}~\bibnamefont
  {Heisenberg}},\ }\href {\doibase 10.1088/1475-7516/2018/10/054} {\bibfield
  {journal} {\bibinfo  {journal} {JCAP}\ }\textbf {\bibinfo {volume} {10}},\
  \bibinfo {pages} {054} (\bibinfo {year} {2018})},\ \Eprint
  {http://arxiv.org/abs/1801.01523} {arXiv:1801.01523 [gr-qc]} \BibitemShut
  {NoStop}%
\bibitem [{\citenamefont {Jackiw}\ and\ \citenamefont
  {Pi}(2003)}]{Jackiw:2003pm}%
  \BibitemOpen
  \bibfield  {author} {\bibinfo {author} {\bibfnamefont {R.}~\bibnamefont
  {Jackiw}}\ and\ \bibinfo {author} {\bibfnamefont {S.}~\bibnamefont {Pi}},\
  }\href {\doibase 10.1103/PhysRevD.68.104012} {\bibfield  {journal} {\bibinfo
  {journal} {Phys. Rev. D}\ }\textbf {\bibinfo {volume} {68}},\ \bibinfo
  {pages} {104012} (\bibinfo {year} {2003})},\ \Eprint
  {http://arxiv.org/abs/gr-qc/0308071} {arXiv:gr-qc/0308071} \BibitemShut
  {NoStop}%
\bibitem [{\citenamefont {Pani}\ and\ \citenamefont
  {Cardoso}(2009)}]{Pani:2009wy}%
  \BibitemOpen
  \bibfield  {author} {\bibinfo {author} {\bibfnamefont {P.}~\bibnamefont
  {Pani}}\ and\ \bibinfo {author} {\bibfnamefont {V.}~\bibnamefont {Cardoso}},\
  }\href {\doibase 10.1103/PhysRevD.79.084031} {\bibfield  {journal} {\bibinfo
  {journal} {Phys. Rev. D}\ }\textbf {\bibinfo {volume} {79}},\ \bibinfo
  {pages} {084031} (\bibinfo {year} {2009})},\ \Eprint
  {http://arxiv.org/abs/0902.1569} {arXiv:0902.1569 [gr-qc]} \BibitemShut
  {NoStop}%
\bibitem [{\citenamefont {Moura}\ and\ \citenamefont
  {Schiappa}(2007)}]{Moura:2006pz}%
  \BibitemOpen
  \bibfield  {author} {\bibinfo {author} {\bibfnamefont {F.}~\bibnamefont
  {Moura}}\ and\ \bibinfo {author} {\bibfnamefont {R.}~\bibnamefont
  {Schiappa}},\ }\href {\doibase 10.1088/0264-9381/24/2/006} {\bibfield
  {journal} {\bibinfo  {journal} {Classical Quantum Gravity}\ }\textbf
  {\bibinfo {volume} {24}},\ \bibinfo {pages} {361} (\bibinfo {year} {2007})},\
  \Eprint {http://arxiv.org/abs/hep-th/0605001} {arXiv:hep-th/0605001}
  \BibitemShut {NoStop}%
\end{thebibliography}%

\end{document}